\documentclass[iop,twocolappendix]{emulateapj}
\pdfoutput=1
\usepackage{nicefrac}
\usepackage{graphicx}
\usepackage{txfonts}
\usepackage{natbib}
\usepackage[scaled]{helvet}
\usepackage{epsfig}
\usepackage{url}
\usepackage[normalem]{ulem}
\usepackage{color,soul}
\usepackage{ltxtable} 
\usepackage{longtable}
\usepackage{array}
\usepackage{bigstrut}

\bibpunct{(}{)}{;}{a}{}{,}
\newcommand{\vsini}{$v\,{\rm sin}\,i$}  

\newcommand{\WBr}{W$_{\text{Br11}}$}

\begin{document}

\title{\bf Outbursts and disk variability in B\MakeLowercase{e} stars}

\author{
Jonathan Labadie-Bartz$^{1}$,
S. Drew Chojnowski$^{2}$,
David G. Whelan$^{3}$,
Joshua Pepper$^{1}$, 
M. Virginia McSwain$^{1}$,
Marcelo Borges Fernandes$^{4}$,
John P. Wisniewski$^{5}$,
Guy S. Stringfellow$^{6}$,
Alex C. Carciofi$^{7}$,
Robert J. Siverd$^{8}$,
Amy L. Glazier$^{3}$,
Sophie G. Anderson$^{3}$,
Anthoni J. Caravello$^{3}$,
Keivan G. Stassun$^{9}$, 
Michael B. Lund$^{9}$, 
Daniel J. Stevens$^{10}$,
Joseph E. Rodriguez$^{11}$, 
David J. James$^{12}$,
Rudolf B. Kuhn$^{13}$
}

\affil{$^1$Department of Physics, Lehigh University, 16 Memorial Drive East, Bethlehem, PA 18015, USA}
\affil{$^2$Apache Point Observatory and New Mexico State University, P.O. Box 59, Sunspot, NM 88349-0059, USA}
\affil{$^3$Department of Physics, Austin College, 900 N. Grand Ave., Sherman, TX 75090, USA}
\affil{$^4$ Observat\'{o}rio Nacional, Rua General Jos\'{e} Cristino 77, 20921-400, S\~{a}o Cristov\~{a}o, Rio de Janeiro, 20921-400, Brazil}
\affil{$^5$Department of Physics \& Astronomy, The University of Oklahoma, 440 W. Brooks St., Norman, OK 73019, USA}
\affil{$^{6}$Center for Astrophysics and Space Astronomy, Department of Astrophysical and Planetary Sciences, University of Colorado, 389 UCB, Boulder, CO 80309-0389, USA}
\affil{$^{7}$Instituto de Astronomia, Geof\'{i}sica e Ci\^{e}ncias Atmosf\'{e}ricas, Universidade de S\~{a}o Paulo, Rua do Mat\~{a}o 1226, Cidade Universit\'{a}ria, 05508-900 S\~{a}o Paulo, SP, Brazil}
\affil{$^{8}$Las Cumbres Observatory Global Telescope Network, 6740 Cortona Drive, Suite 102, Santa Barbara, CA 93117, USA}
\affil{$^{9}$Department of Physics and Astronomy, Vanderbilt University, Nashville, TN 37235, USA}
\affil{$^{10}$Department of Astronomy, The Ohio State University, 140 W. 18th Ave., Columbus, OH 43210, USA}
\affil{$^{11}$Harvard-Smithsonian Center for Astrophysics, 60 Garden St, Cambridge, MA 02138, USA}
\affil{$^{12}$Department of Astronomy, University of Washington, Box 351580, Seattle, WA 98195, USA}
\affil{$^{13}$South African Astronomical Observatory, P. O. Box 9, Observatory 7935, Cape Town, South Africa}

\shorttitle{Outbursts and Disk Variability in Be Stars}

\begin{abstract}
In order to study the growth and evolution of circumstellar disks around classical Be stars, we analyze optical time-series photometry from the KELT survey with simultaneous infrared and visible spectroscopy from the APOGEE survey and BeSS database for a sample of 160 Galactic classical Be stars. The systems studied here show variability including transitions from a diskless to a disk-possessing state (and vice versa), and persistent disks that vary in strength, being replenished at either regularly or irregularly occurring intervals. We detect disk-building events (outbursts) in the light curves of 28\% of our sample. Outbursts are more commonly observed in early- (57\%), compared to mid- (27\%) and late-type (8\%) systems. A given system may show anywhere between 0 and 40 individual outbursts in its light curve, with amplitudes ranging up to $\sim$0.5 mag and event durations between $\sim$2 and 1000 days. We study how both the photometry and spectroscopy change together during active episodes of disk growth or dissipation, revealing details about the evolution of the circumstellar environment. We demonstrate that photometric activity is linked to changes in the inner disk, and show that, at least in some cases, the disk growth process is asymmetrical. Observational evidence of Be star disks both growing and clearing from the inside out is presented. The duration of disk buildup and dissipation phases are measured for 70 outbursts, and we find that the average outburst takes about twice as long to dissipate as it does to build up in optical photometry. Our analysis hints that dissipation of the inner disk occurs relatively slowly for late-type Be stars.

\end{abstract}


\section{Introduction} \label{sec:Intro}
Classical Be stars are rapidly rotating, non-supergiant, B-type stars that have, or have had, Balmer lines in emission in their spectrum \citep{Jaschek1982}. This emission is attributed to a gaseous circumstellar disk in Keplerian orbit. The disk is formed from material ejected from the star, some of which diffuses outwards. Very rapid stellar rotation significantly lowers the gravitational barrier, but an additional mechanism is required to launch material into orbit \citep{Rivinius2013}. Classical Be stars as a class are pulsators, and there is growing evidence that suggests pulsation plays a key role in the mass-loss mechanism \citep{Rivinius1998, Kee2014, Rivinius2016, Baade2016}.

Discrete mass-loss events are a hallmark of classical Be stars, and are sometimes called `flickers' when they occur on short timescales \citep[$<$50 days;][]{Keller2002}, and are otherwise referred to as `outbursts' \citep[e.g.][]{Grundstrom2011,Rivinius1998}. We make no distinction based on the duration, and refer to all such events as `outbursts.' The signature of an outburst is seen in spectroscopy as a sudden enhancement of emission features, and typically manifests photometrically as a rapid brightening of the system (although, for systems viewed nearly edge-on, outbursts cause a net dimming). These events trace the outflow of material into the circumstellar environment. Once ejected, the material is then governed mainly by gravity and viscosity. The \textit{viscous decretion disk} \citep[VDD;][]{Lee1991,Carciofi2011} model is currently the best framework by which to understand the behavior of these disks once they are formed. Recently ejected material orbits the star, circularizing and diffusing outward through viscous forces and orbital phase mixing. Simultaneously, a majority of the ejected mass falls back onto the star. This matter that falls back supplies the outflowing matter the required angular momentum that allows it to attain progressively wider orbits \citep{Kroll1997,Haubois2012}. Line-driven ablation also potentially contributes to disk dissipation \citep{Kee2016}.

This work is focused on studying the outbursts of classical Be stars. We aim to gain a better understanding of how the circumstellar environment evolves as disks grow and dissipate, and also to understand the general outburst properties of this sample of Galactic Be stars. We do this by analyzing optical light curves, and infrared and optical spectroscopy. From light curve analysis, we measure when outbursts occur, the associated timescales, and their photometric amplitudes. Light curves trace the growth and subsequent dissipation of the inner disk. Hydrogen emission features originating in the circumstellar environment indicate whether or not the star currently possesses a disk. The relative abundance of emitting material can be inferred from the strength of the emission features, and the shape of the line profile traces how the material is distributed in velocity. Near-contemporaneous optical and infrared spectra help us to better understand the radial structure of the disk, since different lines are generally formed at different physical locations in the disk. Monitoring the evolution of these observables reveals information about how the disk is formed and how it evolves over time.

In Sections~\ref{sec:Spec_data} and~\ref{sec:Phot_data} we introduce the surveys, telescopes, and instruments that supply the spectroscopic and photometric data, and describe how the data are analyzed. Section~\ref{sec:formation_location} describes the different regions of the disk probed by these observables. Results are presented in Section~\ref{sec:results}, beginning with a detailed analysis of four individual systems, and concluding with a more global view of outbursts in the sample. The conclusion follows (Section~\ref{sec:conclusion}). Finally, an appendix highlights additional interesting systems.

\section{Spectroscopic Data and Analysis}
\label{sec:Spec_data}
Throughout this work, we use spectroscopy from different telescopes and instruments, and for different purposes. Here we describe the origin of our spectroscopic data, and the methods by which they are analyzed. 

\subsection{The Apache Point Observatory Galactic Evolution Experiment survey}
The Apache Point Observatory Galactic Evolution Experiment (APOGEE) employs a 300 fiber spectroscopic instrument characterized by high resolution ($R$ $\sim$ 22,500), high signal-to-noise ratio (S/N $>$100), and $H$-band near-infrared (NIR) coverage \citep[1.51 - 1.70 $\mu$m;][]{Majewski2015,Wilson2010}. The instrument is attached to the Sloan Digital Sky Survey 2.5 m telescope \citep{Gunn2006}. APOGEE-I, a program in the Sloan Digital Sky Survey III \citep[SDSS-III;][]{Eisenstein2011}, has observed 238 classical Be stars, 128 of which are new discoveries \citep{Chojnowski2015}. The APOGEE Be (ABE) sample includes other types of B-type, emission-line stars (namely, Herbig Ae/Be and B[e] objects), which are not discussed here. We limit our discussion in this paper to only the classical Be stars, hereafter referred to as ``Be stars.'' The APOGEE data presented here is from the twelfth data release of SDSS-III \citep{Alam2015}.

Disk emission features typically have two peaks on either side of the line center, one being blue-shifted (the violet peak), and the other red-shifted (the red peak). These peaks encode information about the kinematic properties of the emitting disk material. The velocity separations of the violet and red emission peaks were measured interactively (visually) for all Be star spectra with well-defined Brackett (Br) series line profiles. In interpreting the profiles of the H-Br lines, the models of optically thin lines from \citet{Hummel1992} were largely relied on. In the case of Be stars with Keplerian disks, measuring the peak separations for optically thin lines gives a lower limit to twice the projected rotational velocity (2~{\vsini}) of the stars \citep{Hummel1994}. For systems with double-peaked line profiles, the systemic radial velocities (RVs) were estimated by measuring the position of both the violet and red peaks, and then taking the average of the two peaks as the line center. This was done for each H-Br line in a given spectrum, and the individual H-Br RVs were averaged to get a single RV for each spectrum. In addition, the equivalent width (EW; defined to be positive in absorption and negative in emission) of the Br11 line (\WBr) was measured via direct summation of a 100 {\AA} window centered on Br11, with typical errors of $\sim$0.32 {\AA}. All of these measurements and more detailed explanations of the methods used are provided in \citet{Chojnowski2017}, including both star-averaged and individual spectrum quantities. In this paper, we focus mainly on the Br11 line (centered at 16811 {\AA}), since it tends to be the strongest in the series, but similar trends are seen in the other Br lines. 

\subsection{APO + ARCES}
Due to a large fraction of the ABE star sample lacking spectral type information in the literature, we obtained high-resolution optical spectra of stars of interest using the Apache Point Observatory (APO) 3.5m telescope and the Astrophysical Research Consortium Echelle spectrograph \citep[ARCES;][]{Wang2003}. In each exposure, the ARCES instrument covers the full optical spectrum (3,500--10,000 {\AA}) at a resolution of $R\sim$ 31,500, recording the light in 107 orders on a 2048x2048 SITe CCD. We used standard echelle data reduction techniques in IRAF\footnote{IRAF is distributed by the National Optical Astronomy Observatories, which are operated by the Association of Universities for Research in Astronomy, Inc., under cooperative agreement with the National Science Foundation.}, including bias subtraction, scattered light and cosmic-ray removal, flat-field correction, and wavelength calibration via thorium-argon lamp exposures. The orders were then continuum normalized, trimmed so as to allow a 10 {\AA} overlap between orders, and merged into a single one-dimensional spectrum. Exposure times were estimated with the goal of achieving an S/N of at least 50 at 4500 {\AA}. The OB spectral atlas of \citet{Walborn1990} provides an appropriate set of standard stars to which ARCES spectra were compared visually.

\subsection{AO long-slit spectrograph}
A number of our stars were also targeted using a low-resolution long-slit spectrograph at Adams Observatory (AO), Austin College. We used a grating with 1200 grooves per millimeter that disperses the light to 0.54 {\AA} per pixel in the wavelength range 3850 - 4950 {\AA}. The slit size is matched to a two pixel width, and the resolution ($\lambda / \Delta \lambda$) varies between 3000 - 4500 across the spectrum.  Data reduction procedures were written in Python, and are explained in depth in \citet{Whelan2017}.

\subsection{The Be Star Spectra (BeSS) database}
The BeSS database\footnote{http://basebe.obspm.fr} is a catalog that not only attempts to include all known Be stars, but also contains spectra for about half of the Be stars listed therein \citep{Neiner2005}. Dozens of observers, both professional and amateur, have collectively submitted over 100,000 spectra to the BeSS database. These data come from a large variety of telescopes and instruments, and are therefore of inhomogeneous quality, depending on the expertise of the observer and the equipment used. However, each spectrum is subject to a quality check for format and scientific validity by the BeSS administrators before being incorporated into the database. Although any wavelength regime is allowed, optical spectra are by far the most common type of submission. When available, we consider spectra from the BeSS database in our analysis. In particular, we focus on the H$\alpha$ line, which is a popular observable of Be stars, and is covered in a majority of BeSS spectra. This informs us about the status of these disks at different epochs.

\subsection{Optical spectroscopy and estimating spectral type}
Whenever possible, we incorporate optical spectroscopy into our analysis. This allows us to estimate spectral types and to monitor the status of the disk at different epochs and at different wavelengths. This is especially informative when the $H$-band spectra, optical spectra, and photometry have overlapping epochs.

With the exception of B and Be stars hotter than B1, for which the presence and strength of He~{\sc ii}~$\lambda$~4685 and the ratios of Si~{\sc iii}/Si~{\sc iv} are used for spectral typing, spectral classification of the majority of B and Be stars relies on the relative strengths of the He~{\sc i} lines versus those of Mg~{\sc ii}~$\lambda$~4481 and Si~{\sc ii}~$\lambda$~4128-4130. For instance, the relative strengths of the He~{\sc i}~$\lambda$~4471 {\AA} and Mg~{\sc ii}~$\lambda$~4481 {\AA} lines offer a relatively good proxy for temperature, although rotation is known to play a role in interpreting their relative strengths \citep[{\it e.g.},][ and references therein]{Gray2009}. Luminosity class is largely determined by the widths of the hydrogen Balmer and metallic absorption lines. Spectral classifications of stars are historically done by comparison to a set of spectra of known spectroscopic standard stars. \citet{Morgan1973}, for example, provides one of the most complete sets of spectroscopic standard stars. A number of complications arise when assigning spectral classifications to Be stars. They are very rapidly rotating and therefore have broad spectral features. This introduces difficulties, especially since rapid rotation can alter the relative depths of lines with different intrinsic widths \citep{Gray2009}. Rapid rotation adds further complications besides line broadening. Be stars bulge outward near the equator (due to their rapid rotation) and therefore have a substantially higher surface gravity and temperature at the poles compared to the equatorial region. The inclination angle of the star then influences perceived line strengths. Line damping is yet another issue adding to the difficulty of classifying Be star spectral types. This effect arises from the filling in of absorption lines due to flux from the circumstellar disk, making the absorption lines appear weaker than they actually are. Furthermore, photospheric lines can have significant contributions from the disk, in addition to the continuum line damping. As the amount of material in the disk is often varying, so too do these effects change over time. 

Because of these difficulties, spectral types for Be stars must be considered carefully. All stars for which we present new spectral classifications with temperature hotter than B1 have clear He~{\sc ii}~$\lambda$~4685 absorption line detection, as well as luminosity-sensitive lines like O~{\sc ii}. We expect our new classifications to be accurate to within $\pm$0.5 in temperature class for stars earlier than B2, thanks in part to features such as the Si~{\sc iii} lines at 4552, 4567, and 4571 {\AA}. For stars later than B2, an uncertainty of $\pm$1 in spectral type is typical, but some cases (\textit{e.g.} shell stars) have larger uncertainties of $\pm$2. These uncertainties are appropriate for the newly reported spectral types presented here, but it is also prudent to apply similar levels of caution to spectral types reported in the literature for Be stars, especially when these are done in any sort of automated way.

Considering the uncertainties in the reported spectral types, it is useful to adopt coarse bins in stellar temperature. For stars that have not yet been spectral typed in this work, a designation from the literature is adopted. Following the convention of \citet[][hereafter ``LB17'']{Labadie-Bartz2017}, we consider ``early-type" Be stars as those with spectral types earlier than B4, ``mid-type" Be stars have spectral types including B4, B5, and B6, and ``late-type" Be stars have spectral types including B7 and later. Stars without a specific spectral type (e.g. a spectral type of ``Be'') are considered ``unclassified.'' Despite the difficulties in assigning a specific temperature class to Be stars, they are still reliably cast into these ``early-,'' ``mid-,'' and ``late-type'' designations (although we can only be certain of this for stars newly spectral typed in this work).

\section{Photometric Data and Analysis}
\label{sec:Phot_data}
The Kilodegree Extremely Little Telescope (KELT) is a photometric exoplanet transit survey using two small-aperture (42 mm) wide-field (26$^{\circ}$ x 26$^{\circ}$) telescopes, with a northern location at Winer Observatory in Arizona in the United States, and a southern location at the South African Astronomical Observatory near Sutherland, South Africa. The KELT survey covers over 70$\%$ of the sky, obtaining photometric precision of $\sim$1$\%$ for $\sim$4.4 million objects in the magnitude range of $7 < V < 13$ with baselines of up to 10 years. 

The effective passband that KELT uses is roughly equivalent to a broad $R$-band filter. KELT light curves use JD (TT) as the time system. The long baseline combined with a typical cadence of 30 minutes and high photometric precision makes the KELT dataset a valuable resource for studying variable stars across a range of timescales and magnitudes. This is particularly true for Be stars, which are variable on timescales from hours to decades. See LB17 for more information about the utility of KELT light curves for studying Be stars, and \citet{Pepper2007,Pepper2012} for more details about the KELT survey.

The large pixel scale of KELT ($\sim$23$^{\prime\prime}$) can result in multiple objects casting light onto the aperture used to extract a light curve for a given target, especially in crowded fields in the Galactic plane. This can have two different effects. One is that contaminating sources may broadcast a signal that is then expressed in the light curve of the target object, despite this signal having no origin in the target. The other is that contaminating light from nearby sources will act to dampen genuine signals that originate in the target. Despite being in or near the Galactic plane, the majority of our targets (including all of those discussed in detail) are significantly brighter than any nearby neighbors. Also, the photometric signature of outbursts is not easily imitated by impostors. We make the assumption that the outbursts we detect are attributed to the Be star target in all cases. All stars in this sample are confirmed to be within the range of spectral types where the Be phenomenon is found, and are confirmed to have emission consistent with classical Be stars (seen in their APOGEE spectra). The exception are the `quiescent' systems (labeled like ``ABE-Q01''), which are previously known Be stars that show no signs of emission in their APOGEE spectra. This is not to say that these quiescent systems are not Be stars, but just that they do not happen to have a detectable disk at the times they were visited by APOGEE. The main consequence of the large pixel scale of KELT is the dilution of the amplitude of the Be star variability, caused by the presence of nearby (and typically much fainter) sources. Inspection of the sky in the vicinity of each of our targets reveals that this contamination is a minor issue in the large majority of cases. The most severe consequence of this might be the suppression of outburst amplitudes to the point of non-detection, but this is a realistic scenario in only a small number of particularly contaminated sources. Although it is difficult to exactly quantify the degree of dilution, this effect is small in practice, since nearly all of the active, outbursting Be stars in this sample are only marginally blended. In other words, we assume that all detected outbursts are attributed to the target Be star, and that the photometric amplitudes are only minimally suppressed.

The catalog of ABE stars was cross-matched to the KELT catalog, and 171 Be stars were found to exist in both data sets. However, five of these light curves have saturation effects that can greatly complicate the analysis, and an additional six have been found to not be classical Be stars \citep{Chojnowski2017}. Our sample then consists of 160 Be stars, all of which have KELT light curves and multiple APOGEE spectra, with about half also having optical spectra. Of these, 120 are observed by KELT-North, 33 are observed by KELT-South, and 7 are observed by both KELT-North and KELT-South (the joint field J06). These light curves are available for download in machine-readable format in the .tar.gz package

Outbursts are a common feature in the light curves of Be stars and are typically characterized by a monotonic increase in the brightness of the system (the `rising phase,' where the disk grows), followed by a decay back toward baseline (the `falling phase,' where the disk dissipates). \citet{Rivinius1998} and \citet{Huat2009} refer to these as the `outburst' and `relaxation' phases. In our terminology, these two distinct phases comprise a single outburst. In shell Be stars (which are simply Be stars viewed nearly equator-on), this trend is reversed, and an outburst begins with a fading of the system, followed by a return toward baseline. This reversal is solely a consequence of the inclination angle \citep{Haubois2012,Sigut2013}. At low to intermediate inclination angles, the presence of a disk increases the brightness of the system. At high inclination angles (nearly equator-on), a disk will cause the net brightness of the system to decrease, as stellar flux in the optical is both absorbed and scattered out of our line of sight \citep{Haubois2014}. At an inclination angle of $\sim70^{\circ}$, there is effectively no net change in the optical continuum flux during an outburst, as the disk absorbs and scatters approximately the same amount of flux that it emits along our line of sight \citep{Haubois2012}.

\begin{figure}[!ht]
\centering\epsfig{file=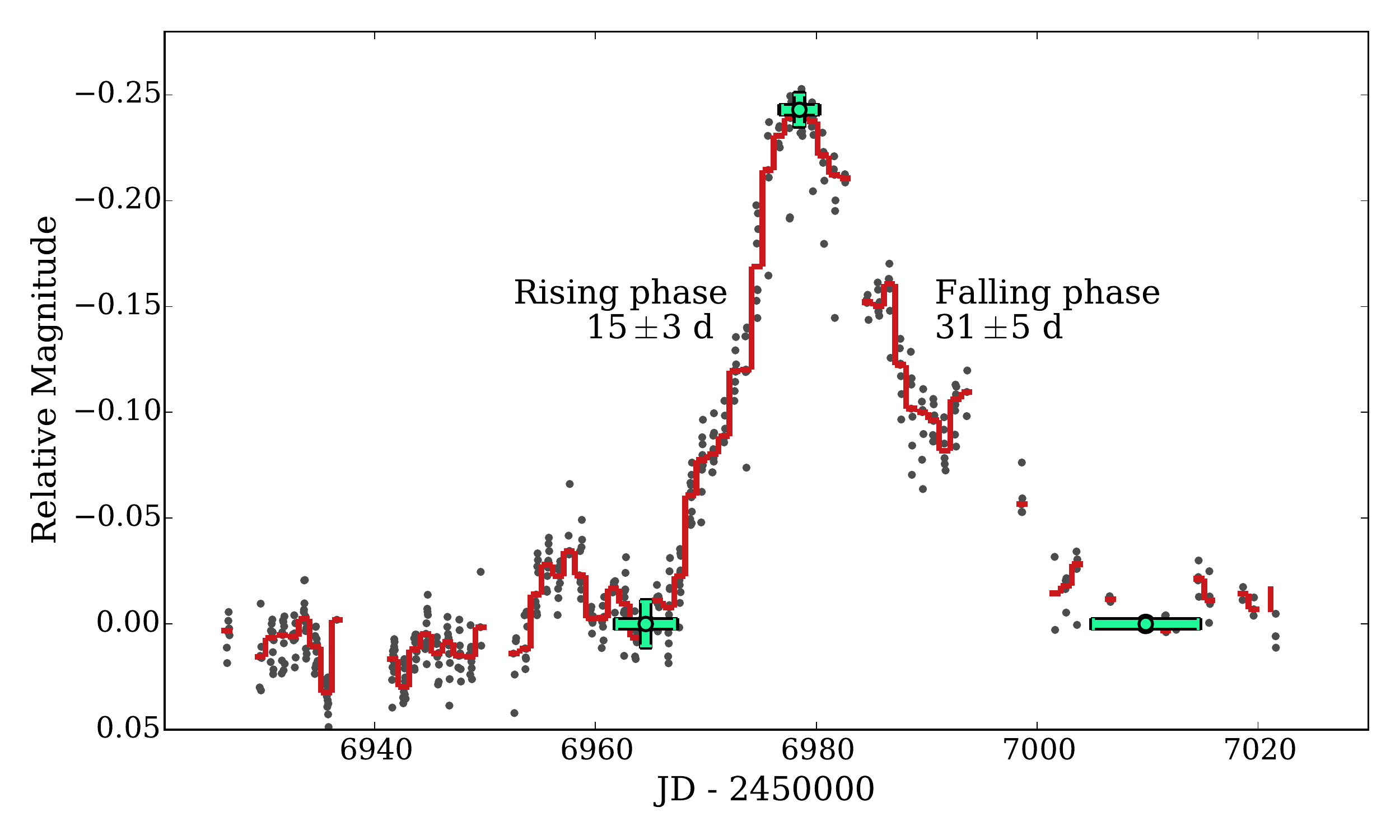,clip=,width=0.99\linewidth}
\caption{A typical outburst, as seen in photometry for the star ABE-105. The rising phase begins at JD=2456964 $\pm$ 3, and continues to rise up to a maximum brightness at JD=2456979 $\pm$ 2, 15 days later. Then, the falling phase ensues, lasting 33 days until JD=2457010 $\pm$ 5. The system brightens by a maximum of 0.246 $\pm$ 0.013 magnitudes. The green points and error bars represent the adopted values and uncertainties in these measurements.}
\label{fig:Outburst_example}
\end{figure}

Light curves for all stars in this sample were inspected visually for the presence of outbursts, which are identified by their morphology. The number of discrete outburst events for each object is tallied whenever tractable, and is used to calculate the rate at which outbursts occur for a given system (outbursts per year). We also measure the duration of the initial rising phase and the subsequent falling phase, as well as the photometric amplitude of the event, by a close visual inspection of the light curve. This can only be done for `well-behaved' outbursts that are sufficiently sampled by KELT observations. It is sometimes unclear exactly when an outburst starts, reaches peak brightness, and ends. We account for this by including uncertainties, which again are measured visually. Figure~\ref{fig:Outburst_example} shows how a typical outburst appears in a KELT light curve. The duration of the rising and falling phases are shown, and the amplitude is clear. The error bars represent the uncertainty in the timing and amplitude. These uncertainties vary from event to event, but the average values are about 4 and 7 days for the duration of the rising and falling phases, respectively, and about 0.015 mag in amplitude.

\section{Formation loci of observables} \label{sec:formation_location}
\begin{figure}[!ht]
\centering\epsfig{file=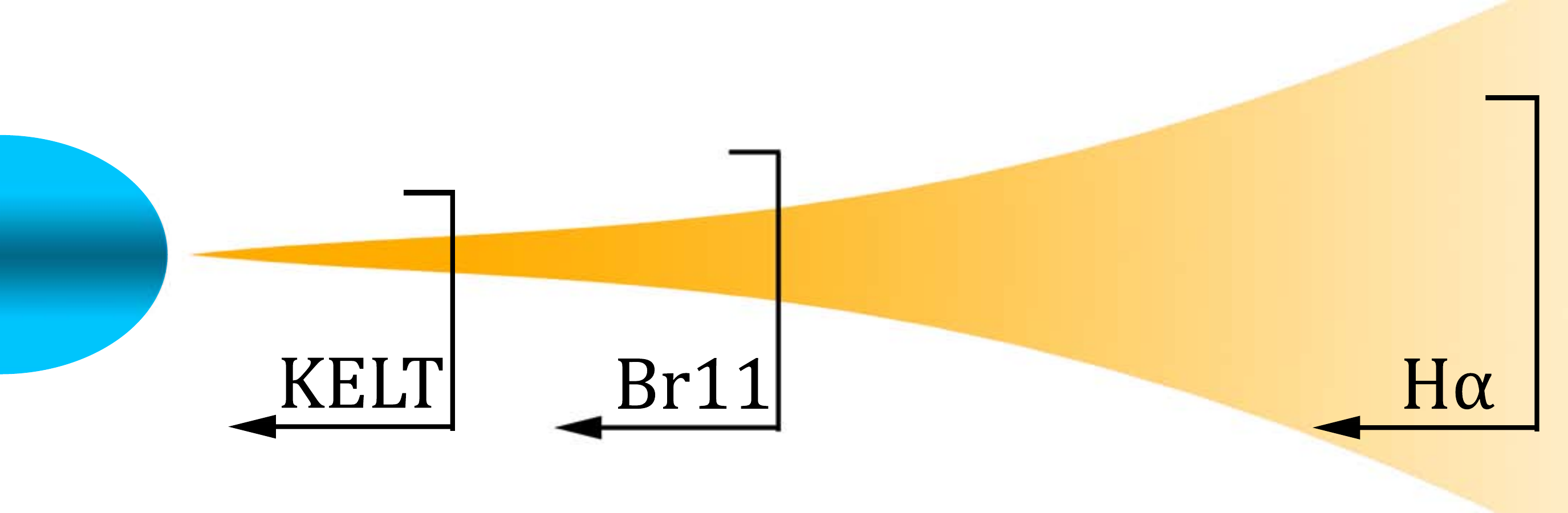,clip=,width=0.99\linewidth}
\caption{Simplified schematic view of a Be star with a flared disk. The approximate extent of the regions from which our three main observables arise are marked.}
\label{fig:Formation_loci}
\end{figure}

Throughout this work, we deal mainly with three different observables -- visible continuum photometry (KELT), the Brackett series in the NIR (APOGEE), and visible spectroscopy (mainly the H$\alpha$ line; ARCES, AO, and BeSS). These three observables are sensitive to different parts of a Be star disk. This idea, and relevant model predictions, is presented and discussed in \citet{Carciofi2011}, which serves as a useful reference for estimating the disk regions probed by the observables used in this work. KELT photometry primarily probes the inner $\sim$1--2 R$_{*}$ of the disk, as measured out from the stellar equator. The NIR Br11 line probes the disk at larger radii, out to $\sim$2--6 R$_{*}$ \citep[][]{Chojnowski2015}. H$\alpha$ traces an even larger area of the disk, out to $\sim$5--15 R$_{*}$ \citep{Rivinius2013}, or greater. \citet{Slettebak1992} find that H$\alpha$ emission arises in the range of 7--19 R$_{*}$, on average. Observations of Be stars at longer wavelengths (\textit{e.g.} millimeter or radio) reveal disks that extend out to many tens or hundreds of stellar radii \citep[\textit{e.g.}][]{Klement2017}. These more extended regions of the disk are largely inaccessible to the modes of observation used in this work. The exact extent that our observables probe depends on many factors, including the stellar flux, inclination angle, and the distribution of material in the disk (which varies with time). Despite these complications, it remains generally true that the KELT, Br11, and H$\alpha$ observations probe what we refer to as the ``inner,'' ``mid,'' and ``outer'' disk areas. We stress that these regions are not rigidly defined, as applied in this work. However, they are useful constructs when considering different types of data taken at similar times for a given system. This scheme is qualitatively illustrated in Figure~\ref{fig:Formation_loci}. 

For all non-shell systems with both optical and NIR spectra, there is a greater separation between the violet and red peaks in the Br11 line compared to H$\alpha$. Multiple factors contribute to this. When part of a disk is optically thick in some line (which is not uncommon for H$\alpha$), non-coherent scattering broadening can act to decrease the peak separation \citep{Hummel1994}. This effect influences both the peak separation and the emission-line profile. The orbital velocity of particles in the disk decreases with distance, so emission originating at larger disk radii will also contribute to a smaller peak separation.

Variability in the disk tends to occur most rapidly in the innermost regions, with timescales increasing with radius. In part because H$\alpha$ probes a much greater area of the disk relative to the other observables, the H$\alpha$ line often exhibits the most dramatic disk signatures, with emission features sometimes exceeding 10 times the continuum level.

\section{Results} \label{sec:results}
First, we discuss a few specific systems where spectroscopic observations coincide with photometric outbursts, revealing information about these episodes through their variability. Additional illustrative examples are included in the appendix. We then look at the general properties of outbursts across the entire sample. 

\subsection{Individual systems} \label{sec:indv_otbs}
There are four systems (ABE-098, -138, -A01, and -026) for which we provide a detailed analysis. For these stars, multiple APOGEE spectra are collected immediately before, during, or shortly after an outburst as seen in the KELT data, giving us snapshots of the circumstellar environment, with additional context provided by the light curves. Multiple archival spectra covering the H$\alpha$ line are available from the BeSS database for ABE-138, -A01, and -026, giving us even more information about the changing circumstellar environment over a long baseline. Considering the different preferential formation loci of the various observables (as discussed in \S \ref{sec:formation_location}), a picture emerges of the disk both growing and dissipating from the inside outward.

\subsubsection{ABE-098 ($=$ BD$+$63 1955 $=$ HD 219523)} \label{sec:indv_otbs_098}
\begin{figure}[!ht]
\centering\epsfig{file=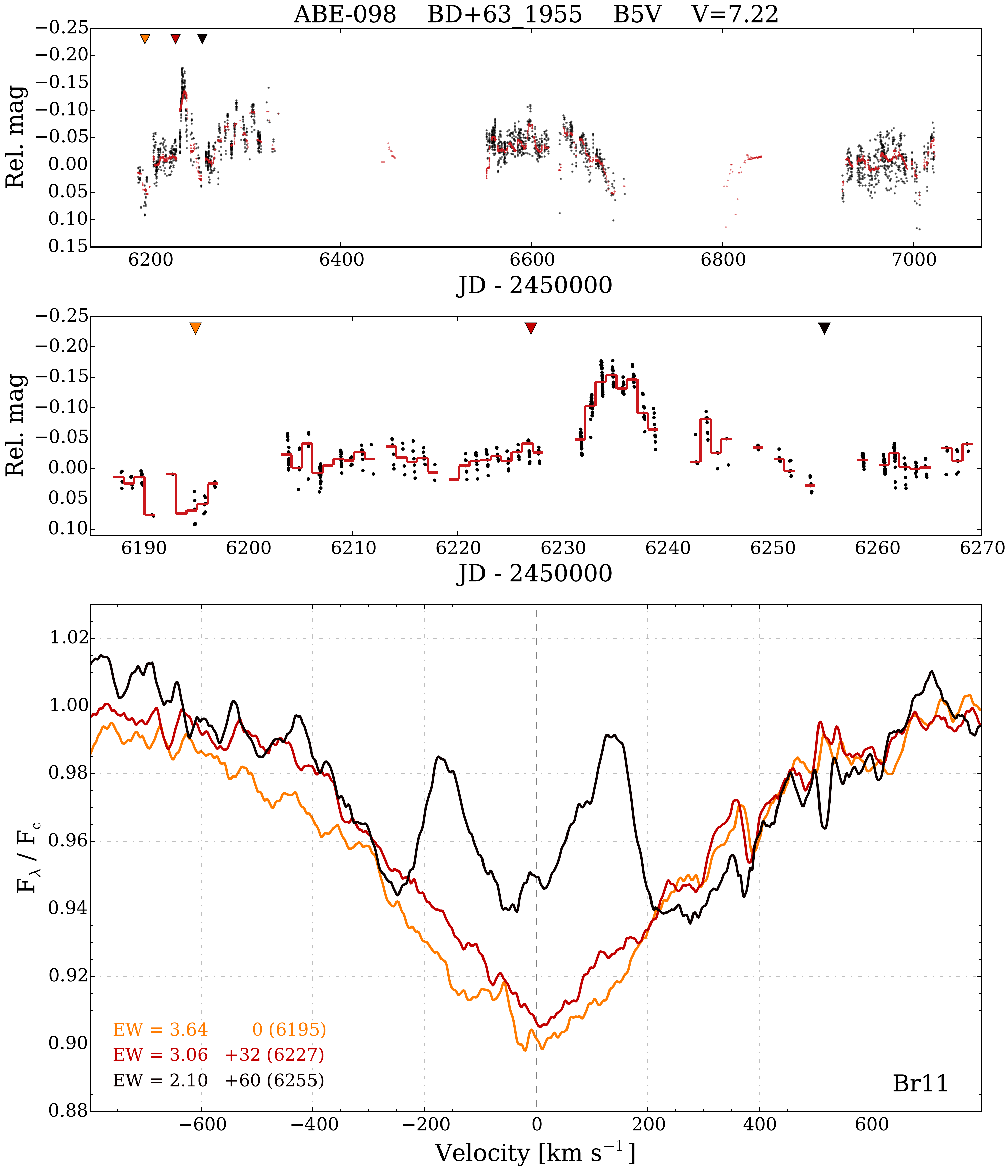,clip=,width=0.99\linewidth}
\caption{Top: full KELT light curve (black), with red points showing the data after applying a low-pass filter. Middle: zoom-in of an outburst. The red lines show the nightly median. Bottom: APOGEE spectra, showing the Br11 line. Triangles in the upper two panels indicate epochs of APOGEE observations. We estimate a rising time of 6 $\pm$ 2 days, a falling time of 14 $\pm$ 4 days, and an amplitude of 0.14 $\pm$ 0.018 mag for this outburst.}
\label{fig:Outburst_098}
\end{figure}

ABE-098 has a spectral type of B5V (this work, from an ARCES spectrum). This is a system with two APOGEE spectra taken prior to the onset of a photometric outburst and one spectrum taken just following the falling phase. Figure~\ref{fig:Outburst_098} shows the KELT light curve in the top panel, with a zoom-in on the region surrounding the outburst shown in the panel below. Downward-pointing triangles indicate epochs of the three APOGEE spectra, and the corresponding Br11 lines are shown in the bottom panel. These spectra bracket the outburst nicely, giving us information about the circumstellar environment both before the rising phase and after the falling phase. The first spectrum has \WBr = 3.64 {\AA}, and does not show obvious double-peaked emission at JD$_{0}$ = 2456195. There is a slight bump just to the violet side of the absorption core, which may be caused by noise, pulsation, or some circumstellar material, but there is no substantial Br11-emitting disk at this time. The next spectrum (JD$_{0}$ + 32 days) also shows no evidence of a disk. There is then a four-day observing gap, followed by a photometric outburst. At the epoch of the third spectrum, JD$_{0}$ + 60 days, the Br11 line shows a very clear disk signature, with \WBr = 2.10 {\AA} (and $\Delta v_p$ = 304 km s$^{-1}$). At this point, the brightness of the system in the KELT passband has relaxed back to baseline.

This sequence demonstrates that photometric outbursts correspond to the injection of stellar material into the circumstellar environment, some of which settles into a disk. It also provides evidence for the `inside-out' clearing of Be star disks that is predicted by the VDD model. Since the brightness of the system in the KELT bandpass has returned to baseline (within observational errors) by the epoch of the final spectrum, there is no substantial inner disk at this time (otherwise there would be some photometric excess). Yet, the emission feature unambiguously shows the presence of a disk. Since the Br11 line probes the disk out to greater radii than does the KELT optical continuum photometry, the evidence for a disk in the Br11 line and absence of a photometric excess in KELT data implies that the inner-most region of the disk is sparse, while some farther out Br11-emitting material remains. The Br11 emission in the final spectrum shows a violet-to-red peak (V/R) ratio that is slightly less than unity, suggesting some asymmetry in the disk. This may indicate that the circumstellar material has not yet been thoroughly mixed. 

Put simply, prior to the outburst, there is no disk. During the rising phase, the inner disk grows. The falling phase shows the inner disk dissipating. During the falling phase, some amount of material has migrated radially outward and is seen in the Br11 emission feature.

\subsubsection{ABE-138 ($=$ V1448 Aql $=$ HD 180126)}
Similar to the previous example, this Be star has multiple spectra in the vicinity of an outburst. Photometry shows an outburst with a short two-day rising phase, which is sampled by APOGEE spectra at the onset of the rising phase, and at peak brightness. These spectra give us a glimpse into the changes in the circumstellar environment that accompany the rising phase of this outburst.

\begin{figure}[!ht]
\centering\epsfig{file=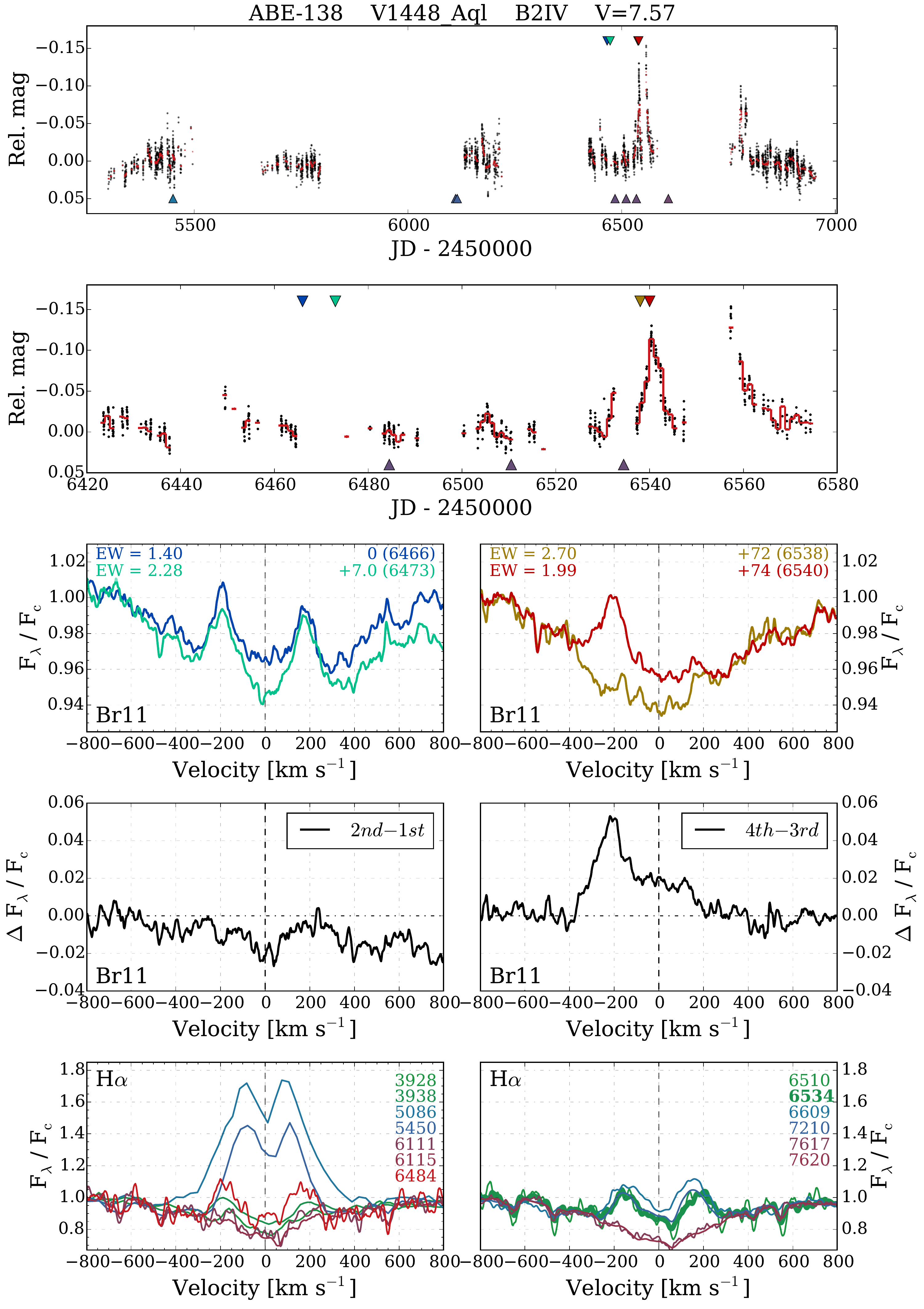,clip=,width=0.99\linewidth}
\caption{First row: raw KELT light curve in black, with red points showing the data after applying a low-pass filter. Colored downward-pointing triangles correspond to epochs of APOGEE observations, and upward-pointing triangles indicate BeSS observations. Second row: zoom-in on the region above, highlighting the outburst near JD-2450000 = 6540. The red lines show the nightly median of the photometric data. Third row: the Br11 line of the first (left) and second (right) groupings of APOGEE spectra. The value of \WBr is shown in the upper-left corner. The number of days since the first APOGEE spectrum is printed in the upper-right corner, with the JD-2450000 date in parentheses. Fourth row: difference spectra between the first (left) and the final (right) pairs of APOGEE spectra. Fifth row: H$\alpha$ spectra from the BeSS database, with the JD-2450000 dates in the upper right. }
\label{fig:Outburst_138}
\end{figure}

Stellar parameters for ABE-138 are described by \citet{Fremat2006}, who found $T_{\rm eff}$ = 20,000 $\pm$ 1,500 K, log~$g$ = 3.80 $\pm$ 0.10 (c.g.s), {\vsini} = 243 $\pm$ 20 km s$^{-1}$, an inclination angle between 39$^{\circ}$ and 60$^{\circ}$, and a spectral type of B2 IV (nothing in our AO spectrum suggests that this estimate is inaccurate). The full KELT light curve is shown in the upper panel of Figure~\ref{fig:Outburst_138}, with downward-pointing triangles indicating the epochs of the four APOGEE observations, which occur in two groupings. Upward-pointing triangles indicate epochs of H$\alpha$ measurements from BeSS spectra. In the next panel, a portion of the light curve that includes the outburst near JD-2450000 = 6540 is shown. This baseline also includes four APOGEE and three BeSS spectra. In the next row, the left (right) panel shows the Br11 lines from the first (second) grouping of APOGEE spectra, with the difference spectra plotted in the row below. The bottom panels show the H$\alpha$ line from 13 BeSS spectra, spanning about 10 years, and with emission that varies in strength and sometimes disappears completely. Multiple outbursts are apparent in the light curve, and three of the four APOGEE spectra, and 8 out of the 13 BeSS spectra, show emission. This is a classical Be star at an intermediate inclination angle, with mass-loss episodes that are irregularly spaced and of varying amplitudes, and a disk that appears and disappears, and varies in strength.

The first group of APOGEE spectra (at JD$_{0}$ = 2456465, and JD$_{0}$ + 7 days) shows clear double-peaked Br11 emission, indicating the presence of a disk at these epochs. There is no obvious photometric excess at this time. It is possible that the Br11-emitting disk can be traced back to a recent outburst which may have occurred some time during the observing gap between JD-2450000 = 6440 -- 6450. The photometry immediately after this gap shows a slight brightness enhancement before returning to baseline, possibly suggesting the tail end of an outburst. The majority of the change in \WBr~ is likely due to continuum normalization issues (particularly on the red side of the line), but the decreasing emission in the line core, and the change in the V/R ratio (from V/R $>$ 1 to V/R $\approx$ 1) is most likely real. The mean peak separation in these two measurements is 369 km s$^{-1}$.

The second grouping of spectra is valuable, as both are collected during the rising phase of an outburst. The first of these, taken at JD$_{0}$ + 72 days, shows only a very weak disk signature with \WBr = 2.70 {\AA}. Two days later, at the peak brightness of the outburst (JD$_{0}$ + 74  days), there is clearly emission, and \WBr = 1.99 {\AA}. Using the values for \WBr ~and magnitude at the times of the final two spectra, we calculate $\Delta$\WBr/$\Delta t$ = -0.36 {\AA} d$^{-1}$, and $\Delta$mag/$\Delta t$ = -0.04 mag d$^{-1}$. It is possible that the strength of the Br11 emission in the final spectrum is somewhat suppressed due to line damping.

The circumstellar environment probed by the Br11 line is highly asymmetric at the epoch of the final APOGEE spectrum, with 84.3$\%$ (15.7$\%$) of the enhancement between the final two APOGEE spectra originating from the violet (red) side. Rapid cyclic variations in the ratio of the strength of the violet and red peaks (V/R) are seen in other classical Be stars during outbursts \citep[\textit{e.g.} $\mu$ Cen;][]{Rivinius1998}, which can be explained if the outflow of material is not axisymmetric. The so-called \v{S}tefl frequencies are sometimes detected during outburst events and are interpreted as tracing large-scale gas- ircularization \citep[e.g.][]{Stefl2000,Baade2016}. Through shearing and viscosity, a non-axisymmetric outflow will evolve toward a symmetric disk over time, according to the VDD model. 

From the eight H$\alpha$ measurements with well-defined emission peaks, we measure the peak separation and find the mean value for $\Delta v_p$ = 297 km s$^{-1}$, with a standard deviation of 66 km s$^{-1}$. This is lower than the mean peak separation of the Br11 line, where $\Delta v_p$ = 369 km s$^{-1}$. Relative to Br11, the smaller peak separation for H$\alpha$ is likely caused by H$\alpha$-emitting material at larger radii, having a relatively slow orbital velocity. Non-coherent scattering broadening may also play a role if a portion of the disk is optically thick to the H$\alpha$ line at the epoch of any BeSS observations.

One H$\alpha$ spectrum is made bold in Figure~\ref{fig:Outburst_138}. With an epoch of JD-2450000 = 6535, this spectrum is taken just four days prior to the third APOGEE spectrum. It is clear that a typical disk signature is seen in H$\alpha$ at this epoch, but only a very weak (if any) disk signature exists in the Br11 line four days later. This is naturally explained by the VDD model, which predicts Be star disks both growing, and dissipating, from the inside outward. Between the first and third APOGEE spectra, the ``inner'' to ``mid'' disk has dissipated (to the point of non-detection in KELT and APOGEE data), but the ``outer'' disk remains largely intact, as evidenced by the three H$\alpha$ double-lined emission profiles observed in this time span (at JD-2450000 = 6484, 6510, and 6534). The ``outer'' disk also eventually dissipates, as there is no sign of H$\alpha$ emission at JD-2450000 = 7617.

\subsubsection{ABE-A01 ($=$ MWC 5 $=$ BD+61 39)} \label{sec:A01}
\begin{figure}[!ht]
\centering\epsfig{file=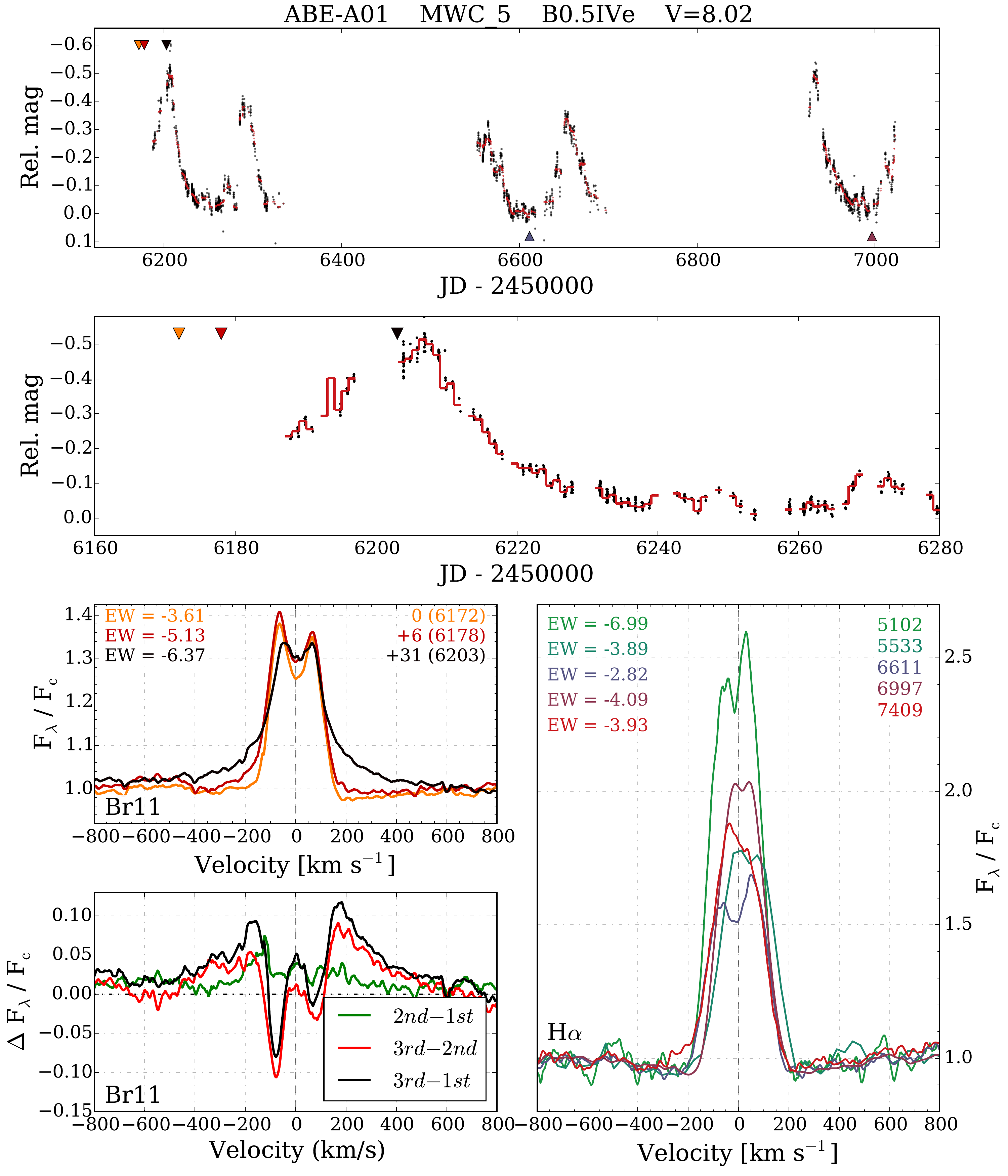,clip=,width=0.99\linewidth}
\caption{Top: raw KELT light curve, with downward (upward) pointing triangles indicating epochs of APOGEE (BeSS) observations. Middle: zoomed-in view of the first outburst. Bottom: The left two panels show the Br11 line of the three APOGEE spectra, with colors corresponding to the epoch of observation and the colored triangle markers in the light curve plots (upper), and the differences between these (lower). The right panel shows the H$\alpha$ line from five BeSS spectra.}
\label{fig:Outburst_A01}
\end{figure}

Like the previous example of ABE-138, APOGEE observed ABE-A01 during the rising phase of an outburst. \citet{Morgan1953} assign this star a spectral type of B0.5IV (nothing in our AO spectrum suggests this is inaccurate). Figure~\ref{fig:Outburst_A01} shows the full KELT light curve, a zoomed-in view of the outburst and the epochs of APOGEE observations, the Br11 line profile of the three APOGEE spectra and their differences, and five H$\alpha$ spectra from the BeSS database. In all spectra, from both APOGEE and BeSS, we infer the presence of a disk, even when spectra are taken near photometric minimum (e.g. H$\alpha$ at JD-2450000 = 6611 and 6997). 

As the rising phase of the first outburst progresses and the system becomes brighter, there is a growing amount of emitting material, as evidenced by the increasingly negative values for the Br11 line EW (\WBr = -3.61, -5.13, and -6.37 {\AA}, in chronological order). The central depression partially fills in, and the line profile edge becomes less sharp. The bulk of the increasing emission arises in the wings of the line profile. The growing emission wings can be attributed primarily to electron scattering. As the density in the inner disk rises, an increase in the electron scattering of line photons causes the emission wings to grow, an effect that becomes stronger as the amount of circumstellar material (and free electrons) increases. The rising brightness in the light curve likewise indicates a growing inner disk.

A decrease in peak separation ($\Delta v_p$ = 128, 130, and 108 km s$^{-1}$, in chronological order) is seen in the final Br11 line. This may be explained in part by the outer edge of the Br11-emitting disk moving out to larger radii, where material is orbiting at lower velocities. Also, as the disk builds up and becomes more dense, an increase in the optical thickness in the Br11 line may act to decrease the peak separation through non-coherent scattering of line photons. 

Because the strength of the Br11 line is measured relative to the local continuum flux, an increase in the continuum level will serve to suppress the apparent strength of the line emission. Because of this, it is likely that the Br11 line is increasing in absolute strength more dramatically than it is increasing in its strength relative to the continuum flux (which is what is plotted in Figure~\ref{fig:Outburst_A01}). Regardless of this effect, the increase in the emission wings still indicates a growing inner disk.

By taking the difference between spectra (lower-left panel of Figure~\ref{fig:Outburst_A01}), we can more clearly see how the Br11 emission-line profile is changing. Beyond highlighting the growth of the wings, the difference spectra allow us to compare the contributions from the red and violet halves of the line profile. Considering the difference between the second and first spectra, 57.7\% of the enhancement in \WBr ~comes from the violet half of the line, and 42.3\% from the red half. In the difference between the third and first spectra, 46.3\% of the \WBr ~enhancement comes from the violet half of the line, and 53.7\% from the red half. Between the third and second spectra, 32.0\% of the \WBr ~enhancement comes from the violet half, and 68.0\% from the red half. Asymmetry in the changes of the Br11 line implies asymmetries in the disk as it grows during this rising phase.

\begin{figure}[!ht]
\centering\epsfig{file=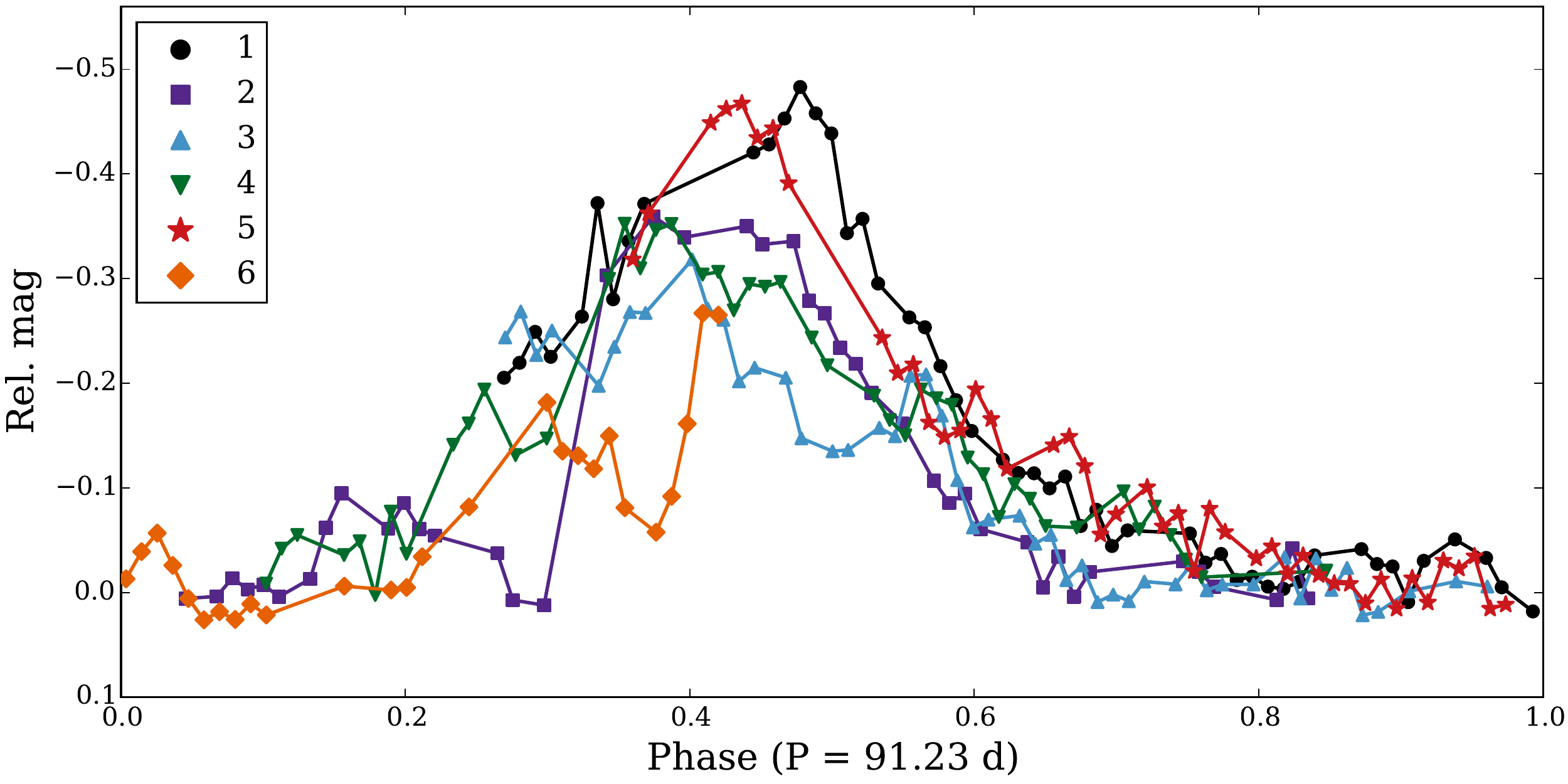,clip=,width=0.99\linewidth}
\caption{Light curve for ABE-A01 phased to a period of 91.23 days, showing that the outbursts occur with some regularity. Markers indicate the nightly median magnitude after outlier removal. The different colors and markers correspond to the six individual outbursts seen in the raw light curve in Figure~\ref{fig:Outburst_A01}.}
\label{fig:SRO_A01}
\end{figure}

There are six discrete high-amplitude outbursts in the KELT light curve (see the top panel of Figure~\ref{fig:Outburst_A01}), but not all of these are sampled fully. The beginning of the rising phase is missing for the first, third, and fifth outbursts, and the final outburst is only observed during part of the initial rising phase. There is also other variability interspersed, with shorter timescales and lower amplitudes. These six major outbursts occur with some regularity, although they are not strictly periodic, varying somewhat in their amplitude and morphology. Their similarities in shape and timing are apparent when the light curve is phased to a period of 91.23 days, as shown in Figure~\ref{fig:SRO_A01}.

Prompted by the interesting show of repeating outbursts, the photometric data for ABE-A01 were subjected to a frequency analysis, in order to search for signals with periods shorter than three days. This upper limit on the periodic signals of interest was chosen based on the typical pulsational periods of Be stars, and the timescales associated with Be star rotation and Keplerian orbits in the region of the circumstellar environment in which KELT photometry is sensitive to \citep{Rivinius2013}. This process requires first removing the high-amplitude variability on longer timescales (including the six major outbursts), which dominate the light curve. In the same manner employed in \citet{Rivinius2016}, a Fourier-based high-pass filter was applied to the photometry, iteratively identifying and removing low-frequency sinusoidal signals. This process results in a detrended light curve with all long-term trends removed, suitable for recovering signals with periods less than three days. The entire detrended light curve was analyzed for periodic signals with a generalized Lomb-Scargle (LS) periodogram \citep{Press1992,Zechmeister2009}, as implemented in the \textsc{Vartools} light curve analysis package \citep{Hartman2012}. 

\begin{figure}[!ht]
\centering\epsfig{file=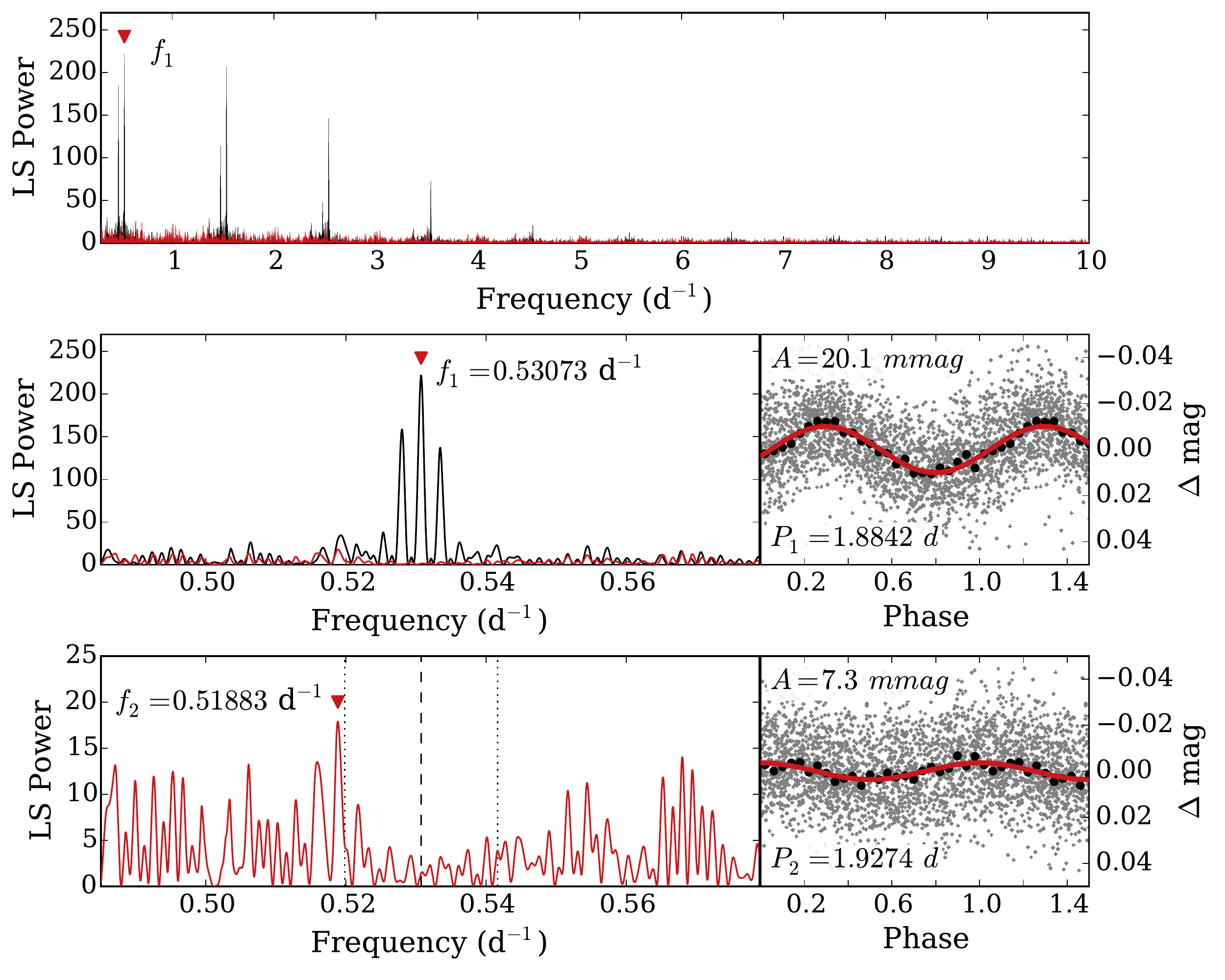,clip=,width=0.99\linewidth}
\caption{Frequency analysis for ABE-A01, after removal of low-frequency variability. In each LS periodogram (top and left two panels), the black curve shows the periodogram, and the red curve shows the periodogram after pre-whitening against the top peak ($f_1$). Top: LS periodogram between 0.3 and 10 day$^{-1}$. Middle: zoomed-in view of the top peak (left) and the light curve phased to this period (right). The gray `+' signs show the KELT data, larger black circles show the data median-binned (with 25 bins in phase), and the red curve is a sinusoidal fit to the median-binned data. Bottom: zoomed-in view of the periodogram in the vicinity of $f_1$ after pre-whitening against $f_1$ (the position of which is shown as a vertical dashed line). The two dotted lines show the positions where we might expect to see another peak, if in fact the outbursts in this system are modulated by a delta frequency. The photometry is then phased to the strongest peak that exists in the vicinity of these two predicted frequencies. It is unclear if this peak is caused by genuine astrophysical variability, or is just a spurious peak caused by noise in the data and/or the sampling rates of KELT.}
\label{fig:Freq_A01_new}
\end{figure}

The results of this analysis are shown in Figure~\ref{fig:Freq_A01_new}. The top panel shows the LS periodogram (black curve) out to a frequency of 10 day$^{-1}$. Higher frequencies were probed as well (up to 500 day$^{-1}$), but the periodogram is featureless beyond 10 day$^{-1}$. A single strong signal is detected at $f_1$ = 0.53073 day$^{-1}$. The other obvious peaks are aliases of this signal, caused mostly by the observing pattern of KELT (daily, monthly, and yearly aliases). The red curve shows the periodogram after pre-whitening against $f_1$. The middle panel shows the periodogram in the immediate vicinity of $f_1$ (left), and the light curve phased to the corresponding period (right). The bottom panel highlights the pre-whitened periodogram, in the same frequency range as the above plot, and identifies the top peak (of the pre-whitened periodogram) in this region.

Six separate and unique portions of the light curve, corresponding to the six major individual outbursts, were analyzed in this manner as well. This was done mainly to study $f_1$ over time. In each portion of the light curve, $f_1$ is recovered at the same frequency (to within $\sim$0.2\%). This signal remains coherent (\textit{i.e.} experiences no shift in phase) throughout the observational baseline and does not appear to significantly vary in amplitude. The photometric signal does not appear to be double- or triple-waved when phased to two or three times the recovered period.

All of the observed features of $f_1$ are consistent with stellar pulsation. The frequency is within the range where pulsation in Be stars is expected, and is similar to the pulsations of the class of slowly pulsating B stars, of which Be stars have been conjectured to be a rapidly rotating and more complicated sub-class of \citep{Aerts2006,Kurtz2015}. The photometric amplitude is high (20.1 mmag), but not unusually so \citep{Balona1995}, especially since pulsation amplitudes tend to be larger in early-type Be stars \citep{Rivinius2013}, which is the case with ABE-A01 (B0.5IVe). The fact that this signal persists throughout the entire observational baseline and remains coherent in phase is also consistent with pulsation.

So-called {\v S}tefl frequencies are sometimes detected in the light curves of Be stars. These signals are caused by asymmetries in the circumstellar material orbiting the star, which can modulate the net observed flux \citep[and also line profiles;][]{Stefl1998,Stefl2000}. The period of the {\v S}tefl frequencies is determined by the orbital period close to the star, and can be of a similar timescale as the periodic signal recovered in ABE-A01. However, there are a number of reasons to doubt this as the cause for $f_1$. {\v S}tefl frequencies are only found in conjunction with pulsational signals of similar, but slightly higher, frequencies \citep[\textit{e.g.}][]{Stefl1998,Baade2016}. There is only one significant frequency in the light curve of ABE-A01. {\v S}tefl frequencies have so far only been detected in photometry for Be stars that have high inclination angles, since a photometric signal only manifests when the density enhancements are projected against the stellar disk \citep[\textit{e.g.}][]{Baade2016}. ABE-A01 is at a low inclination angle. The fact that the signal is coherent (in phase and amplitude) over the entire observational baseline is an argument against the signal being a {\v S}tefl frequency, which are typically transient on timescales much longer than the orbital period in the inner regions of the disk, although exceptions are possible \citep[\textit{e.g.} $\eta$ Cen;][]{Rivinius2003}. We therefore attribute $f_1$ to pulsation.

It has been proposed that combinations of multiple pulsation modes can interact to control the `clock' that dictates the time-variable mass-loss rates of Be stars \citep{Rivinius1998,Kee2014}. Such `combination frequencies' are the preferred explanation for the low-frequency variability seen in the Be star $\eta$ Cen \citep{Baade2016}. \citet{Kurtz2015} argue that nonlinear mode coupling can give rise to combination frequencies in Be stars that have higher amplitudes than the parent frequencies. In this framework, the difference between two pulsation modes can be referred to as the `difference frequency' ($\Delta f$ = $|f_{1}$ - $f_{2}|$, where $f_{1}$ and $f_{2}$ are the frequencies of two pulsation modes). This $\Delta f$ then describes the approximate frequency of the outbursts themselves.

We apply this idea to ABE-A01. Because the outbursts are approximately evenly spaced, we can measure their period (91.23 days) and therefore the frequency with which they occur (0.01096 day$^{-1}$), which we suppose is the difference frequency. We then have $\Delta f$ = 0.01096 day$^{-1}$, and $f_{1}$ = 0.53073 day$^{-1}$. We are motivated to search for a second pulsation mode, which should occur with a frequency $f_{2}$ = $f_{1}$ $\pm$ $\Delta f$. The two solutions for $f_{2}$ are 0.51977 day$^{-1}$ and 0.54169 day$^{-1}$. The bottom panel in Figure~\ref{fig:Freq_A01_new} shows our attempt to search for the signature of a second pulsation mode, $f_{2}$. This panel shows the pre-whitened periodogram, with the position of $f_{1}$ indicated by a vertical dashed line, and our predictions for possible expected values of $f_{2}$ marked by vertical dotted lines. After pre-whitening against $f_{1}$, there are no peaks with substantial power. However, there is a weak peak near one of the frequencies expected for $f_{2}$ (see the bottom panel in Figure~\ref{fig:Freq_A01_new}). Although it is possible that this signal is astrophysical and represents a pulsation mode in the star, this can neither be confirmed nor ruled out with the available data. It should be noted that ABE-A01 is viewed at a low inclination angle, which can cause certain non-radial pulsation modes to have a very low photometric amplitude, due to azimuthal averaging. The lack of additional strong peaks (besides $f_{1}$) in the periodogram computed from the KELT light curve therefore does not imply that the star oscillates in only one mode.

\subsubsection{ABE-026 ($=$ V438 Aur $=$ HD 38708)}
\begin{figure}[!ht]
\centering\epsfig{file=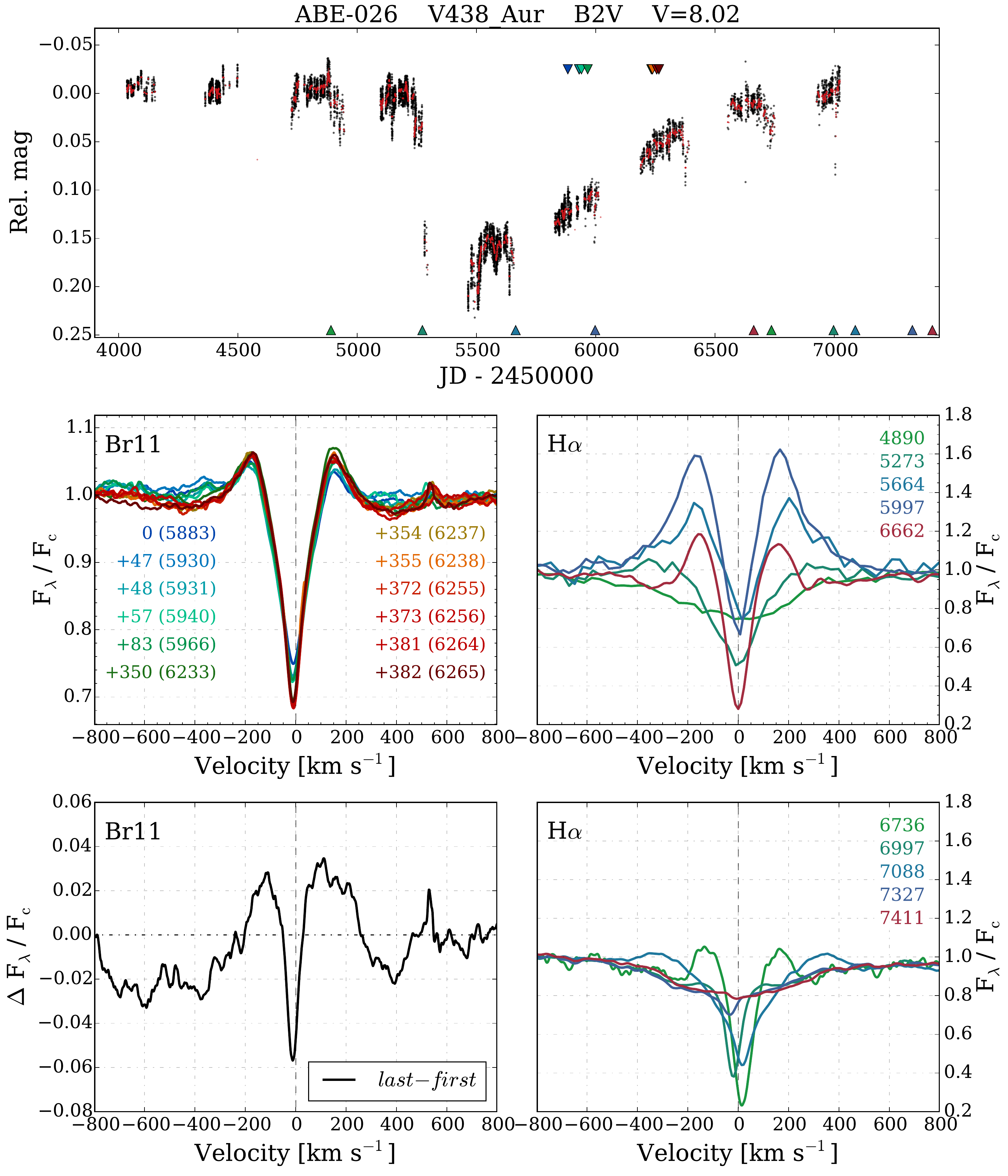,clip=,width=0.99\linewidth}
\caption{Top: KELT light curve, with downward (upward) pointing triangles indicating epochs of APOGEE (BeSS) spectra. Middle left: Br11 line of 12 APOGEE spectra. Middle and bottom right: BeSS spectra, centered on H$\alpha$. Bottom left: difference between the final and initial APOGEE spectra. Spectroscopic epochs are indicated in the same manner as in Figure~\ref{fig:Outburst_138}.}
\label{fig:Outburst_026}
\end{figure}

Among all of the stars in this sample, ABE-026 has perhaps the most dramatic outburst. Viewed nearly edge-on, this is an excellent example of a shell star, where the growth of a disk causes the system to appear fainter, and also results in deep shell absorption in the hydrogen lines. Figure~\ref{fig:Outburst_026} shows the KELT light curve, 12 APOGEE spectra taken over 382 days, and 10 H$\alpha$ spectra from BeSS, with a baseline of nearly 3000 days. The first four years of KELT data show little variability, and the first BeSS spectrum (at JD-2450000 = 4890) shows a broad absorption line with no evidence of emission or shell absorption. At the very end of the fourth season in the KELT light curve (approximately JD-2450000 = 5280), the system begins to rapidly dim, indicating the onset of an outburst. A spectrum from BeSS was serendipitously taken during this rising phase\footnote{Although the system is dimming, we still refer to this as the rising phase of the outburst, since the disk is growing.} at JD-2450000 = 5273, showing a deep absorption core and broad emission wings, with a large peak separation, indicative of a high-density inner disk and a small H$\alpha$-emitting region, both facts consistent with a forming disk. As the long falling phase of this outburst ensues, the brightness of the system relaxes toward baseline, and the H$\alpha$ line continues to evolve. The emission wings evolve toward a smaller peak separation, suggesting that the size of the H$\alpha$-emitting region continues to grow outwards. Although an increasing optical depth can also cause $\Delta v_p$ to shrink, we do not expect this to be a major contributing factor since the disk is dissipating (becoming more diffuse), and not building up. The final season of KELT data shows the system back at its baseline brightness. A weak disk in H$\alpha$ is present at JD-2450000 $=$ 7327, but has disappeared by JD-2450000 $=$ 7411. The disk has completed its life cycle in these observational modes, persisting for between 2047 and 2131 days.

It is likely that in addition to the major disk buildup phase near JD-3450000 = 5280, some relatively minor outbursts take place before the disk has completely dissipated in photometry and in H$\alpha$. There is enhanced light curve activity near JD-2450000 $=$ 5500--5700, and 6700, perhaps signifying further mass loss. The enhancement in the high-velocity wings of H$\alpha$ at JD-2450000 $=$ 7088 points to the addition of some new disk material between JD-2450000 $=$ 6997--7088.

ABE-026 was visited 12 times by APOGEE over a 382 day span during the falling phase of the outburst. The Br11 line of each spectrum shows a deep, narrow, absorption core, with roughly symmetric emission wings. The difference between the final and the initial spectrum is shown in the lower-left panel of Figure~\ref{fig:Outburst_026}. The absorption core becomes deeper, and the emission wings enhanced, as the falling phase of the outburst progresses. There is an obvious increase in the optical continuum flux during this spectroscopic sequence, and it is possible that the NIR continuum flux in the vicinity of the Br11 line is likewise changing. Therefore, variability in the Br11 line profile relative to the local continuum should be treated with caution.

Part of the reason this outburst has such a long falling time in its light curve is that we are not simply seeing some effect of the inner disk, which is generally the case for non-shell Be stars. Rather, we are mainly seeing the effect of stellar continuum photons being absorbed and scattered out of our line of sight by the intervening gas. So, even after the inner disk has dissipated, we still see a flux decrement because the outer disk continues to partially obscure the star.

After the disk life cycle shown in Figure~\ref{fig:Outburst_026}, this system experiences another episode of disk growth and dissipation. Although we have no photometric data covering this second cycle at present, the KELT survey is ongoing, and further data reduction will likely reveal at least some of this event. H$\alpha$ measurements from BeSS show the system evolving from a diskless state at JD-2450000 $=$ 7411, to having a shell profile with emission wings at JD-2450000 $=$ 7648. A third spectrum (ARCES; JD-2450000 = 7795) shows no disk signature in H$\alpha$. This sequence of three spectra spans 384 days, which is the maximum lifespan of the disk in this episode. This is much shorter than the $\sim$2100 day lifespan of disk created by the the first outburst.

\begin{figure}[!ht]
\centering\epsfig{file=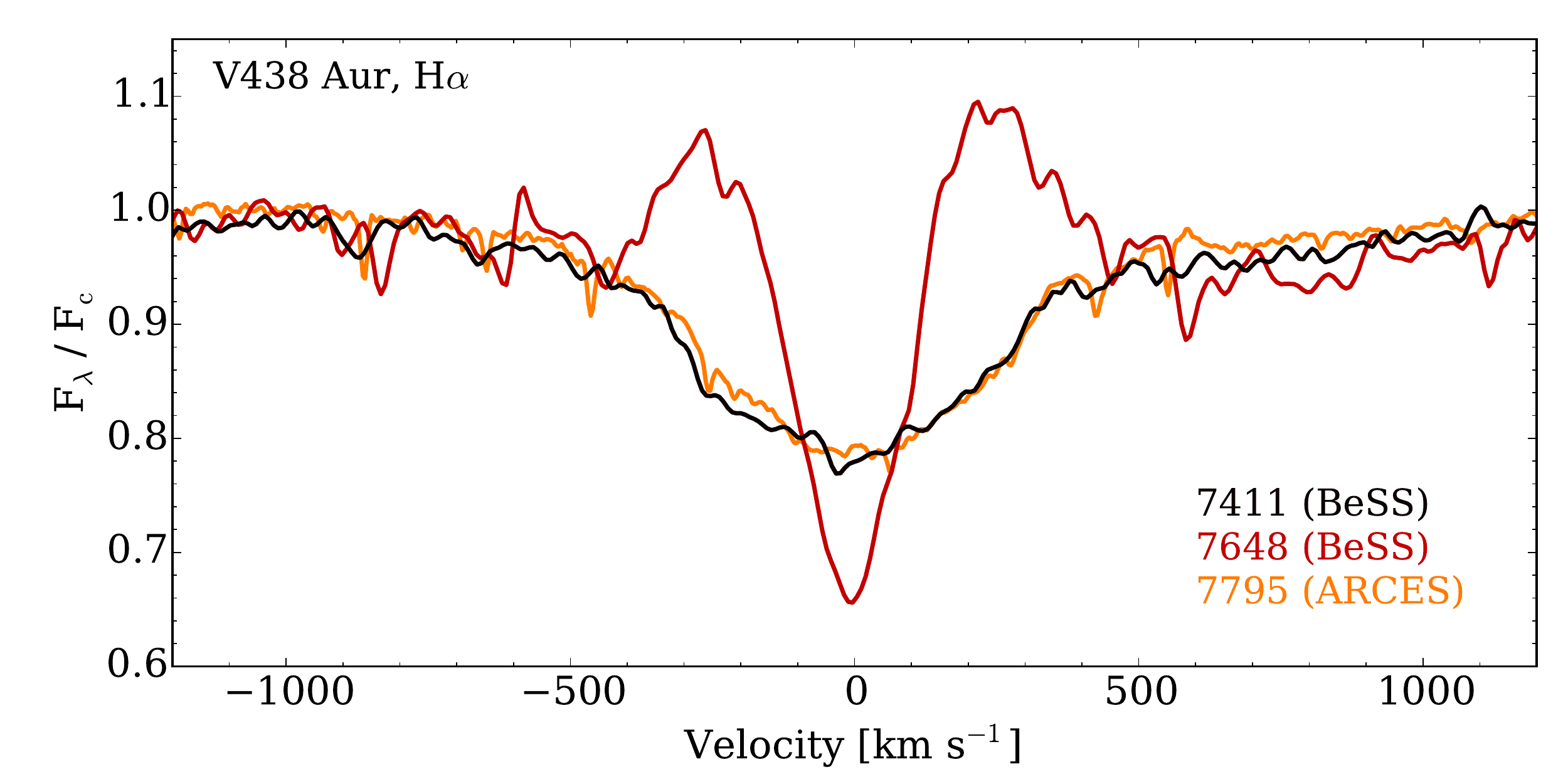,clip=,width=0.99\linewidth}
\caption{Spectra showing a new disk life cycle for ABE-026.}
\label{fig:Outburst_026_b}
\end{figure}

\subsection{Global outburst properties}

\subsubsection{Outburst rates and photometric amplitudes}
We detect one or more outbursts in 28\% of our sample, being more often seen in early-type Be stars (57\%), compared to mid (27\%) and late types (8\%). See Table~\ref{tbl:variable_fractions} for more details. These fractions are similar to the incidence rates of Be star light curves showing outbursts in LB17 (51\%, 20\%, and 5\% of early, mid, and late types, respectively), where a larger sample of 510 Be stars was studied with KELT light curves. There are 32 Be stars among the 160 analyzed in this work that also exist in the LB17 sample. These results reflect the trend of earlier Be stars being more variable in general, while later-type systems tend to have more stable disks that last for longer times \citep[\textit{e.g.}][]{Hubert1998,Rivinius2013}. This may be because early-type Be stars are intrinsically more prone to mass-loss episodes, or it could be that the observational signature of outbursts in Be stars of later spectral types is often too small to be detected with KELT. Evidence for the latter of these points is emerging, in that cooler Be stars create disks with surface densities that are too low to leave an obvious observational signature in the visible continuum flux \citep{Vieira2017}. The observables used in this paper are generally formed within the inner $\sim$20 $R_{*}$ of the disk. The lack of any disk signature in optical photometry, or in the Br11 or H$\alpha$ lines does not mean that no disk exists, as there may be material farther out than the region in which a given observable is sensitive to.

\begin{figure}[!ht]
\centering\epsfig{file=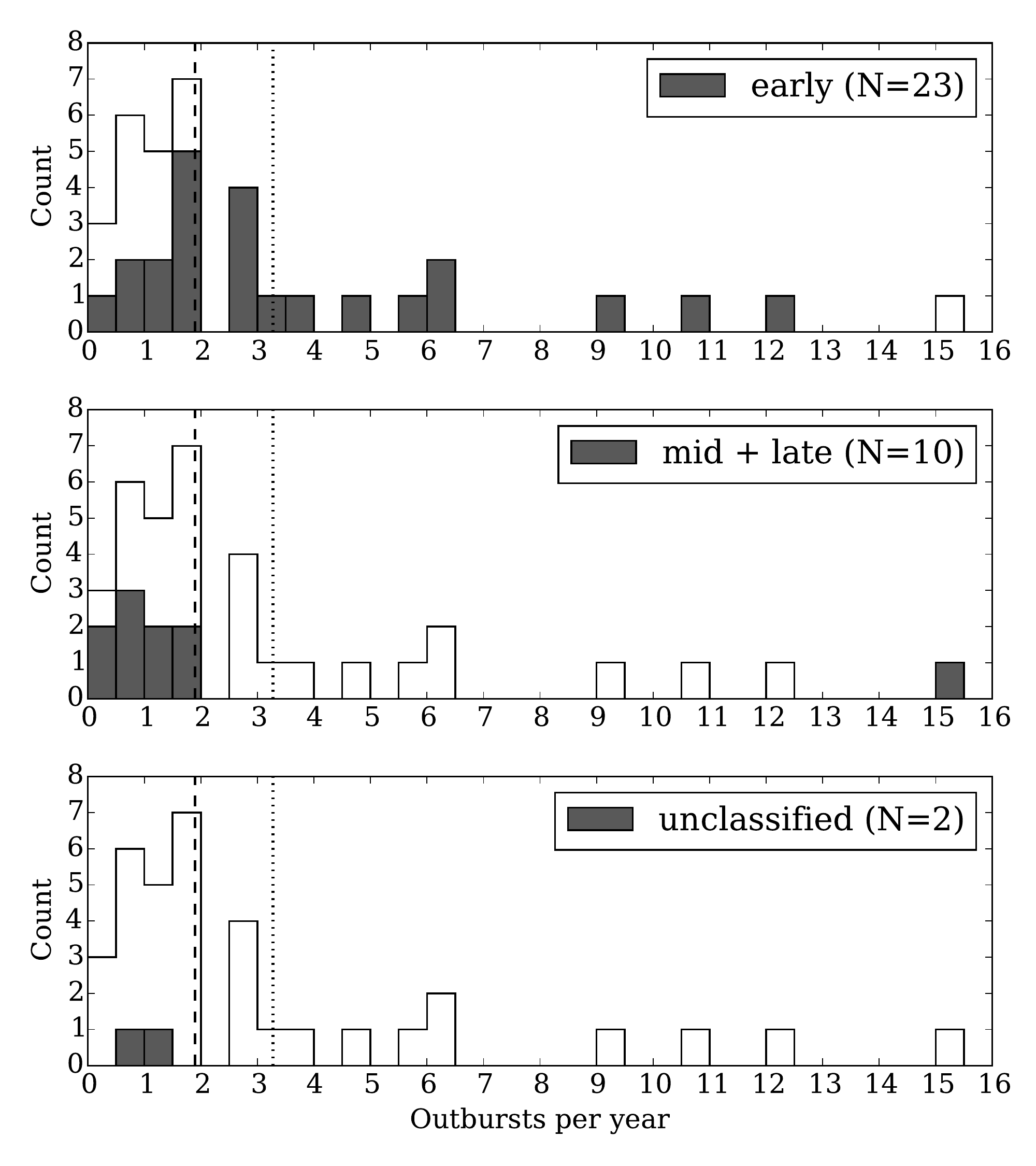,clip=,width=0.99\linewidth}
\caption{Distribution of outburst rates for early (top), mid + late (middle), and unclassified (bottom) spectral types, calculated for all systems showing one or more outbursts, so long as the number of outbursts is well-defined. The solid line making up the envelope of the distribution in all three panels includes all stars, regardless of spectral type. The vertical dashed line denotes the median, and the vertical dotted line the mean, of this distribution as a whole. See Table~\ref{tbl:spec-table} for the number of outbursts seen in each system.}
\label{fig:hist_of_outburst_rates}
\end{figure}

For systems with one or more outbursts, the outburst rates (outbursts per year) are calculated in the same fashion as in LB17, by dividing the total number of detected outbursts in a given system by its observational baseline, less the seasonal gaps, and are shown in Figure~\ref{fig:hist_of_outburst_rates}. Some systems show outbursts, but have complex light curves where the total number of outbursts is not clear. These are not included in Figure~\ref{fig:hist_of_outburst_rates} (since the discrete number of outbursts cannot be counted), but are included in the fractions shown in Table~\ref{tbl:variable_fractions} (because they do show at least one outburst). Because outbursts are more commonly observed in early-type stars, these dominate Figure~\ref{fig:hist_of_outburst_rates}. There is a wide spread in the outburst rates across the sample, with many systems showing zero or a small number of outbursts, while others experience tens of outbursts throughout their observational baseline. The median and mean of this distribution is 1.9 and 3.3 outbursts year$^{-1}$, respectively. For any given system, an outburst with a larger amplitude generally corresponds to a larger mass ejection episode. With this in mind, it is not necessarily the case that stars with high outburst rates have proportionally higher mass-loss rates, since a single large outburst can eject more material than many small outbursts. Table~\ref{tbl:spec-table} includes the number of outbursts for each system. 

\begin{table}
 \centering
 \caption{Fractions showing outbursts}
 \label{tbl:variable_fractions}
 {\renewcommand{\arraystretch}{1.25}
 \begin{tabular}{|c|c|c|c|c|}
    \hline
 \textit{All} & \textit{Early} & \textit{Mid} & \textit{Late} & \textit{Unclassified}\\
\hline
28$\%$ (\(\nicefrac{44}{160}\))& 57$\%$ (\(\nicefrac{31}{54}\)) & 27$\%$ (\(\nicefrac{6}{22}\)) & 8$\%$ (\(\nicefrac{5}{61}\))  & 9$\%$ (\(\nicefrac{2}{23}\)) \\
\hline
 \end{tabular}
 }
 \begin{flushleft}
  \footnotesize Fraction of stars showing one or more outbursts detectable in their KELT light curve, according to spectral type. The category `all' includes early-, mid-, and late-type stars, as well as those without a specific classification. 
  \end{flushleft}  
\end{table}

\begin{figure}[!ht]
\centering\epsfig{file=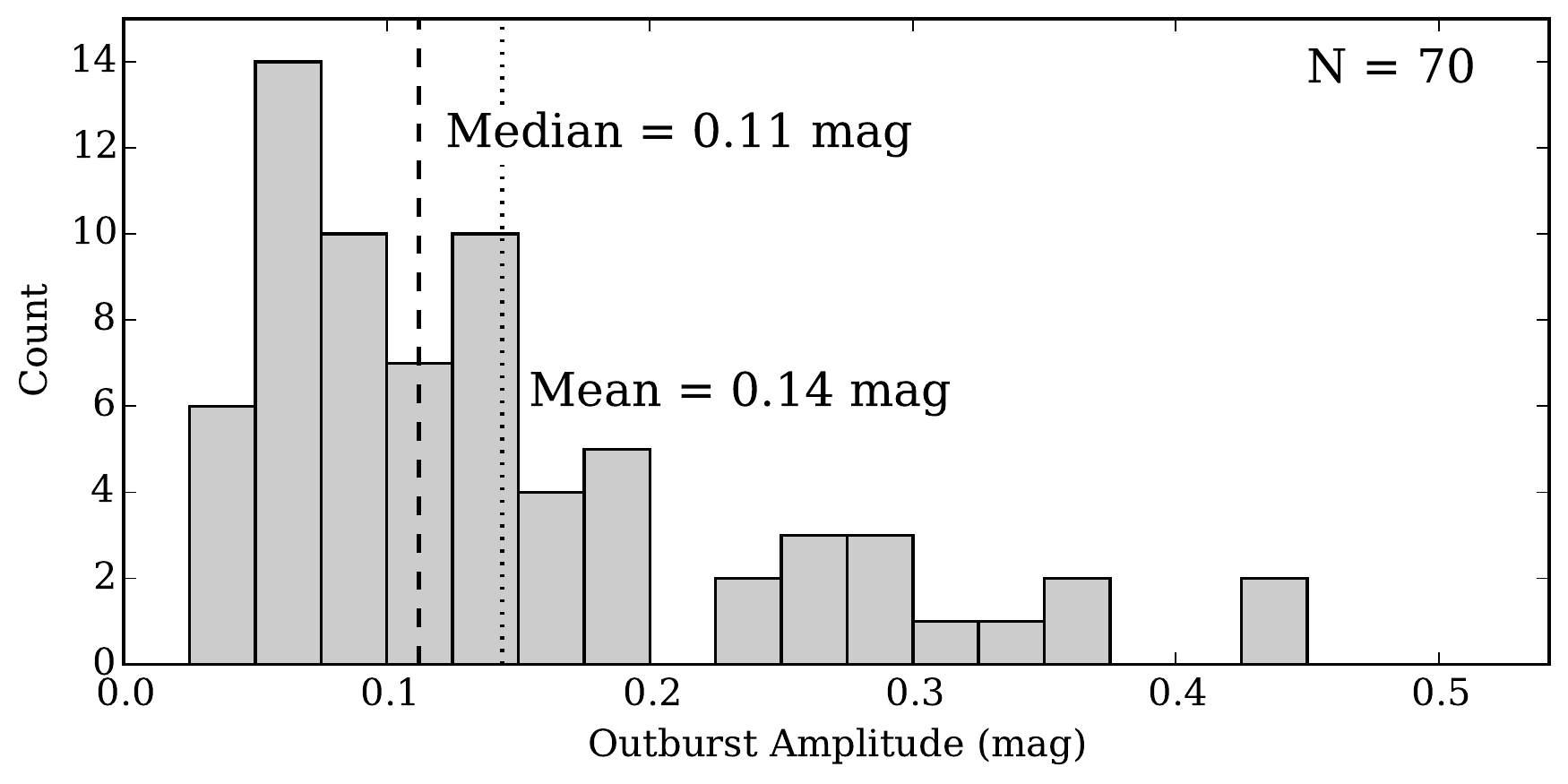,clip=,width=0.99\linewidth}
\caption{Histogram of outburst amplitudes. The amplitude in the KELT passband depends strongly on the inclination angle of the system, which is not taken into account here.}
\label{fig:Outburst_Amps}
\end{figure}

There are 70 outbursts that are reasonably well-defined in our light curve data, where we measure the photometric amplitude. The distribution of these amplitudes, in the approximately broad $R$-band filter employed by KELT, is shown in Figure~\ref{fig:Outburst_Amps}. It should be kept in mind that outburst amplitude depends strongly on inclination angle, which is not taken into account here. Half a magnitude is an approximate upper limit on the amplitude of Galactic Be star variability in the KELT passband (see also LB17). Table~\ref{tbl:otb-tbl} in the appendix includes information for each outburst that was measured.

\subsubsection{Correlations between falling and rising times}
\begin{figure}[!ht]
\centering\epsfig{file=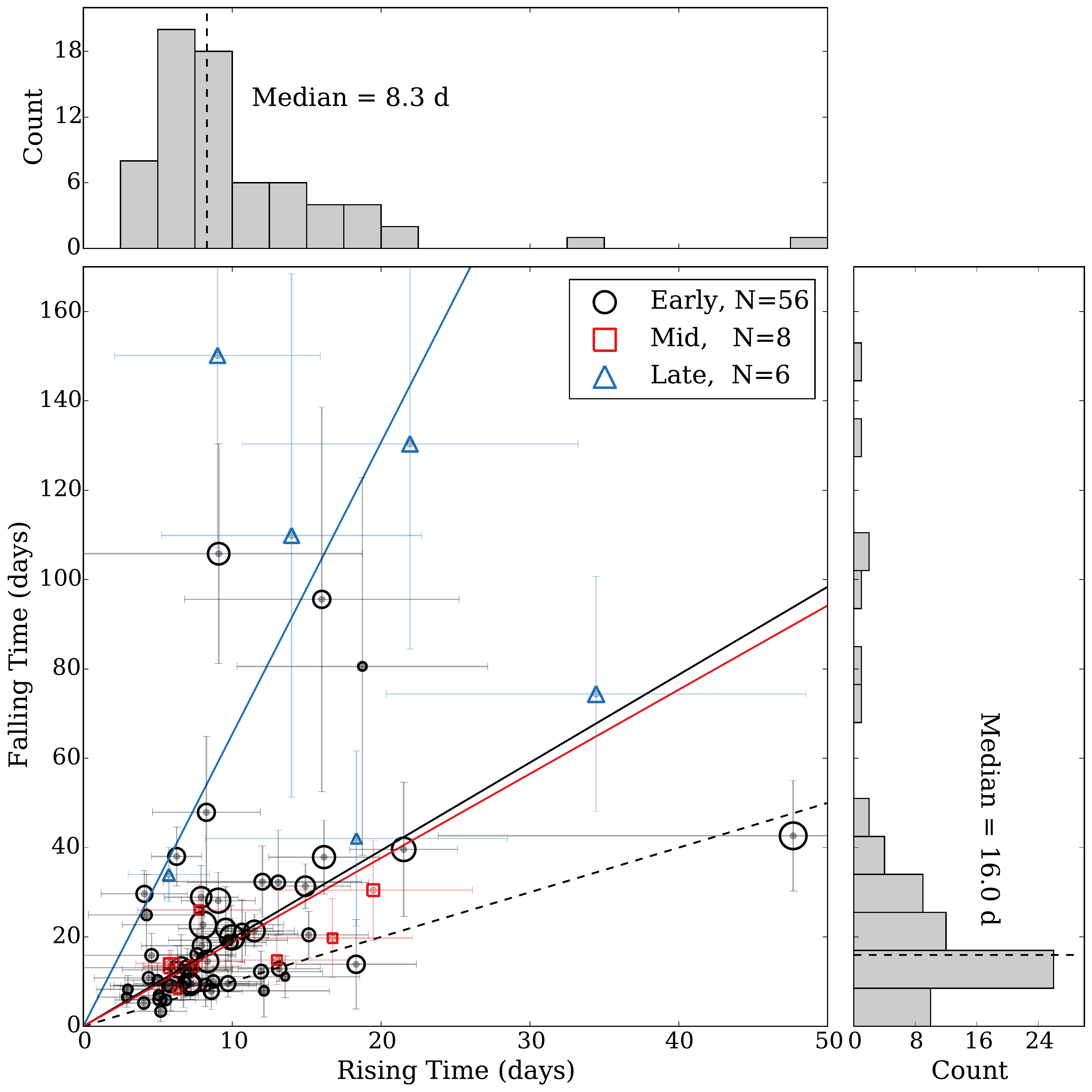,clip=,width=0.99\linewidth}
\caption{Each point represents the duration of the rising and falling times for 70 outbursts in 24 unique stars (18 early, 4 mid, and 2 late), with the marker size proportional to the photometric amplitude. The dashed line has a slope of 1. Circles, squares, and triangles correspond to early-, mid-, and late-type stars, respectively. The black, red, and blue lines begin at the origin and are lines of best fit to the early-, mid-, and late-type groups. Their slopes are 1.97 (early), 1.88 (mid), and 6.54 (late). The top (right) panel shows a histogram of the rising (falling) times. Table~\ref{tbl:otb-tbl} contains the information used to make this plot.}
\label{fig:Outburst_properties}
\end{figure}

Whenever possible, the photometric amplitude and duration of the rising and falling phases are measured for an outburst captured in KELT photometry. These three quantities can only be measured for outbursts that are reasonably well-defined, and are thoroughly sampled in the light curve data, such as the example in Figure~\ref{fig:Outburst_example}. There are 24 objects with such events (18 early, 4 mid, and 2 late types), from which we measure 70 outbursts in total. Figure~\ref{fig:Outburst_properties} shows the correlation between rising time and falling time for 70 well-behaved outbursts lasting 300 days or less. Although longer outbursts are seen in a few cases, they are relatively rare and have large uncertainties. Considering events shorter than 300 days allows us to focus on outbursts of roughly comparable magnitude. The dashed line has a slope of 1. The majority of points fall above this line, having longer falling, compared to rising, times, and those that do not are close (within measurement uncertainties). The median rising time is 8.3 days, and the median falling time is 16.0 days. The median of the ratios of falling time to rising time for this collection is $\sim$2. A best-fit line to each group has a slope of 1.97, 1.88, and 6.54 for early-, mid-, and late-type stars. This fit takes into account measurement uncertainties by assigning each point a weight proportional to one over the square of the error (in both the $x$- and $y$-directions). Although the sample size is too small to draw any definite conclusions and there is significant scatter, these results suggest that, for a single event, relative to the rising time, the inner disk dissipates quickly for hotter stars (early and mid types), and more slowly for cooler late-type systems.

The observed trend of the rising time being shorter than dissipation is a prediction of the VDD model. During disk buildup, the photometric variability is governed mainly by the timescales of matter redistribution over the inner disk, while during dissipation (\textit{i.e.} when mass injection into the disk has significantly slowed or stopped), the inner disk is instead fed by matter re-accreting back from the entire disk, with naturally longer timescales. When an outburst occurs, the falling phase will proceed more slowly if there is a pre-existing disk. This `mass reservoir effect' means that the dissipation timescales depend not only on the outburst being considered, but also on the previous history of mass injection into the disk \citep{Ghoreyshi2017}. This effect is stronger when the disk is more massive. Lacking sufficient knowledge of the mass-injection history of the disks in this sample, we make no attempt to correct for this effect, and report the values measured from the photometric data with no regard to whether or not a disk already exists prior to the outburst.

Nevertheless, there are a few systems with spectroscopic data showing the lack of a pre-existing disk immediately before the time of outburst, meaning that there is essentially no mass reservoir effect acting in these cases. ABE-098 (B5V), as discussed in Section~\ref{sec:indv_otbs_098} has no Br11-emitting disk immediately before the outburst that occurs near JD-2450000 = 6230, which has a ratio of falling to rising time of about 2.7. ABE-154 and -162 both have a spectral type of B8, and each experiences an outburst that is preceded by an APOGEE spectrum that shows no substantial Br11-emitting disk. So, for these events, we expect no interference from the mass reservoir effect. The outburst in ABE-162 has a ratio of falling to rising time of about 2.7. While the outburst that is bracketed by APOGEE spectra in ABE-154 is not fully sampled, the falling phase is many times (perhaps up to 10 times) longer than the rising phase. We have no knowledge of the status of the disk in the vicinity of the other outbursts in ABE-154, but they have similarly long falling, relative to rising, phases, and a similar baseline brightness. These two stars (ABE-154 and -162) are responsible for all six of the late-type outbursts depicted in Figure~\ref{fig:Outburst_properties}. Even though this sample size is small and the scatter large, the relatively large slope of the fit to falling over rising time for late-type outbursts is likely not heavily influenced by the mass reservoir effect.

The aforementioned late-type systems stand in stark contrast to the early-type systems that dominate Figure~\ref{fig:Outburst_properties}. The majority of early-type stars with measured outbursts have substantial disks at all observed epochs, particularly ABE-164 (B0Ve, three measured outbursts), -184 (B1Ve, 15 measured outbursts), -A01 (B0.5IVe, five measured outbursts), -A03 (B1Ve, five measured outbursts), and -A26 (B1 II/IIIe, 18 measured outbursts). Plots for these can be found in the Appendix (except for ABE-A01, discussed in Section~\ref{sec:A01}). Together, these five systems contribute 46 of the 56 measured outbursts for early-type systems. These systems exhibit some of the strongest spectroscopic disk signatures among the sample. Although the strength of these features varies over time, they never approach a diskless state. All of the observed outbursts for these systems then occur while there is presumably already a substantial disk. Therefore, we expect the mass reservoir effect to `interfere,' lengthening the time over which the photometric dissipation phase takes place. Without a pre-existing disk, it is reasonable to assume that a single typical outburst in an early-type star would have a smaller ratio of falling to rising time compared to the best-fit slope of 1.97 measured in this sample, seen in Figure~\ref{fig:Outburst_properties}.

\subsubsection{Regularly repeating outbursts} \label{sec:SRO}
\begin{figure*}[!ht]
\centering\epsfig{file=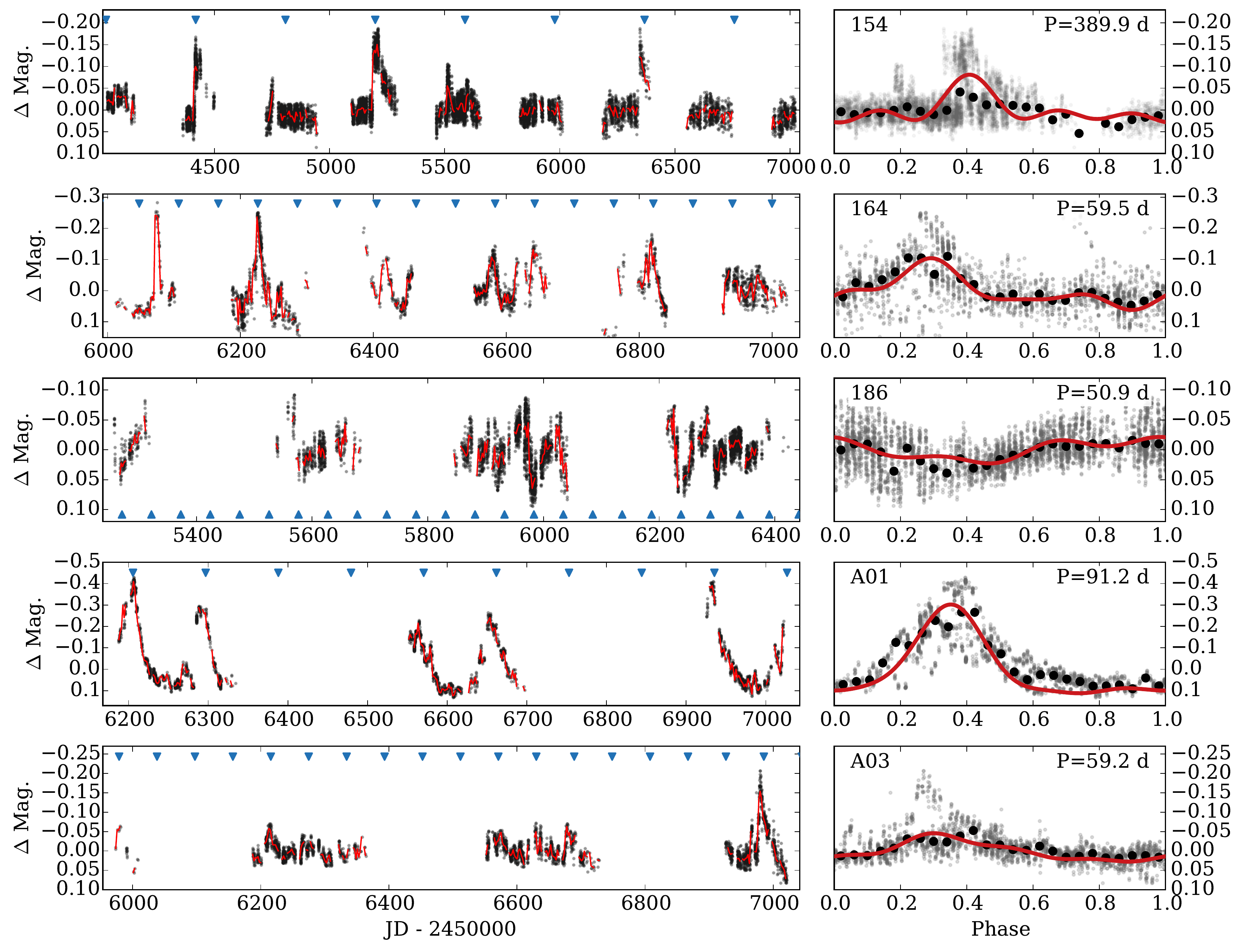,clip=,width=0.99\linewidth}
\caption{Plots for the five systems that show regularly repeating outbursts. Left: raw light curve (black) with binned data in red. Right: Light curve phased to the period that best describes the pattern of outbursts in the raw data. The larger black circles show the data median-binned with 25 bins in phase, and the red curve is a three-term Fourier fit, to guide the eye. The downward-pointing triangles in the plots of the raw light curves correspond to the ephemerides of the outbursts (\textit{i.e.} if these outbursts were strictly periodic, each triangle would mark the peak of an event). ABE-186 (middle row) is a shell star \citep{Chojnowski2015}, where outbursts cause a dimming. For this reason, the triangles point upwards.}
\label{fig:SRO}
\end{figure*}

There are five systems in our sample that show repeating outbursts, similar to ABE-A01 (see Section~\ref{sec:A01}). These are shown in Figure~\ref{fig:SRO}, where the left column displays the full KELT light curve, and the right column plots the data phased to the period that best describes the timing of the outbursts. Although these events repeat, they are not strictly periodic. Instead, there are sometimes variations in the timing, amplitude, and shape of the outbursts. Cycles may be skipped, and additional outbursts sometimes occur, apparently at random. A total of 35 Be stars with repeating outbursts are identified in LB17. There is some overlap between the present sample and that used in LB17. Both ABE-A01 and -A03 were identified as having repeating outbursts in LB17. Future work is planned to analyze the stellar properties and photometric behavior of all systems having repeating outbursts for which we have sufficient data, with a focus on links between pulsation and outbursts. However, a detailed analysis of these systems is beyond the scope of this work.

\section{Conclusion and further work}
\label{sec:conclusion}

We have analyzed optical light curves for 160 classical Be stars to study the disk creation process, and to monitor the evolution and demise of these disks once formed. These discrete episodes of disk creation leave a generally consistent imprint on a light curve, rising from baseline to a peak brightness, then falling back toward baseline on a relatively longer timescale. The frequency of occurrence, amplitude, and associated timescales can vary greatly, not only from one star to the next, but also for any given system. In most cases, outbursts occur with no discernible pattern. However, there are some systems that experience outbursts that repeat at a nearly regular rate and with similar amplitudes. ABE-A01 is one such example, with four others shown in Figure~\ref{fig:SRO}. In this sample, we find 44 stars (28\%) to have at least one outburst detected in their KELT light curve. On average, the duration of the falling phase is about twice that of the rising phase for early- and mid-type stars, and larger for late-type stars (see Figure~\ref{fig:Outburst_properties}). Amplitudes up to $\sim$0.5 mag (in a wide $\sim R$-band filter) are seen (Figure~\ref{fig:Outburst_Amps}). A higher degree of photometric variability is seen in early-type stars, which are more likely to have at least one detectable outburst compared to their cooler counterparts (Table~\ref{tbl:variable_fractions}). 

KELT light curves are generally sensitive to only the innermost region of Be star disks, giving us clues as to how the circumstellar environment closest to the star changes. By including time-series spectroscopy of the hydrogen Brackett series from the NIR APOGEE survey, and H$\alpha$ spectra from the BeSS database, we have many `snapshots' of the circumstellar environment. This allows us to unambiguously infer the presence of a disk, and also to measure its strength, projected velocity profile, and any asymmetries. By leveraging spectra taken during active outburst phases, we showed that the circumstellar environment can be quite asymmetric during disk growth (ABE-138, -A01). The material settles into a more axisymmetric configuration over time, according to the predictions of the VDD model. Another advantage of combining optical photometry and the Br11 and H$\alpha$ lines is that these observables probe different regions of the disk. Instances where these different observational modes are measured near-contemporaneously show evidence of the Be star disk both growing and also dissipating from the inside outward, in agreement with theoretical expectations. 

The Be stars in this sample experience outbursts that occur over a wide range of timescales -- days (\textit{e.g.} ABE-138), weeks (\textit{e.g.} ABE-098), months (\textit{e.g.} ABE-A01), and years (\textit{e.g.} ABE-026; and ABE-082, -160 in the Appendix for non-shell stars). Other works have similarly observed apparent outbursts over a large range in time. Analysis of Kepler data has shown aperiodic Be star variability with amplitudes below 10 mmag and timescales of days to weeks, which may indicate small-scale outbursts \citep{Balona2015,Rivinius2016}. On the other hand, some Be stars experience outbursts with timescales of many hundreds or thousands of days and greater. These longer outbursts can leave a qualitatively different light curve signature, where the rising phase is followed by a plateau in brightness. This plateau can persist for thousands of days, and even up to decades, so long as the disk-feeding mechanism remains active. In this scenario, the inner disk continues to be fed by mass ejected from the star, but at some point becomes saturated, and the addition of new material does not increase its brightness. These ``more complete'' disk-building events are modeled in \citet{Haubois2012}, and are observed in some Be stars \citep[\textit{e.g.} $\omega$ CMa;][]{Ghoreyshi2017}. Be stars lose mass in episodes of vastly varying intensity and duration. With this in mind, the outbursts explored in this work address only part of the story of Be star mass loss.

Future work is planned to analyze the periodic variability in the light curves of the Be stars in this sample, searching for signs of pulsation, binarity, and other signals. Relationships, or lack thereof, between the periodic behavior and the outburst behavior of these systems will be explored. We also aim to measure the amplitude and the rising and falling times of outbursts in a much larger sample of Be stars for which we have KELT light curves, in the same manner employed in this work. The present sample of 70 outbursts from 24 individual Be stars described here is an early step toward understanding correlations between outburst rising and falling timescales, amplitudes, and stellar spectral types. Improving on and extending these methods to a significantly larger sample will allow for more quantitative statements regarding the distribution of outburst properties and their dependence on stellar spectral type.

\section{Acknowledgements}
This work has made use of the BeSS database, operated at LESIA, Observatoire de Meudon, France: http://basebe.obspm.fr. We especially thank the following observers who have submitted data to BeSS that we have used in this publication: Christian Buil, Joan Guarro Fl\'{o}, St\'{e}phane Ubaud, Thierry Garrel, Olivier Garde, Tim Lester, Andrew Wilson, Dong Li, Jean-No\"{e}l Terry, Stephane Charbonnel, Michel Pujol, Thierry Lemoult, Olivier Thizy, Coralie Neiner, Andr\'{e} Favaro, Nico Montigiani, and Massimiliano Mannucci. We thank the anonymous referee for providing comments that improved this work. M.V.M. acknowledges support from NSF grant AST-1109247 and a Class of 1961 Professorship from Lehigh University. Work by D.J.S. was partially supported by NSF CAREER Grant AST-1056524. J.L.-B. acknowledges institutional support from Lehigh University and a Sigma Xi Grant-in-Aid of research and thanks Dr. Jon E. Bjorkman for many illuminating conversations. This research has made use of the SIMBAD database, operated at CDS, Strasbourg, France. This research has made use of NASA's Astrophysics Data System. Funding for SDSS-III has been provided by the Alfred P. Sloan Foundation, the Participating Institutions, the National Science Foundation, and the U.S. Department of Energy Office of Science. The SDSS-III Web site is http://www.sdss3.org/. SDSS-III is managed by the Astrophysical Research Consortium for the Participating Institutions of the SDSS-III Collaboration including the University of Arizona, the Brazilian Participation Group, Brookhaven National Laboratory, Carnegie Mellon University, University of Florida, the French Participation Group, the German Participation Group, Harvard University, the Instituto de Astrofisica de Canarias, the Michigan State/Notre Dame/JINA Participation Group, Johns Hopkins University, Lawrence Berkeley National Laboratory, Max Planck Institute for Astrophysics, Max Planck Institute for Extraterrestrial Physics, New Mexico State University, New York University, Ohio State University, Pennsylvania State University, University of Portsmouth, Princeton University, the Spanish Participation Group, University of Tokyo, University of Utah, Vanderbilt University, University of Virginia, University of Washington, and Yale University. This work is, in part, based on observations obtained with the Apache Point Observatory 3.5-meter telescope, which is owned and operated by the Astrophysical Research Consortium.

\appendix

\section{Additional Examples of Outbursts with Coincident Spectra}

Here we highlight some interesting cases, where simultaneous photometry and spectroscopy reveal changes in the star + disk system. This selection shows diversity in light curve and emission-line variability, demonstrating episodes of disk growth and creation, inside-out disk clearing, and outwardly migrating disks.

Table~\ref{tbl:otb-tbl} provides information for the 70 well-defined outbursts where the rising and falling times, and the photometric amplitudes are measured. Each of these is an event where the brightness of the system increases first, then falls back to baseline. This is partly because systems with low to intermediate inclination angles ($i\lesssim 70^{\circ}$) are more common than those with higher inclination angles, and also because outbursts that cause a net brightening tend to be more well-defined in the KELT data compared to their inverted counterparts. Table~\ref{tbl:otb-tbl} includes the date of the beginning and end of each rising and falling phase, and the baseline and peak brightness associated with each outburst event (in the KELT passband). Table~\ref{tbl:spec-table} includes each object in this sample, reporting the ABE-ID, a common identifier, $V$-band magnitude, a spectral type, the corresponding reference, the $T_{eff}$ class (early, mid, late, or unclassified), number of APOGEE visits, the KELT field, the first and last dates of KELT observations, and the number of detected outbursts. Spectral type references of ``New'' indicate that an object was discovered to be a Be star through inspection of APOGEE spectra, and is announced in \citet{Chojnowski2015}. These systems have neither an available literature spectral type nor has an optical spectrum been acquired with APO-ARCES or the AO long-slit spectrograph.

\vspace{5mm}
\subsection{ABE-027}
\begin{figure}[!ht]
\centering\epsfig{file=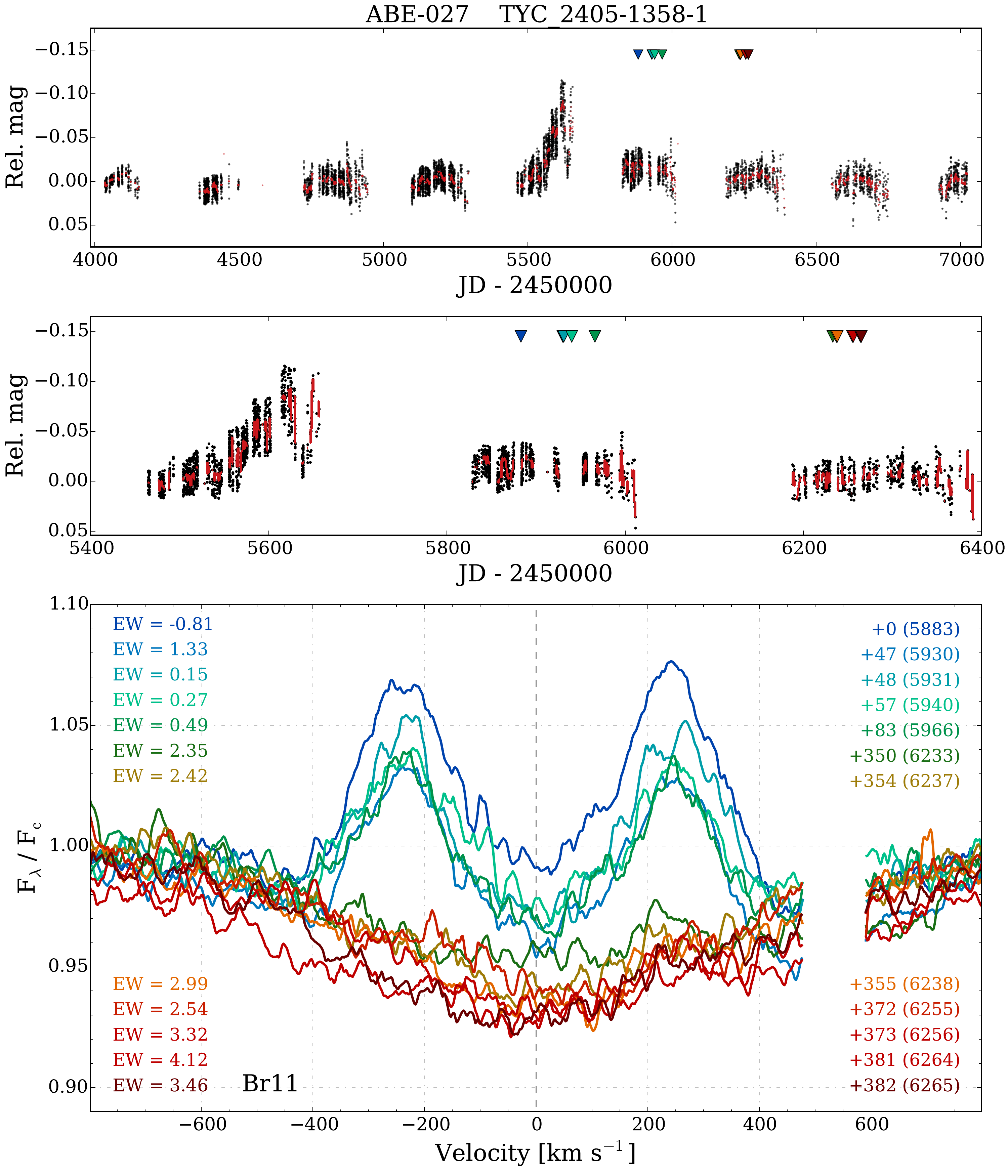,clip=,width=0.99\linewidth}
\caption{Same as Figure~\ref{fig:Outburst_098}, but for star ABE-027.}
\label{fig:OTB_027}
\end{figure}

A photometric outburst with an unusual morphology occurs near JD - 2450000 $=$ 5600. A slow rising phase is followed by a much shorter falling phase, but the system beings to brighten again before the falling phase is complete. A gap in coverage prevents a better understanding of this event. In the season following the outburst, there is a very slight photometric excess, and Br11 shows a clear disk (Figure~\ref{fig:OTB_027}). A second grouping of APOGEE spectra, taken $\sim$300 days after the first group, shows no sign of emission, and there is no longer any detectable photometric excess.

\subsection{ABE-082}
\begin{figure}[!ht]
\centering\epsfig{file=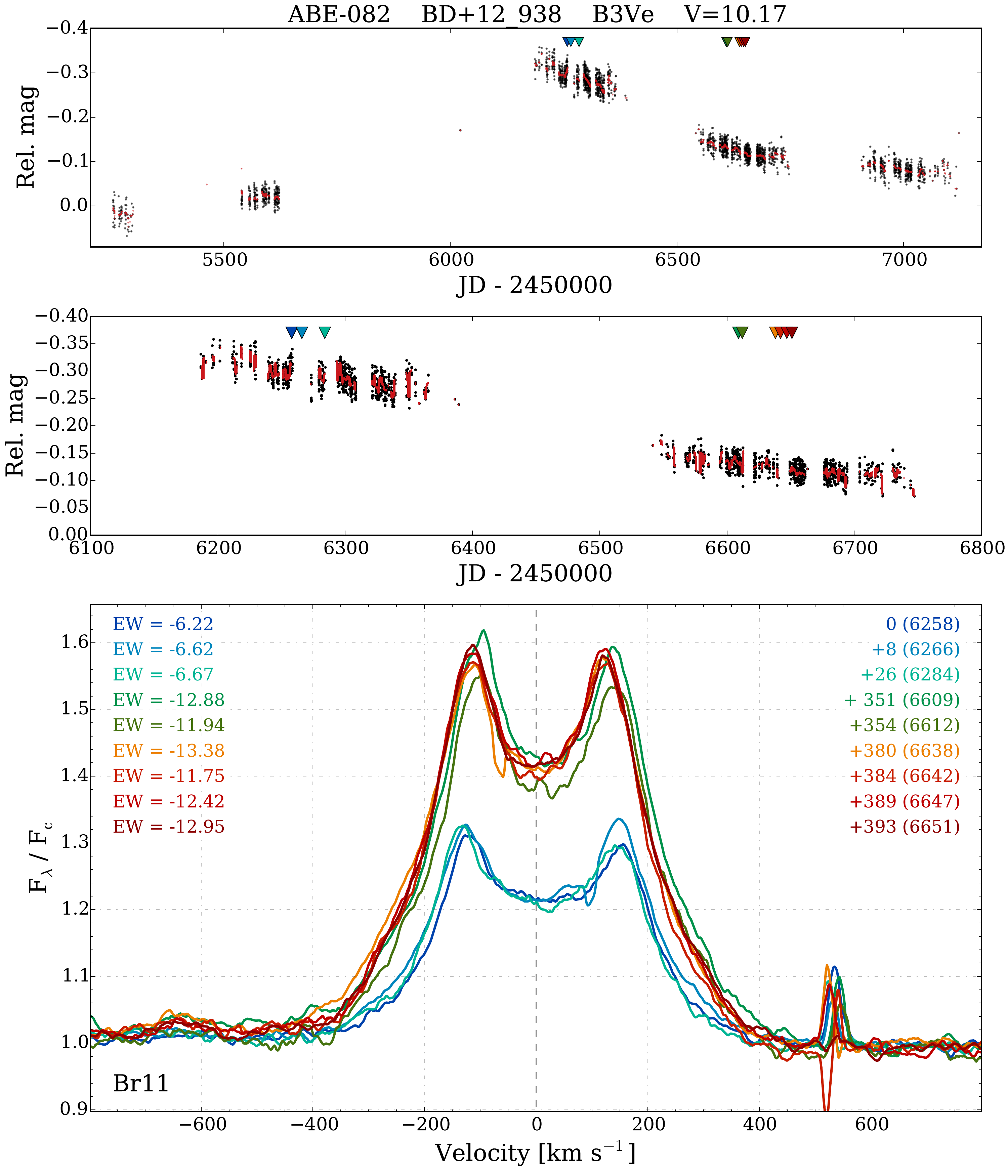,clip=,width=0.99\linewidth}
\caption{Same as Figure~\ref{fig:Outburst_098}, but for star ABE-082.}
\label{fig:OTB_082}
\end{figure}

The large photometric amplitude and long falling timescale apparent in the light curve of ABE-082 imply that a large amount of mass is ejected in this outburst, although the rising phase occurs during a gap in coverage (see Figure~\ref{fig:OTB_082}). The first grouping of three APOGEE spectra have an average \WBr $= -$6.50 {\AA} and an average Br11 $\Delta v_p$ $=$ 273.2 km s$^{-1}$. The second grouping of six spectra, taken about a year later, shows much stronger emission with an average \WBr $= -$12.05 {\AA}, and a lower peak separation, with the average Br11 $\Delta v_p$ $=$ 239.1 km s$^{-1}$. The diminished peak separation from the first to the second group of spectra indicates that the preferential formation radius has increased, and that the disk is dissipating. This is corroborated by the decreasing visible continuum flux in the KELT light curve. It is perhaps unexpected that the strength of the Br11 line appears to increase as the disk is dissipating. Because the strength of the emission line is measured here relative to the local continuum, the apparent increase in strength is best explained not by an absolute increase in the amount of Br11 line emission, but rather by a decrease in the strength of the local continuum. This example demonstrates the need for caution when interpreting relative line strengths for systems with significant continuum variability.

\subsection{ABE-154}
\begin{figure}[!ht]
\centering\epsfig{file=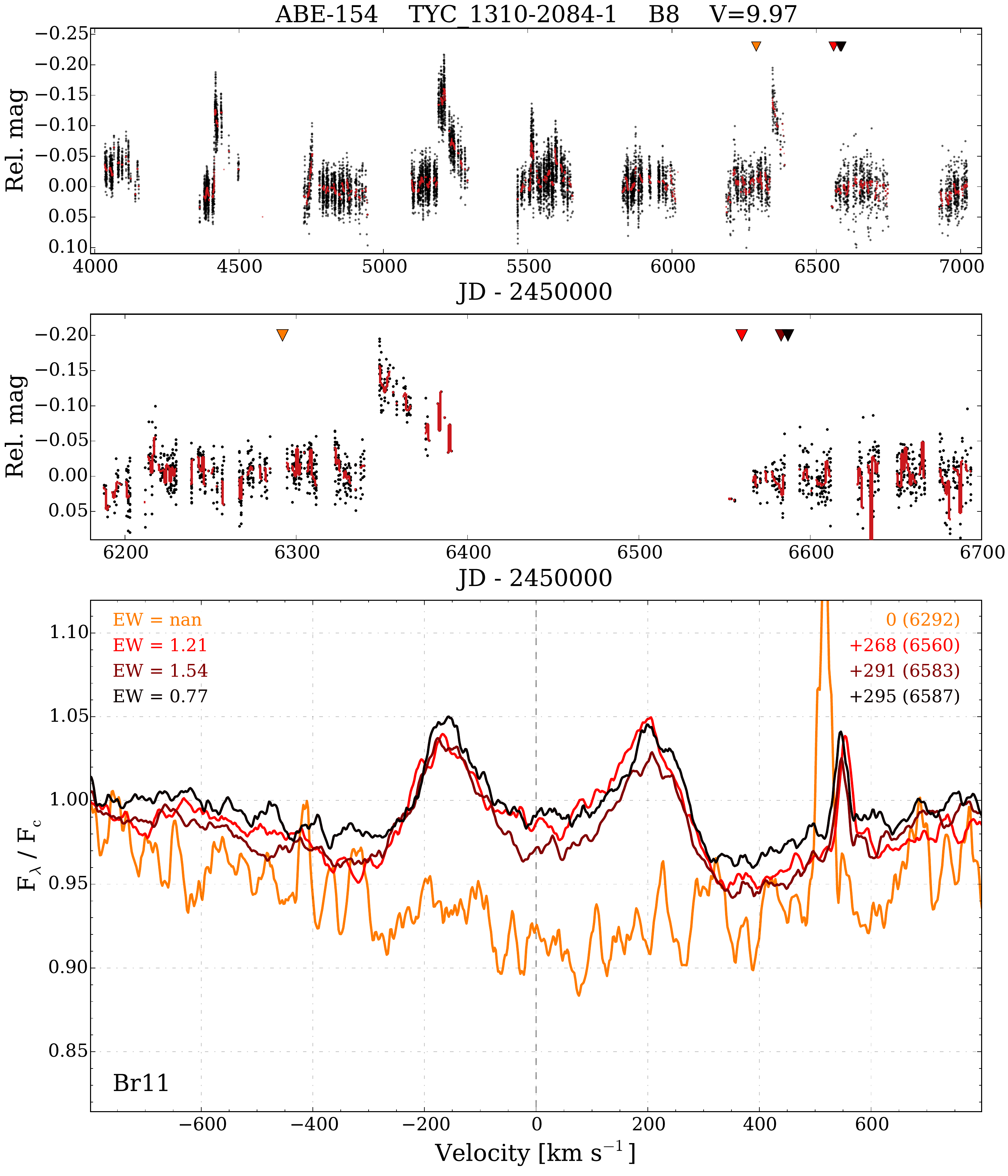,clip=,width=0.99\linewidth}
\caption{Same as Figure~\ref{fig:Outburst_098}, but for star ABE-154.}
\label{fig:OTB_154}
\end{figure}

This system has six detected photometric outbursts, one of which is bracketed by APOGEE spectra, as seen in Figure~\ref{fig:OTB_154}. Prior to the outburst, there is no sign of a disk. The Br11 line $\sim$200 days after the outburst shows a clear disk signature, and there is no photometric excess. Although the coverage of the falling phase of this outburst is incomplete, it is clearly many times longer than the rising phase. This is the case with all observed outbursts in this late-type star. 

\subsection{ABE-160}
\begin{figure}[!ht]
\centering\epsfig{file=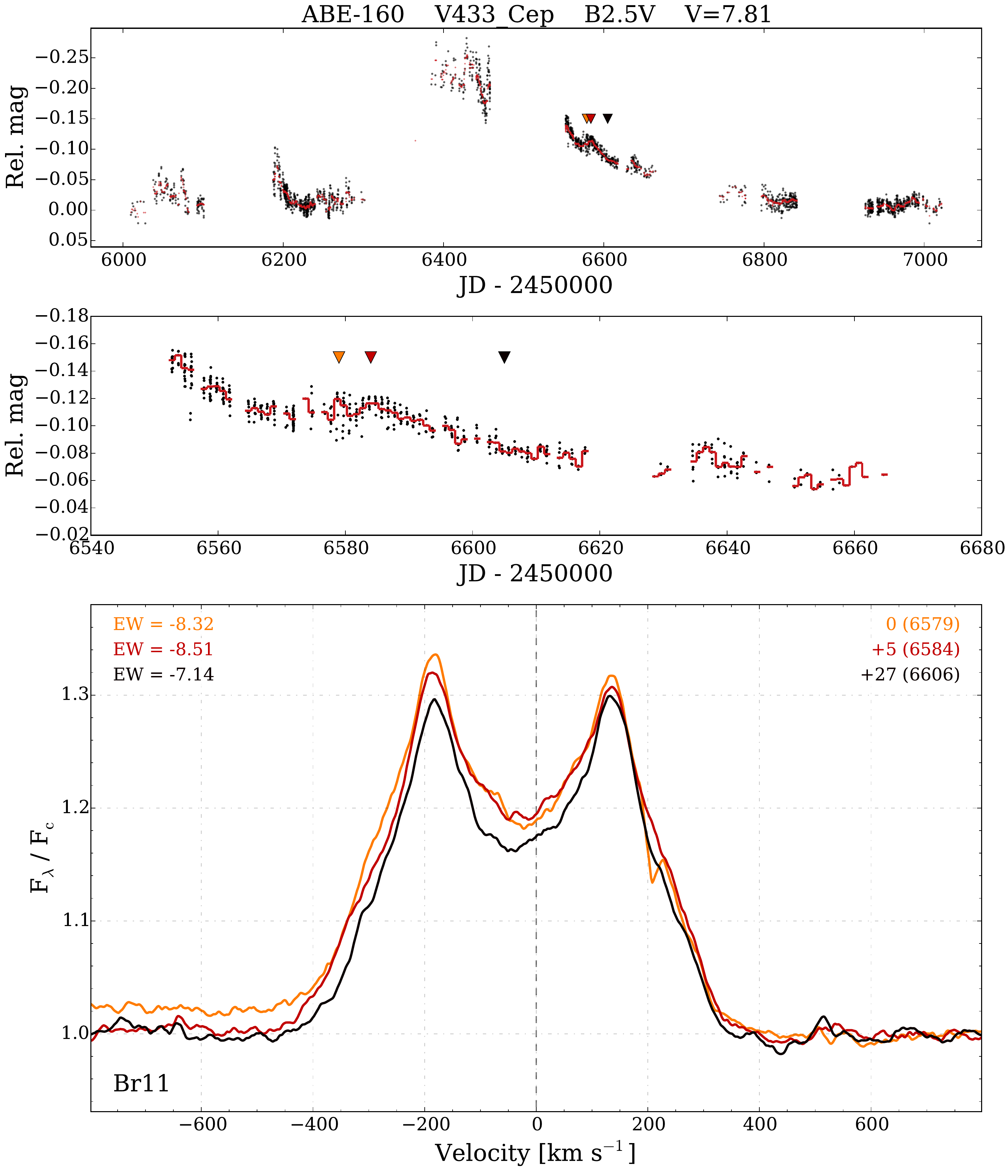,clip=,width=0.99\linewidth}
\caption{Same as Figure~\ref{fig:Outburst_098}, but for star ABE-160.}
\label{fig:OTB_160}
\end{figure}

This system has one large photometric outburst (reaching peak brightness near JD-2450000 $=$ 6400), as well as a few smaller ones (see Figure~\ref{fig:OTB_160}). Spectra taken during the falling phase of the major outburst show a clear disk signature. The emission strength relative to the local continuum decreases slightly during the 27 day APOGEE baseline. 

\subsection{ABE-162}
\begin{figure}[!ht]
\centering\epsfig{file=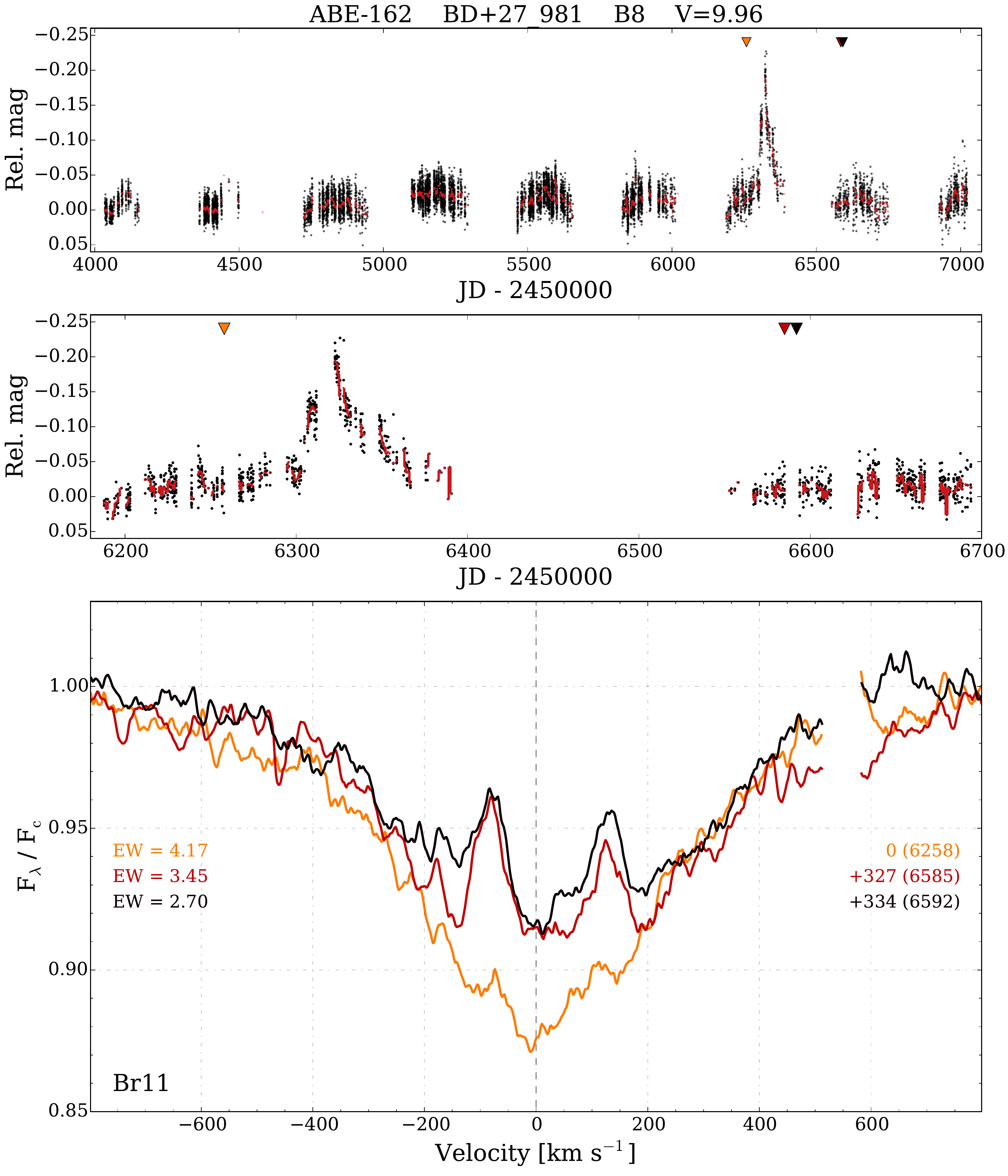,clip=,width=0.99\linewidth}
\caption{Same as Figure~\ref{fig:Outburst_098}, but for star ABE-162.}
\label{fig:OTB_162}
\end{figure}

Similar to ABE-154, there is a clear outburst in photometry seen in this late-type star. A spectrum taken prior to the outburst shows no sign of a disk in Br11, while spectra taken after the outburst do indicate the presence of a disk, even after the system has returned to its baseline brightness (see Figure~\ref{fig:OTB_162}). 

\subsection{ABE-164}
\begin{figure}[!ht]
\centering\epsfig{file=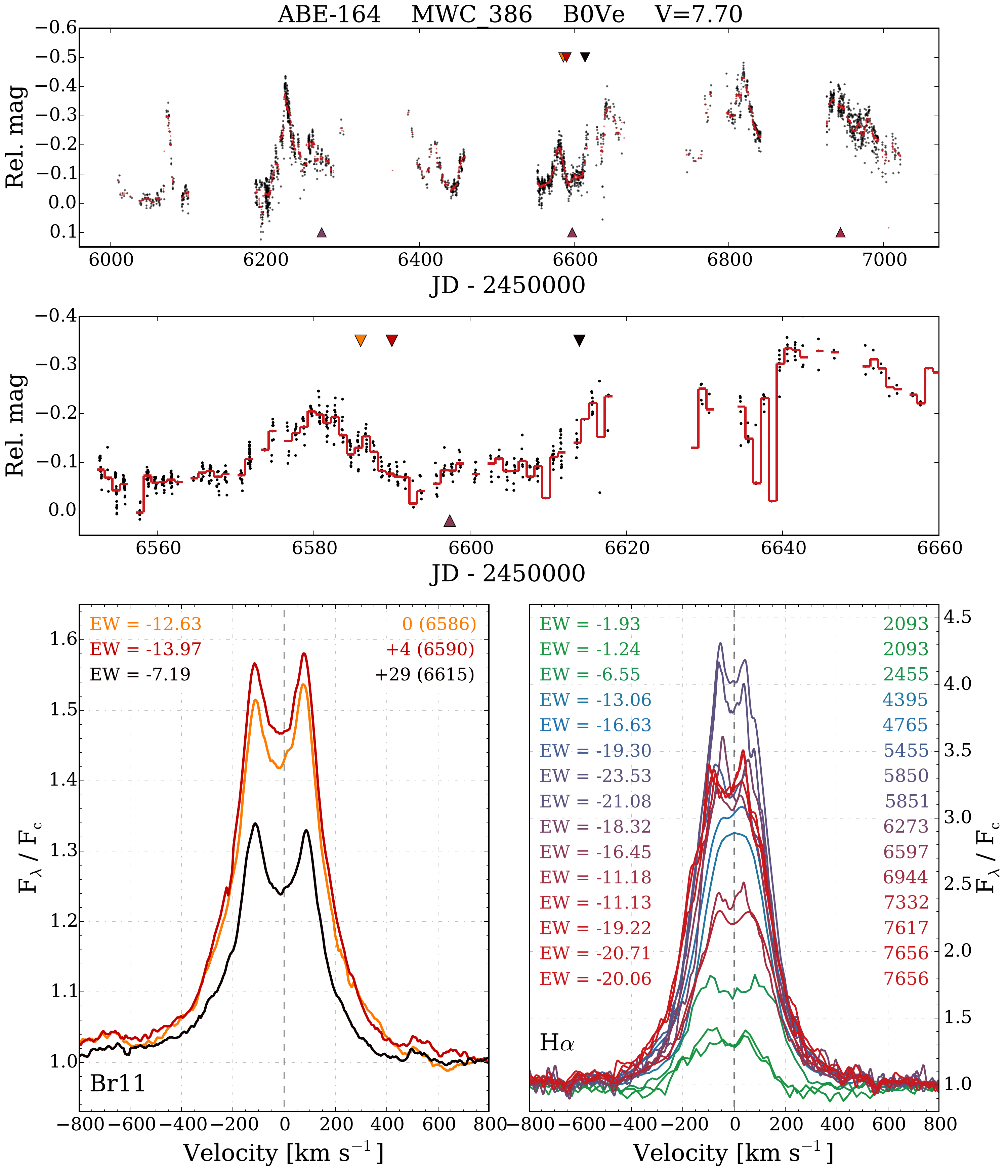,clip=,width=0.99\linewidth}
\caption{Same as Figure~\ref{fig:Outburst_138}, but for star ABE-164.}
\label{fig:OTB_164}
\end{figure}

This early-type system is very active in photometry, spending virtually no time in a photometrically quiescent state. Fifteen H$\alpha$ measurements from BeSS span 5563 days, and all show the presence of a disk, which varies significantly in strength. The H$\alpha$ EW spans an order of magnitude, ranging between $-$1.24 and $-$23.53. The line strength reaches to over four times the continuum level. Three Br11 measurements all show a strong disk signature (see Figure~\ref{fig:OTB_164}).

\subsection{ABE-167}
\begin{figure}[!ht]
\centering\epsfig{file=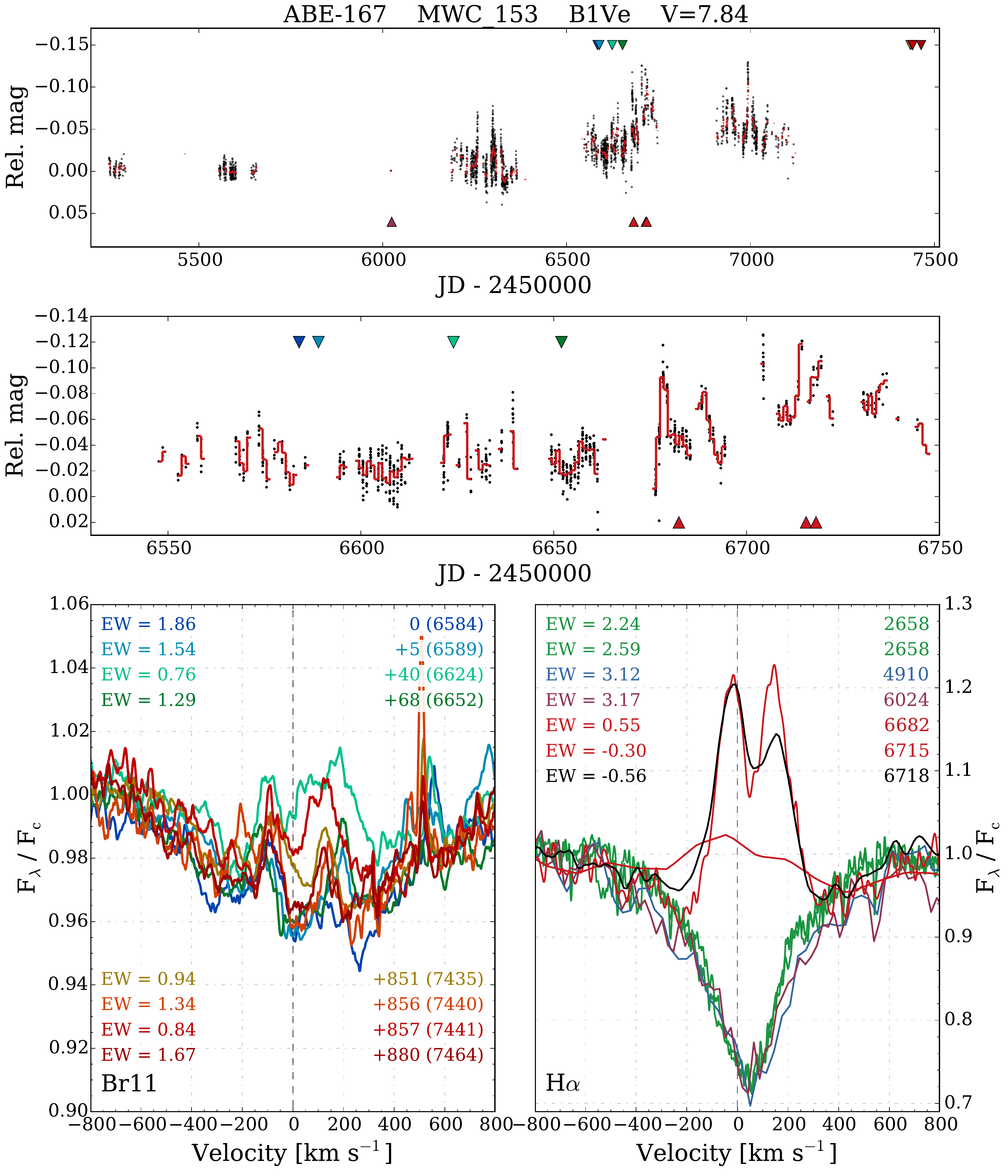,clip=,width=0.99\linewidth}
\caption{Same as Figure~\ref{fig:Outburst_138}, but for star ABE-167. The sharp feature in the lower-left panel near velocity = 500 km s$^{-1}$ is a detector artifact.}
\label{fig:OTB_167}
\end{figure}

Four BeSS spectra from between JD-2450000 = 2658 -- 6024 show no sign of a disk. Some activity in the KELT light curve begins near JD-245000 = 6250, with sparse photometric coverage prior to this point. As the system becomes slightly brighter, the Br11 line shows variability, indicating activity in the circumstellar environment. By JD-2450000 = 6715, the brightness is markedly above baseline and H$\alpha$ is clearly in emission. This system appears to grow a disk from many closely spaced low-amplitude mass-loss events, rather than from a singular well-defined event (see Figure~\ref{fig:OTB_167}). The BeSS spectrum taken at JD-2450000 = 6682 is of low resolution, but does indicate the presence of emitting material. The final two BeSS spectra show clear double-peaked H$\alpha$ emission. A significant change in the V/R ratio is apparent between the second to last BeSS spectrum (JD-2450000 = 6715) having V/R $\approx$ 1, and the final BeSS spectrum (JD-2450000 = 6718) showing clear asymmetry, with V/R $>$ 1. With just three days between these two spectra, the rapid change in the V/R ratio likely has its origins in an asymmetrical inner disk that is still in the process of circularizing. This hypothesis is supported by the relatively high photometric state near these epochs (implying a relatively dense inner disk), as well as the high level of photometric activity (implying active episodes of mass loss).

\subsection{ABE-176}
\begin{figure}[!ht]
\centering\epsfig{file=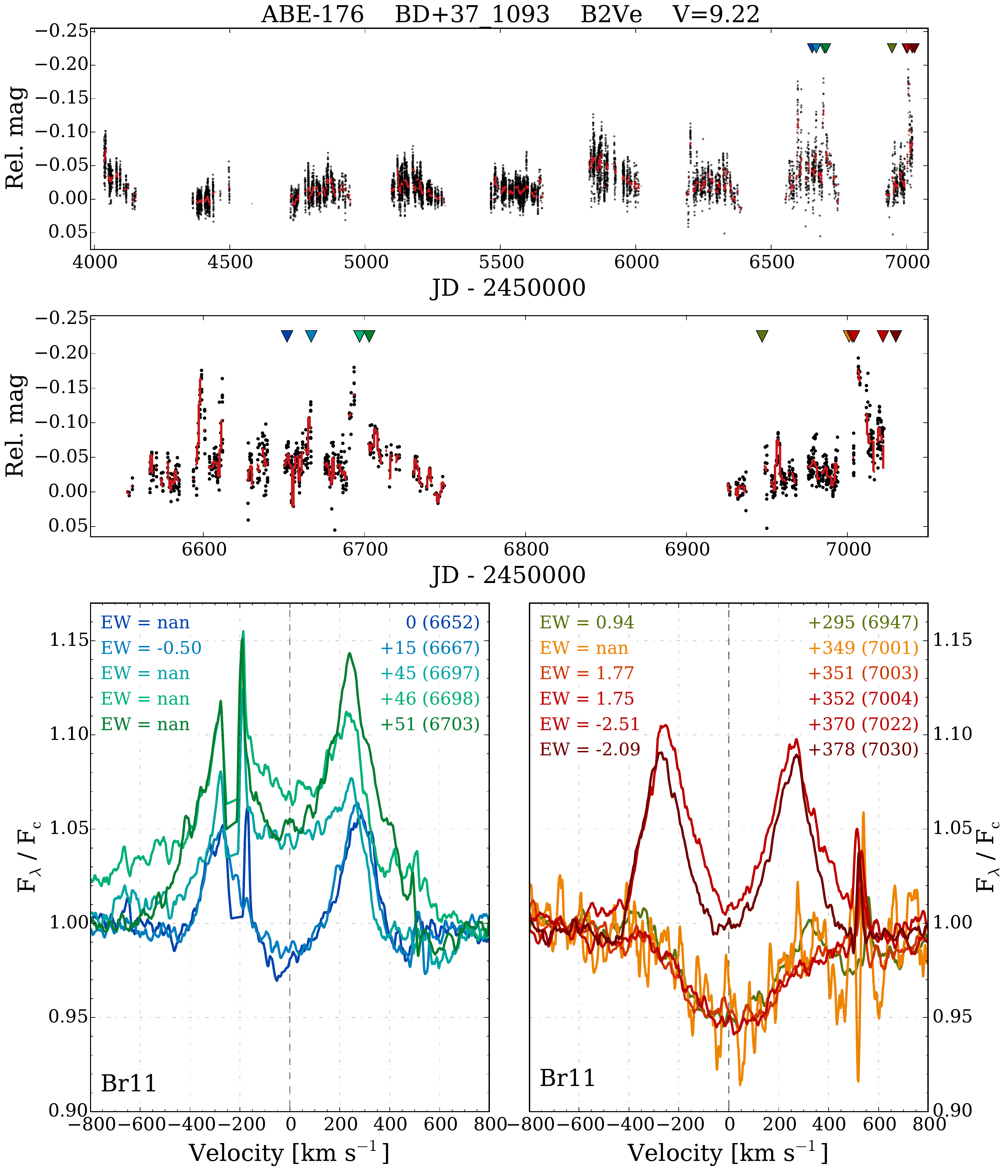,clip=,width=0.99\linewidth}
\caption{Same as Figure~\ref{fig:Outburst_098}, but for star ABE-176. The feature at the violet peak of the emission lines plotted in the bottom-left panel is a detector artifact and is not astrophysical.}
\label{fig:OTB_176}
\end{figure}

There are two groupings of APOGEE spectra, both with simultaneous photometry. During the first grouping, the double-peaked emission increases in strength, seemingly associated with increased photometric activity. The disk has subsequently dissipated by the beginning of the second grouping, as the next four spectra (at JD$_{0}$ +295, +349, +351, and +352 days) show no disk. The final two spectra are preceded by a photometric outburst and clearly show the presence of a disk (see Figure~\ref{fig:OTB_176}).

\subsection{ABE-184}
\begin{figure}[!ht]
\centering\epsfig{file=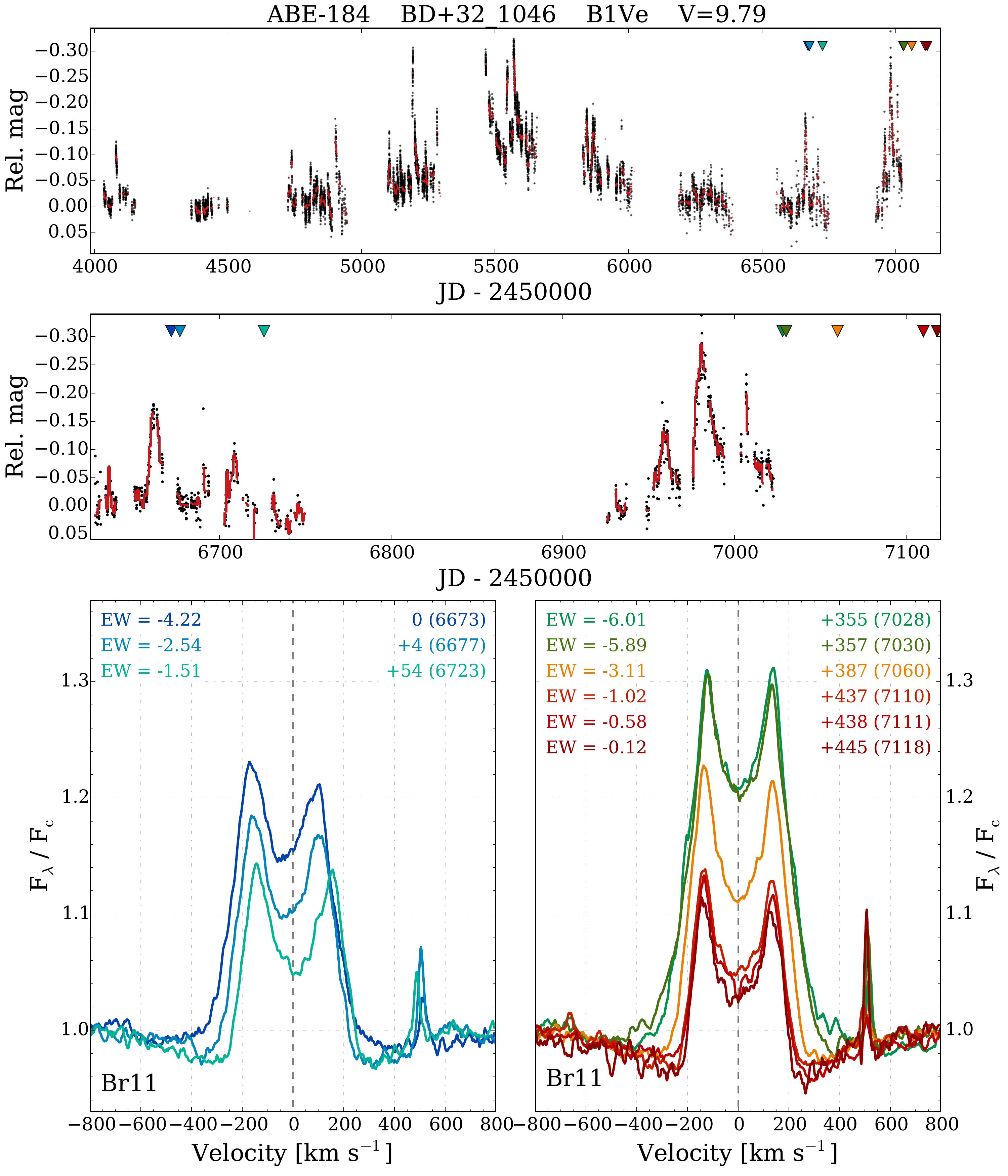,clip=,width=0.99\linewidth}
\caption{Same as Figure~\ref{fig:Outburst_138}, but for star ABE-184.}
\label{fig:OTB_184}
\end{figure}

This system is highly active in photometry and is often in an outbursting state. A clear disk is present in all spectroscopic epochs, varying in strength. There is appreciable RV variation, which is especially apparent in the first three spectra (see Figure~\ref{fig:OTB_184}). One of the most RV-variable objects in the APOGEE Be star sample, this is identified as a possible binary in \citet{Chojnowski2017}.

\subsection{ABE-187}
\begin{figure}[!ht]
\centering\epsfig{file=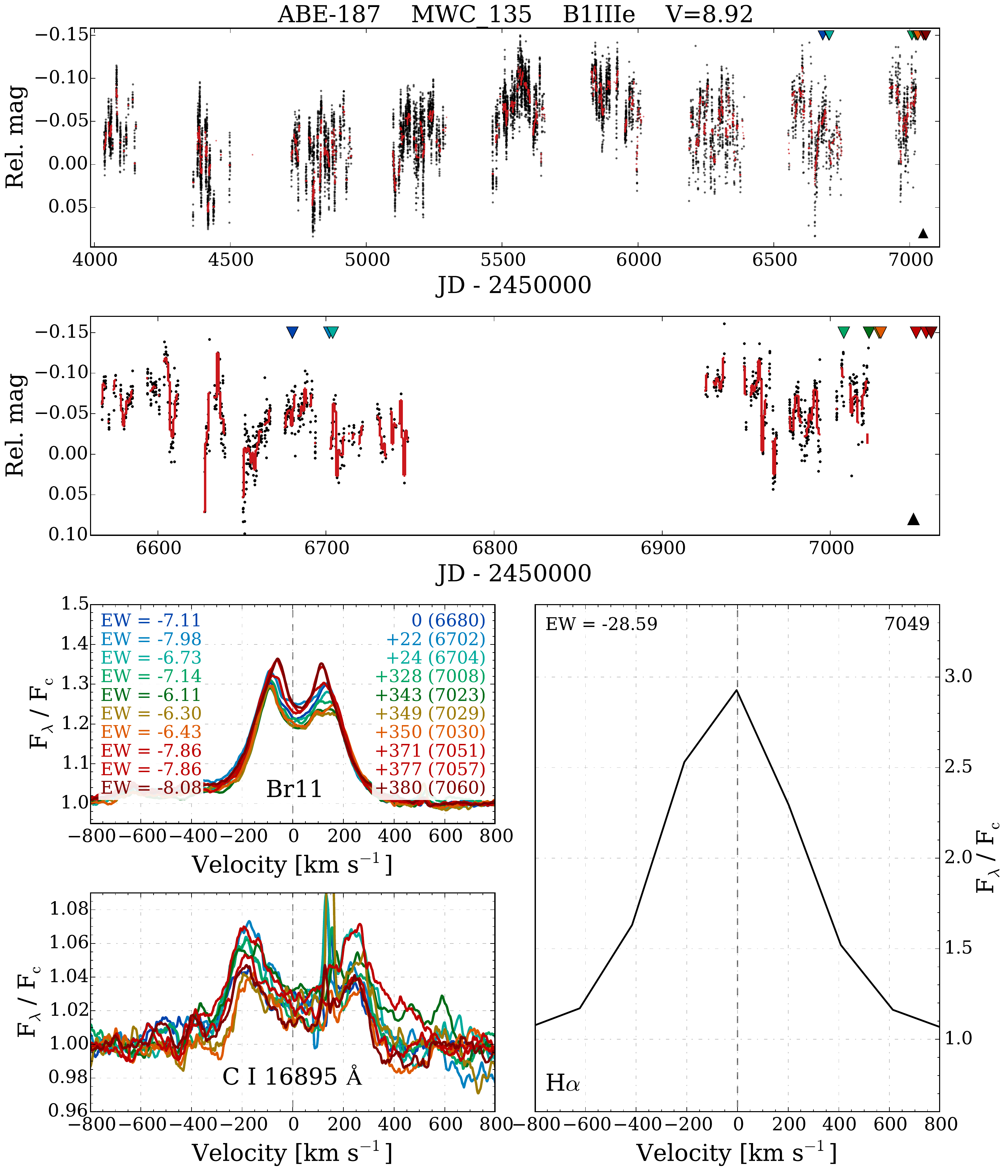,clip=,width=0.99\linewidth}
\caption{Same as Figure~\ref{fig:Outburst_138}, but for star ABE-187. The H$\alpha$ line is from a low-resolution BeSS spectrum, and its features are not well-defined.}
\label{fig:OTB_187}
\end{figure}

The light curve of this system lacks well-defined outbursts, but there is stochastic variability that persists throughout the entire observational baseline. All 10 APOGEE spectra show a clear disk, as does the single low-resolution spectrum from BeSS. The Br11 line is variable, but within a well-defined ``envelope.'' The lack of variability in the Br11 envelope implies that no significant changes in the inner disk occur. A strong C {\sc i} 16895 feature is present in all APOGEE spectra. The very large peak separation, relative to the Br11 line, indicates that it is formed in the circumstellar environment close to the star. This seems to suggest that there is not a large gap between the star and disk, possibly indicating that the disk is fed nearly continuously (see Figure~\ref{fig:OTB_187}). 

\subsection{ABE-A03}
\begin{figure}[!ht]
\centering\epsfig{file=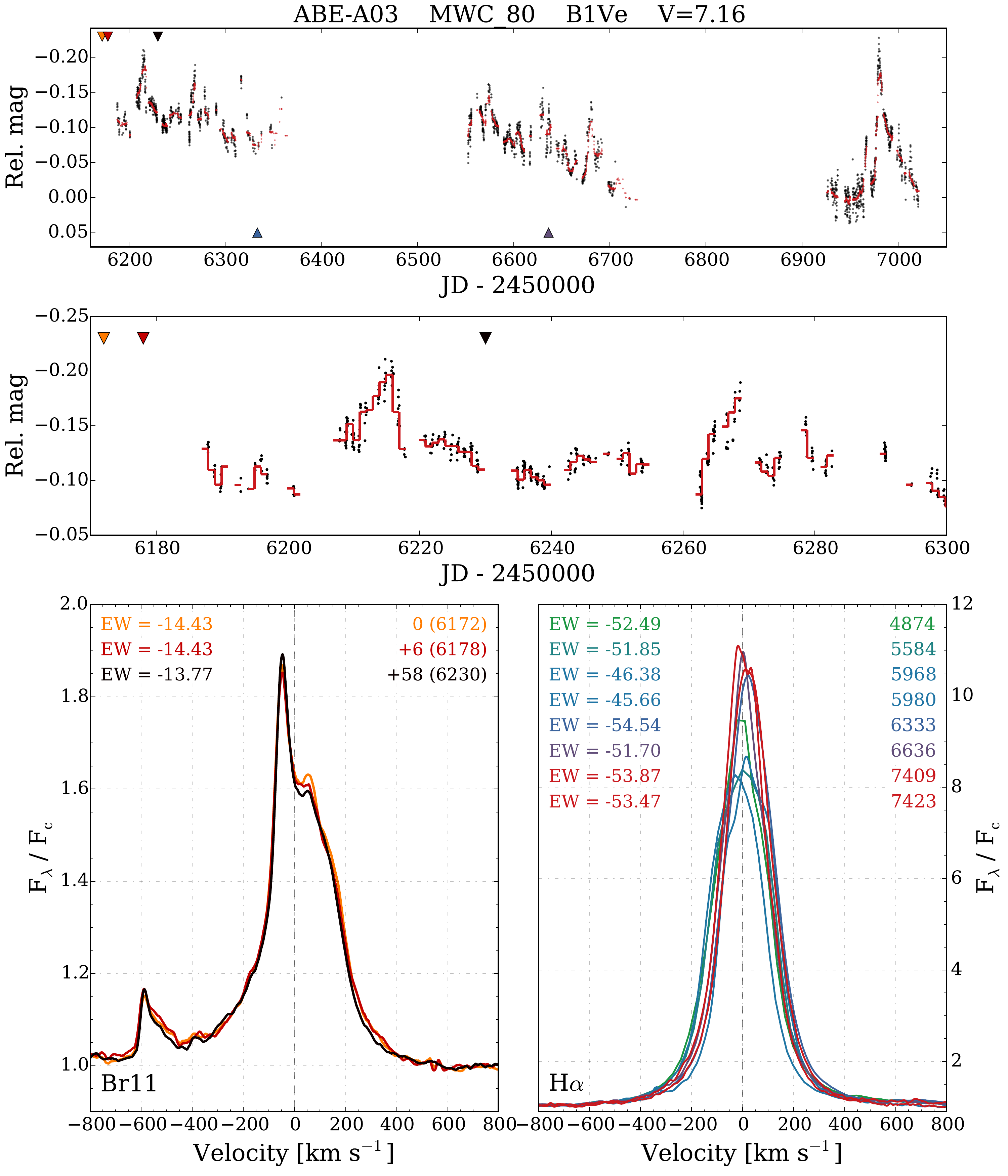,clip=,width=0.99\linewidth}
\caption{Same as Figure~\ref{fig:Outburst_138}, but for star ABE-A03.}
\label{fig:OTB_A03}
\end{figure}

This system has remarkably strong and persistent single-peaked H$\alpha$ emission (see Figure~\ref{fig:OTB_A03}). The Br11 line is also strong, with an interesting profile showing a strong violet enhancement at all epochs. The line changes little over the 58 day APOGEE baseline, despite the presence of an outburst about 15 days prior to the final spectrum. The light curve shows gradual dimming over the three KELT seasons, with many outbursts interspersed.

\subsection{ABE-A26}
\begin{figure}[!ht]
\centering\epsfig{file=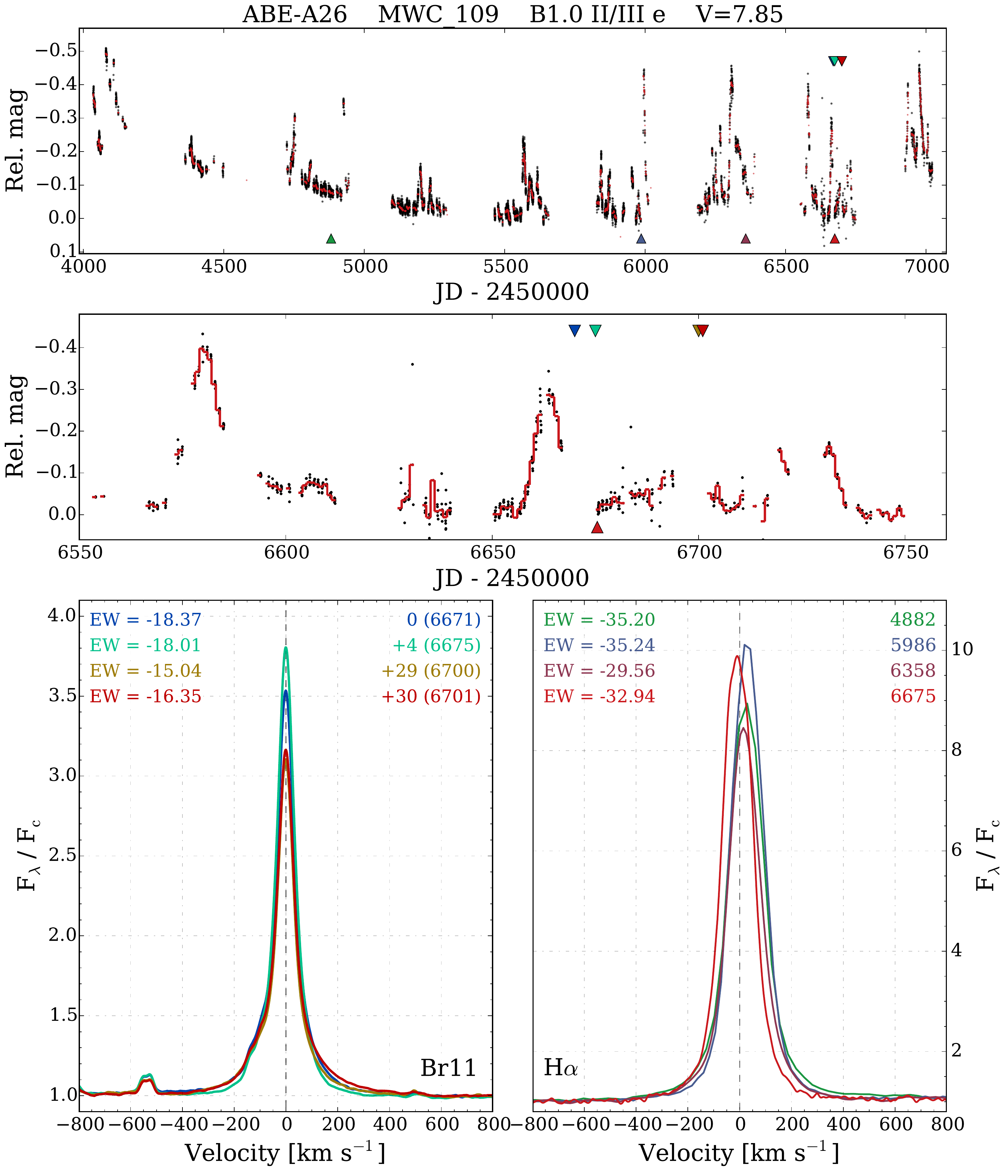,clip=,width=0.99\linewidth}
\caption{Same as Figure~\ref{fig:Outburst_138}, but for star ABE-A26.}
\label{fig:OTB_A26}
\end{figure}

A high level of activity is obvious in the light curve of this system (see Figure~\ref{fig:OTB_A26}). Many high-amplitude outbursts occur in rapid succession. This is viewed at a very low inclination angle, as all spectra show single-peaked emission profiles. The disk is remarkably strong compared to other systems in this sample, with the Br11 emission peak reaching to nearly four times the continuum, and H$\alpha$ reaching a peak around 10 times the continuum level. 

\newpage

\begin{longtable*}{p{0.3cm}cccccccc}
\tablecaption{Well-sampled and well-defined outbursts\label{tbl:otb-tbl}} 

\tabletypesize{\scriptsize}
\tablehead{ 
\colhead{ABE} & \colhead{Rise}    &  \colhead{Fall}  &  \colhead{Amp.} & \colhead{Rising Phase} & \colhead{Peak}         & \colhead{Falling Phase} & \colhead{Baseline}   & \colhead{Maximum}   \\
\colhead{ID}  &  \colhead{Time}   & \colhead{Time}   &      (mag)            & \colhead{Begins}       & \colhead{Brightness}   & \colhead{Ends}          & \colhead{Brightness} & \colhead{Brightness} \\
              &  \colhead{(days)} & \colhead{(days)} &                     & (JD-2450000)           & \colhead{(JD-2450000)} & \colhead{(JD-2450000)}  & \colhead{(mag)}      & \colhead{(mag)}     
}
\startdata
006 & 10 & 19 & 0.057 & 6157.9 $\pm$ 1 & 6167.6 $\pm$ 4 & 6186.9 $\pm$ 3 & 9.274 $\pm$ 0.005 & 9.217 $\pm$ 0.011\\ 
006 & 16 & 96 & 0.181 & 6510.4 $\pm$ 3 & 6526.4 $\pm$ 9 & 6622.0 $\pm$ 42 & 9.314 $\pm$ 0.006 & 9.133 $\pm$ 0.011 \bigstrut[b] \\ 
\hline 
010 & 11 & 21 & 0.034 & 6958.4 $\pm$ 3 & 6969.3 $\pm$ 1 & 6990.0 $\pm$ 5 & 10.002 $\pm$ 0.002 & 9.968 $\pm$ 0.003 \bigstrut \\ 
\hline 
019 & 11 & 21 & 0.138 & 6936.6 $\pm$ 2 & 6947.2 $\pm$ 2 & 6968.5 $\pm$ 7 & 8.088 $\pm$ 0.011 & 7.950 $\pm$ 0.010 \bigstrut \\ 
\hline 
020 & 12 & 46 & 0.059 & 5682.3 $\pm$ 6 & 5694.3 $\pm$ 2 & 5739.9 $\pm$ 10 & 12.008 $\pm$ 0.008 & 11.950 $\pm$ 0.008 \bigstrut \\  
\hline 
025 & 17 & 20 & 0.055 & 5114.1 $\pm$ 5 & 5130.9 $\pm$ 2 & 5150.6 $\pm$ 9 & 9.661 $\pm$ 0.005 & 9.607 $\pm$ 0.005\bigstrut[t] \\ 
025 & 13 & 15 & 0.058 & 5228.8 $\pm$ 4 & 5241.8 $\pm$ 2 & 5256.6 $\pm$ 5 & 9.656 $\pm$ 0.005 & 9.598 $\pm$ 0.008\\ 
025 & 19 & 30 & 0.083 & 5523.8 $\pm$ 6 & 5543.3 $\pm$ 3 & 5573.7 $\pm$ 11 & 9.630 $\pm$ 0.005 & 9.546 $\pm$ 0.008\\ 
025 & 8 & 26 & 0.049 & 5833.6 $\pm$ 3 & 5841.4 $\pm$ 3 & 5867.4 $\pm$ 5 & 9.628 $\pm$ 0.003 & 9.579 $\pm$ 0.006 \bigstrut[b] \\ 
\hline 
027 & 169 & 415 & 0.096 & 5483.9 $\pm$ 23 & 5653.3 $\pm$ 44 & 6068.0 $\pm$ 148 & 10.149 $\pm$ 0.008 & 10.053 $\pm$ 0.011 \bigstrut \\ 
\hline 
033 & 7 & 13 & 0.068 & 6188.9 $\pm$ 3 & 6196.2 $\pm$ 2 & 6209.7 $\pm$ 3 & 9.744 $\pm$ 0.006 & 9.677 $\pm$ 0.007 \bigstrut[t] \\ 
033 & 6 & 14 & 0.062 & 6211.3 $\pm$ 2 & 6217.1 $\pm$ 2 & 6231.2 $\pm$ 2 & 9.742 $\pm$ 0.004 & 9.680 $\pm$ 0.008 \bigstrut[b] \\ 
\hline 
082 & 161 & 1281 & 0.317 & 5863.5 $\pm$ 199 & 6024.1 $\pm$ 172 & 7304.9 $\pm$ 192 & 11.075 $\pm$ 0.019 & 10.758 $\pm$ 0.051 \bigstrut \\ 
\hline 
098 & 6 & 14 & 0.137 & 6228.6 $\pm$ 1 & 6234.5 $\pm$ 1 & 6248.1 $\pm$ 3 & 7.588 $\pm$ 0.006 & 7.451 $\pm$ 0.017 \bigstrut[t] \\ 
098 & 6 & 8 & 0.048 & 6592.5 $\pm$ 1 & 6598.8 $\pm$ 2 & 6607.1 $\pm$ 2 & 7.569 $\pm$ 0.003 & 7.521 $\pm$ 0.003 \bigstrut[b] \\ 
\hline 
105 & 15 & 31 & 0.235 & 6963.6 $\pm$ 3 & 6978.5 $\pm$ 2 & 7009.8 $\pm$ 5 & 9.407 $\pm$ 0.010 & 9.172 $\pm$ 0.007 \bigstrut\\ 
\hline 
138 & 5 & 6 & 0.101 & 6535.3 $\pm$ 2 & 6540.5 $\pm$ 1 & 6546.5 $\pm$ 2 & 8.588 $\pm$ 0.008 & 8.488 $\pm$ 0.008 \bigstrut[t] \\ 
138 & 5 & 16 & 0.096 & 6550.9 $\pm$ 4 & 6555.5 $\pm$ 3 & 6571.4 $\pm$ 4 & 8.592 $\pm$ 0.008 & 8.495 $\pm$ 0.012 \bigstrut[b] \\ 
\hline 
154 & 9 & 150 & 0.143 & 4408.1 $\pm$ 4 & 4417.1 $\pm$ 5 & 4567.3 $\pm$ 43 & 10.896 $\pm$ 0.006 & 10.753 $\pm$ 0.010 \bigstrut[t] \\ 
154 & 18 & 42 & 0.068 & 4735.1 $\pm$ 4 & 4753.5 $\pm$ 9 & 4795.6 $\pm$ 17 & 10.898 $\pm$ 0.006 & 10.830 $\pm$ 0.011\\ 
154 & 22 & 130 & 0.148 & 5183.9 $\pm$ 5 & 5205.9 $\pm$ 10 & 5336.3 $\pm$ 45 & 10.873 $\pm$ 0.009 & 10.726 $\pm$ 0.017\\ 
154 & 6 & 34 & 0.083 & 5507.3 $\pm$ 2 & 5513.0 $\pm$ 2 & 5547.0 $\pm$ 6 & 10.885 $\pm$ 0.004 & 10.802 $\pm$ 0.010\\ 
154 & 14 & 110 & 0.140 & 6332.4 $\pm$ 5 & 6346.3 $\pm$ 7 & 6456.2 $\pm$ 58 & 10.883 $\pm$ 0.005 & 10.744 $\pm$ 0.013 \bigstrut[b] \\ 
\hline 
160 & 91 & 419 & 0.204 & 6316.3 $\pm$ 48 & 6407.2 $\pm$ 44 & 6826.3 $\pm$ 123 & 8.249 $\pm$ 0.021 & 8.046 $\pm$ 0.026 \bigstrut \\ 
\hline 
162 & 34 & 74 & 0.162 & 6287.3 $\pm$ 13 & 6321.7 $\pm$ 6 & 6396.1 $\pm$ 26 & 10.162 $\pm$ 0.014 & 10.000 $\pm$ 0.010 \bigstrut \\ 
\hline 
164 & 8 & 15 & 0.289 & 6064.9 $\pm$ 2 & 6073.2 $\pm$ 2 & 6087.7 $\pm$ 5 & 7.480 $\pm$ 0.014 & 7.191 $\pm$ 0.017 \bigstrut[t] \\ 
164 & 22 & 40 & 0.341 & 6205.2 $\pm$ 3 & 6226.7 $\pm$ 2 & 6266.3 $\pm$ 15 & 7.463 $\pm$ 0.021 & 7.122 $\pm$ 0.020\\ 
164 & 13 & 13 & 0.125 & 6567.4 $\pm$ 2 & 6580.5 $\pm$ 2 & 6593.4 $\pm$ 2 & 7.431 $\pm$ 0.010 & 7.306 $\pm$ 0.010 \bigstrut[b] \\ 
\hline 
165 & 64 & 310 & 0.098 & 6485.8 $\pm$ 46 & 6549.5 $\pm$ 37 & 6859.3 $\pm$ 48 & 8.532 $\pm$ 0.013 & 8.434 $\pm$ 0.021 \bigstrut \\ 
\hline 
176 & 124 & 265 & 0.046 & 5708.8 $\pm$ 69 & 5832.9 $\pm$ 37 & 6098.3 $\pm$ 88 & 9.566 $\pm$ 0.011 & 9.520 $\pm$ 0.016 \bigstrut \\ 
\hline 
184 & 13 & 32 & 0.116 & 4063.0 $\pm$ 3 & 4076.1 $\pm$ 6 & 4108.3 $\pm$ 10 & 10.119 $\pm$ 0.005 & 10.003 $\pm$ 0.023 \bigstrut[t] \\ 
184 & 6 & 6 & 0.066 & 4732.4 $\pm$ 3 & 4738.0 $\pm$ 2 & 4743.9 $\pm$ 3 & 10.098 $\pm$ 0.011 & 10.032 $\pm$ 0.010\\ 
184 & 5 & 7 & 0.065 & 4806.1 $\pm$ 1 & 4811.2 $\pm$ 2 & 4818.1 $\pm$ 4 & 10.116 $\pm$ 0.003 & 10.050 $\pm$ 0.010\\ 
184 & 12 & 32 & 0.152 & 4886.4 $\pm$ 4 & 4898.5 $\pm$ 6 & 4930.9 $\pm$ 6 & 10.135 $\pm$ 0.007 & 9.982 $\pm$ 0.033\\ 
184 & 7 & 10 & 0.092 & 5097.0 $\pm$ 2 & 5103.8 $\pm$ 1 & 5113.6 $\pm$ 3 & 10.083 $\pm$ 0.006 & 9.991 $\pm$ 0.012\\ 
184 & 3 & 7 & 0.050 & 5130.5 $\pm$ 1 & 5133.5 $\pm$ 1 & 5140.0 $\pm$ 2 & 10.096 $\pm$ 0.003 & 10.046 $\pm$ 0.011\\ 
184 & 8 & 29 & 0.256 & 5185.4 $\pm$ 2 & 5193.3 $\pm$ 2 & 5222.2 $\pm$ 7 & 10.084 $\pm$ 0.005 & 9.828 $\pm$ 0.018\\ 
184 & 3 & 8 & 0.056 & 5236.3 $\pm$ 1 & 5239.3 $\pm$ 2 & 5247.6 $\pm$ 3 & 10.073 $\pm$ 0.006 & 10.017 $\pm$ 0.009\\ 
184 & 9 & 10 & 0.095 & 5272.1 $\pm$ 2 & 5280.8 $\pm$ 3 & 5290.8 $\pm$ 2 & 10.058 $\pm$ 0.007 & 9.963 $\pm$ 0.018\\ 
184 & 4 & 30 & 0.157 & 5564.8 $\pm$ 2 & 5568.9 $\pm$ 2 & 5598.5 $\pm$ 5 & 9.982 $\pm$ 0.011 & 9.825 $\pm$ 0.013\\ 
184 & 7 & 13 & 0.108 & 5835.4 $\pm$ 1 & 5842.6 $\pm$ 2 & 5855.2 $\pm$ 3 & 10.058 $\pm$ 0.006 & 9.949 $\pm$ 0.008\\ 
184 & 10 & 19 & 0.101 & 5857.1 $\pm$ 2 & 5867.0 $\pm$ 1 & 5885.8 $\pm$ 8 & 10.062 $\pm$ 0.006 & 9.961 $\pm$ 0.009\\ 
184 & 7 & 13 & 0.149 & 6654.6 $\pm$ 3 & 6661.6 $\pm$ 2 & 6674.6 $\pm$ 4 & 10.101 $\pm$ 0.007 & 9.951 $\pm$ 0.010\\ 
184 & 10 & 10 & 0.134 & 6949.8 $\pm$ 2 & 6959.5 $\pm$ 1 & 6969.1 $\pm$ 3 & 10.120 $\pm$ 0.011 & 9.986 $\pm$ 0.010\\ 
184 & 10 & 22 & 0.234 & 6971.5 $\pm$ 3 & 6981.1 $\pm$ 1 & 7003.0 $\pm$ 9 & 10.076 $\pm$ 0.014 & 9.842 $\pm$ 0.011 \bigstrut[b] \\ 
\hline 
A01 & 48 & 43 & 0.442 & 6159.1 $\pm$ 24 & 6206.8 $\pm$ 3 & 6249.4 $\pm$ 12 & 9.128 $\pm$ 0.029 & 8.687 $\pm$ 0.024 \bigstrut[t] \\ 
A01 & 6 & 9 & 0.079 & 6265.2 $\pm$ 2 & 6270.8 $\pm$ 3 & 6280.0 $\pm$ 3 & 9.163 $\pm$ 0.012 & 9.084 $\pm$ 0.019\\ 
A01 & 9 & 28 & 0.358 & 6280.8 $\pm$ 2 & 6289.9 $\pm$ 2 & 6318.0 $\pm$ 6 & 9.177 $\pm$ 0.013 & 8.819 $\pm$ 0.015\\ 
A01 & 16 & 38 & 0.301 & 6636.4 $\pm$ 3 & 6652.5 $\pm$ 3 & 6690.4 $\pm$ 8 & 9.161 $\pm$ 0.017 & 8.860 $\pm$ 0.011\\ 
A01 & 4 & 5 & 0.077 & 6982.5 $\pm$ 1 & 6986.5 $\pm$ 1 & 6991.7 $\pm$ 1 & 9.197 $\pm$ 0.005 & 9.120 $\pm$ 0.013 \bigstrut[b] \\ 
\hline 
A02 & 19 & 81 & 0.046 & 6644.8 $\pm$ 8 & 6663.5 $\pm$ 3 & 6744.1 $\pm$ 42 & 8.378 $\pm$ 0.004 & 8.332 $\pm$ 0.003 \bigstrut \\ 
\hline 
A03 & 15 & 20 & 0.101 & 6200.0 $\pm$ 4 & 6215.1 $\pm$ 1 & 6235.6 $\pm$ 5 & 8.302 $\pm$ 0.009 & 8.201 $\pm$ 0.006 \bigstrut[t] \\ 
A03 & 4 & 11 & 0.085 & 6310.8 $\pm$ 3 & 6315.2 $\pm$ 3 & 6326.0 $\pm$ 4 & 8.313 $\pm$ 0.005 & 8.229 $\pm$ 0.013\\ 
A03 & 8 & 16 & 0.103 & 6672.4 $\pm$ 2 & 6680.0 $\pm$ 2 & 6696.0 $\pm$ 5 & 8.370 $\pm$ 0.005 & 8.267 $\pm$ 0.007\\ 
A03 & 5 & 3 & 0.071 & 6962.0 $\pm$ 1 & 6967.2 $\pm$ 1 & 6970.6 $\pm$ 2 & 8.391 $\pm$ 0.007 & 8.320 $\pm$ 0.007\\ 
A03 & 6 & 38 & 0.174 & 6973.7 $\pm$ 1 & 6980.0 $\pm$ 1 & 7018.0 $\pm$ 6 & 8.377 $\pm$ 0.009 & 8.203 $\pm$ 0.013 \bigstrut[b] \\ 
\hline 
A16 & 8 & 48 & 0.178 & 5126.3 $\pm$ 2 & 5134.6 $\pm$ 3 & 5182.5 $\pm$ 17 & 10.582 $\pm$ 0.006 & 10.404 $\pm$ 0.012 \bigstrut[t] \\ 
A16 & 112 & 730 & 0.239 & 5682.4 $\pm$ 5 & 5794.2 $\pm$ 72 & 6524.0 $\pm$ 84 & 10.556 $\pm$ 0.005 & 10.317 $\pm$ 0.058 \bigstrut[b] \\ 
\hline 
A20 & 4 & 25 & 0.067 & 5868.6 $\pm$ 3 & 5872.8 $\pm$ 3 & 5897.7 $\pm$ 9 & 8.053 $\pm$ 0.005 & 7.987 $\pm$ 0.010 \bigstrut \\ 
\hline 
A26 & 9 & 106 & 0.288 & 4066.4 $\pm$ 7 & 4075.5 $\pm$ 7 & 4181.3 $\pm$ 24 & 8.206 $\pm$ 0.016 & 7.918 $\pm$ 0.028 \bigstrut[t] \\ 
A26 & 18 & 14 & 0.184 & 4735.1 $\pm$ 3 & 4753.4 $\pm$ 3 & 4767.3 $\pm$ 10 & 8.304 $\pm$ 0.011 & 8.120 $\pm$ 0.013\\ 
A26 & 12 & 8 & 0.055 & 4798.0 $\pm$ 3 & 4810.2 $\pm$ 3 & 4818.1 $\pm$ 5 & 8.314 $\pm$ 0.005 & 8.259 $\pm$ 0.006\\ 
A26 & 7 & 13 & 0.268 & 4920.2 $\pm$ 4 & 4926.7 $\pm$ 5 & 4939.9 $\pm$ 5 & 8.348 $\pm$ 0.005 & 8.080 $\pm$ 0.019\\ 
A26 & 12 & 12 & 0.117 & 5186.7 $\pm$ 3 & 5198.7 $\pm$ 3 & 5210.9 $\pm$ 3 & 8.393 $\pm$ 0.005 & 8.275 $\pm$ 0.009\\ 
A26 & 8 & 9 & 0.091 & 5227.5 $\pm$ 4 & 5235.7 $\pm$ 2 & 5244.9 $\pm$ 4 & 8.398 $\pm$ 0.010 & 8.307 $\pm$ 0.010\\ 
A26 & 14 & 11 & 0.040 & 5496.2 $\pm$ 4 & 5509.7 $\pm$ 3 & 5520.8 $\pm$ 3 & 8.428 $\pm$ 0.005 & 8.389 $\pm$ 0.009\\ 
A26 & 7 & 9 & 0.041 & 5523.1 $\pm$ 3 & 5529.8 $\pm$ 4 & 5538.9 $\pm$ 3 & 8.425 $\pm$ 0.005 & 8.384 $\pm$ 0.007\\ 
A26 & 8 & 18 & 0.200 & 5560.4 $\pm$ 3 & 5568.4 $\pm$ 3 & 5586.4 $\pm$ 8 & 8.404 $\pm$ 0.006 & 8.204 $\pm$ 0.025\\ 
A26 & 9 & 8 & 0.136 & 5835.6 $\pm$ 2 & 5844.1 $\pm$ 1 & 5851.9 $\pm$ 4 & 8.373 $\pm$ 0.011 & 8.236 $\pm$ 0.018\\ 
A26 & 6 & 9 & 0.097 & 5867.4 $\pm$ 1 & 5873.3 $\pm$ 3 & 5882.3 $\pm$ 5 & 8.389 $\pm$ 0.009 & 8.292 $\pm$ 0.016\\ 
A26 & 5 & 10 & 0.059 & 5970.1 $\pm$ 1 & 5975.1 $\pm$ 2 & 5985.5 $\pm$ 3 & 8.419 $\pm$ 0.008 & 8.361 $\pm$ 0.018\\ 
A26 & 8 & 23 & 0.428 & 5988.8 $\pm$ 4 & 5996.8 $\pm$ 3 & 6019.5 $\pm$ 7 & 8.414 $\pm$ 0.016 & 7.986 $\pm$ 0.024\\ 
A26 & 6 & 13 & 0.146 & 6230.4 $\pm$ 2 & 6236.7 $\pm$ 3 & 6249.4 $\pm$ 4 & 8.357 $\pm$ 0.018 & 8.210 $\pm$ 0.021\\ 
A26 & 7 & 10 & 0.195 & 6261.0 $\pm$ 4 & 6268.2 $\pm$ 2 & 6277.8 $\pm$ 2 & 8.359 $\pm$ 0.009 & 8.164 $\pm$ 0.020\\ 
A26 & 10 & 20 & 0.363 & 6569.9 $\pm$ 2 & 6579.9 $\pm$ 2 & 6599.8 $\pm$ 4 & 8.396 $\pm$ 0.008 & 8.033 $\pm$ 0.018\\ 
A26 & 7 & 9 & 0.290 & 6656.1 $\pm$ 2 & 6663.3 $\pm$ 1 & 6672.7 $\pm$ 3 & 8.410 $\pm$ 0.019 & 8.120 $\pm$ 0.023\\ 
A26 & 11 & 21 & 0.261 & 6963.6 $\pm$ 2 & 6975.1 $\pm$ 2 & 6996.4 $\pm$ 3 & 8.239 $\pm$ 0.011 & 7.977 $\pm$ 0.017 \bigstrut[b]\\ 
\hline 
A32 & 167 & 428 & 0.052 & 5337.7 $\pm$ 83 & 5504.6 $\pm$ 78 & 5932.7 $\pm$ 201 & 10.105 $\pm$ 0.006 & 10.053 $\pm$ 0.011 \bigstrut[t] \\ 
\hline 
\\\caption{Information about the rising and falling times and amplitudes, and their uncertainties, for 70 well-defined and sufficiently sampled outbursts for 24 stars. The values reported for the baseline and maximum brightness are in units of KELT instrumental magnitude, after subtraction of a constant offset (in order to approximate V-band magnitudes).}

\end{longtable*}

\begin{longtable*}{p{0.3cm}lrccccccccc}
\tablecaption{APOGEE + KELT Classical Be Star Sample\label{tbl:spec-table}} 
\tabletypesize{\scriptsize}
\tablehead{ 
\colhead{ABE} & \colhead{STAR} & \colhead{NOMAD}    &  \colhead{Spectral} & \colhead{Spectral Type}    & \colhead{T$_{eff}$} & \colhead{APOGEE}   & \colhead{KELT} & \colhead{First}  & \colhead{Last}  & \colhead{Number of} \\
\colhead{ID}  & \colhead{NAME} & \colhead{$V$ mag.} & \colhead{Type}      & \colhead{Reference}       & \colhead{Class}     & \colhead{visits} & \colhead{field} & \colhead{KELT obs.} & \colhead{KELT obs.}  & \colhead{Outbursts}  
}
\startdata
003 & HR 7757 & 6.55 & B7Ve & ARCES + AO & late & 11 & N11 & 2007-05-29 & 2013-06-13 & 0 & \\ 
004 & MWC 344 & 6.73 & B0IIIe & ARCES & early & 4 & N11 & 2007-05-29 & 2013-06-13 & 0 & \\ 
005 & Hen 3-1876 & 9.70 & OB & 1 & unclassified & 7 & N11 & 2007-05-29 & 2013-06-13 & 0 & \\ 
006 & MWC 615 & 8.08 & B2Ve & 2 & early & 3 & S14 & 2010-04-11 & 2014-10-23 & 3 & \\ 
007 & BD-05 4897 & 9.24 & B8II/III & 3 & late & 3 & S14 & 2010-04-11 & 2014-10-23 & 0 & \\ 
009 & TYC 3586-282-1 & 9.19 & B8 & 4 & late & 15 & N24 & 2012-03-23 & 2014-12-30 & 0 & \\ 
010 & BD+50 3188 & 9.31 & B3IIIe & ARCES & early & 3 & N24 & 2012-03-23 & 2014-12-30 & 3 & \\ 
011 & TYC 3583-670-1 & 9.70 & B3Ve & ARCES & early & 3 & N24 & 2012-03-23 & 2014-12-30 & 0 & \\ 
012 & WISE J205547.33+504028.8 & 10.77 & \nodata & New & unclassified & 3 & N24 & 2012-03-23 & 2014-12-30 & 0 & \\ 
013 & EM* CDS 1038 & 10.43 & B7Ve & ARCES & late & 17 & S13 & 2010-03-19 & 2015-03-19 & 0 & \\ 
014 & V2163 Cyg & 6.93 & B5IVe & ARCES + AO & mid & 20 & N24 & 2012-03-23 & 2014-12-30 & 0 & \\ 
019 & BD+56 3106 & 8.18 & B1IIIe & ARCES & early & 7 & N16 & 2012-05-21 & 2014-12-29 & 2 & \\ 
020 & SS 412 & 10.53 & OB:e & 5 & unclassified & 32 & S13 & 2010-03-19 & 2015-03-19 & 2 & \\ 
023 & BD+44 709s & 10.55 & OB & 6 & unclassified & 14 & N17 & 2012-09-17 & 2014-12-29 & 0 & \\ 
024 & TYC 1846-17-1 & 9.60 & A3 & 4 & late & 13 & N04 & 2006-10-26 & 2014-12-31 & 0 & \\ 
025 & BD+29 981 & 9.16 & B4Ve & ARCES & mid & 12 & N04 & 2006-10-26 & 2014-12-31 & 8 & \\ 
026 & V438 Aur & 8.02 & B2V & ARCES + AO & early & 12 & N04 & 2006-10-26 & 2014-12-31 & 3 & \\ 
027 & TYC 2405-1358-1 & 9.82 & B4V & ARCES & mid & 12 & N04 & 2006-10-26 & 2014-12-31 & 1 & \\ 
028 & MWC 794 & 8.09 & B8Ve & ARCES & late & 13 & N04 & 2006-10-26 & 2014-12-31 & 0 & \\ 
029 & BD+34 1307 & 9.17 & B7Ve & ARCES & late & 13 & N04 & 2006-10-26 & 2014-12-31 & 6 & \\ 
030 & BD+34 1318 & 8.81 & B8shell & ARCES & late & 13 & N04 & 2006-10-26 & 2014-12-31 & 0 & \\ 
032 & SS 453 & 10.20 & Be: & 5 & unclassified & 3 & N24 & 2012-03-23 & 2014-12-30 & 0 & \\ 
033 & BD+55 2936 & 9.25 & B4Ve & ARCES & mid & 3 & N16 & 2012-05-21 & 2014-12-29 & 17 & \\ 
034 & MWC 1085 & 8.79 & B3Ve & 7 & early & 3 & N16 & 2012-05-21 & 2014-12-29 & 0 & \\ 
037 & BD+31 1154 & 9.21 & B8 & 8 & late & 14 & N04 & 2006-10-26 & 2014-12-31 & 0 & \\ 
038 & BD+22 3902 & 10.60 & A3 & 9 & late & 20 & N11 & 2007-05-29 & 2013-06-13 & 0 & \\ 
045 & TYC 3692-1234-1 & 10.32 & B7shell & ARCES & late & 3 & N17 & 2012-09-17 & 2014-12-29 & 0 & \\ 
046 & V353 Per & 9.06 & B0III & ARCES & early & 3 & N17 & 2012-09-17 & 2014-12-29 & 0 & \\ 
047 & BD+37 1271 & 7.31 & B8Ve & ARCES & late & 3 & N04 & 2006-10-26 & 2014-12-31 & 0 & \\ 
048 & BD+42 4162 & 8.92 & B8shell & ARCES & late & 13 & N12 & 2007-06-08 & 2013-06-14 & 0 & \\ 
051 & BD+21 3985 & 9.87 & A0 & 9 & late & 3 & N11 & 2007-05-29 & 2013-06-13 & 4 & \\ 
054 & BD+22 825 & 6.52 & B8Ve & ARCES + AO & late & 12 & N04 & 2006-10-26 & 2014-12-31 & 0 & \\ 
055 & BD+04 1529 & 9.08 & B8Ve & AO & late & 15 & J06 & 2010-03-02 & 2015-05-06 & 0 & \\ 
057 & TYC 4056-415-1 & 9.29 & B5Ve & AO & mid & 3 & N17 & 2012-09-17 & 2014-12-29 & 0 & \\ 
060 & BD+38 1712 & 8.30 & B8shell & ARCES + AO & late & 3 & N05 & 2006-10-27 & 2012-04-22 & 0 & \\ 
062 & TYC 4060-96-1 & 8.40 & \nodata & New & unclassified & 3 & N17 & 2012-09-17 & 2014-12-29 & 0 & \\ 
063 & TYC 158-270-1 & 9.42 & B8III & 10 & late & 15 & S05 & 2010-02-28 & 2015-04-09 & 0 & \\ 
064 & TYC 5126-2325-1 & 10.73 & \nodata & New & unclassified & 3 & S13 & 2010-03-19 & 2015-03-19 & 0 & \\ 
065 & BD-06 4858 & 9.36 & B9IV & 3 & late & 3 & S13 & 2010-03-19 & 2015-03-19 & 0 & \\ 
066 & TYC 5121-940-1 & 10.30 & \nodata & New & unclassified & 3 & S13 & 2010-03-19 & 2015-03-19 & 3 & \\ 
067 & HR 1047 & 5.90 & B8Ve & ARCES + AO & late & 8 & N17 & 2012-09-17 & 2014-12-29 & 0 & \\ 
070 & BD-09 4724 & 9.55 & A0IV & 4 & late & 2 & S13 & 2010-03-19 & 2015-03-19 & 0 & \\ 
073 & BD+54 2887 & 9.54 & A0 & 11 & late & 3 & N16 & 2012-05-21 & 2014-12-29 & 1 & \\ 
074 & BD+38 3568 & 8.82 & B8V & AO & late & 18 & N11 & 2007-05-29 & 2013-06-13 & 0 & \\ 
077 & WISE J044231.14+383046.9 & 10.45 & \nodata & New & unclassified & 6 & N03 & 2006-10-25 & 2013-03-12 & 0 & \\ 
078 & TYC 3975-1585-1 & 10.10 & B8 & 12 & late & 3 & N24 & 2012-03-23 & 2014-12-30 & 0 & \\ 
080 & BD+44 3475 & 9.45 & \nodata & New & unclassified & 3 & N24 & 2012-03-23 & 2014-12-30 & 0 & \\ 
081 & BD+57 21 & 7.52 & B9V & AO & late & 3 & N16 & 2012-05-21 & 2014-12-29 & 0 & \\ 
082 & BD+12 938 & 10.17 & B3Ve & ARCES & early & 9 & S05 & 2010-02-28 & 2015-04-09 & 1 & \\ 
083 & BD+13 976 & 9.99 & A0 & 13 & late & 9 & S05 & 2010-02-28 & 2015-04-09 & 0 & \\ 
084 & MWC 683 & 8.98 & B8Ve & ARCES & late & 3 & N16 & 2012-05-21 & 2014-12-29 & 0 & \\ 
085 & NGC 457 198 & 8.85 & B1.5Vpsh & AO & early & 4 & N16 & 2012-05-21 & 2014-12-29 & 0 & \\ 
086 & TYC 3683-1262-1 & 9.84 & B5Ve & ARCES & mid & 4 & N17 & 2012-09-17 & 2014-12-29 & ? & \\ 
088 & MWC 10 & 6.84 & B8Ve & ARCES & late & 3 & N16 & 2012-05-21 & 2014-12-29 & 0 & \\ 
089 & TYC 4029-428-1 & 9.60 & \nodata & New & unclassified & 3 & N16 & 2012-05-21 & 2014-12-29 & 0 & \\ 
090 & BD+66 64 & 8.59 & B9 & 4 & late & 3 & N16 & 2012-05-21 & 2014-12-29 & 0 & \\ 
094 & MWC 671 & 8.85 & B7Ve & ARCES & late & 3 & N16 & 2012-05-21 & 2014-12-29 & 0 & \\ 
095 & BD+08 1343 & 8.91 & A2 & 8 & late & 3 & S05 & 2010-02-28 & 2015-04-09 & 0 & \\ 
096 & BD+08 1366 & 8.51 & B5Ve & ARCES & mid & 3 & S05 & 2010-02-28 & 2015-04-09 & 0 & \\ 
097 & MWC 488 & 8.50 & B6Ve & ARCES & mid & 3 & N04 & 2006-10-26 & 2014-12-31 & 0 & \\ 
098 & BD+63 1955 & 7.22 & B5V & ARCES + AO & mid & 3 & N16 & 2012-05-21 & 2014-12-29 & 1 & \\ 
099 & BD+27 991 & 8.60 & B6Vne: & 14 & mid & 3 & N04 & 2006-10-26 & 2014-12-31 & 0 & \\ 
102 & TYC 2400-1784-1 & 10.40 & \nodata & New & unclassified & 3 & N04 & 2006-10-26 & 2014-12-31 & 0 & \\ 
105 & BD+50 3189 & 8.65 & B0II & ARCES & early & 15 & N24 & 2012-03-23 & 2014-12-30 & 6 & \\ 
107 & TYC 3617-2074-1 & 10.11 & \nodata & New & unclassified & 3 & N24 & 2012-03-23 & 2014-12-30 & 0 & \\ 
108 & BD+23 1295 & 8.63 & \nodata & New & unclassified & 3 & N04 & 2006-10-26 & 2014-12-31 & 0 & \\ 
109 & BD+25 1244 & 9.73 & A2 & 8 & late & 3 & N04 & 2006-10-26 & 2014-12-31 & 0 & \\ 
111 & AS 332 & 9.64 & Be & 15 & unclassified & 17 & S13 & 2010-03-19 & 2015-03-19 & 0 & \\ 
113 & BD+40 999 & 7.32 & B8IV & AO & late & 3 & N03 & 2006-10-25 & 2013-03-12 & 0 & \\ 
128 & HIP 91591 & 8.82 & B8Ve & 16 & late & 6 & S13 & 2010-03-19 & 2015-03-19 & 0 & \\ 
129 & GSC 05692-00540 & 10.45 & B7 & 17 & late & 6 & S13 & 2010-03-19 & 2015-03-19 & 0 & \\ 
130 & GSC 05692-00399 & 10.51 & B7 & 17 & late & 6 & S13 & 2010-03-19 & 2015-03-19 & 0 & \\ 
131 & BD-07 4647 & 9.64 & B5 & 17 & mid & 6 & S13 & 2010-03-19 & 2015-03-19 & 0 & \\ 
132 & BD-07 4630 & 8.96 & B9 & 17 & late & 6 & S13 & 2010-03-19 & 2015-03-19 & 0 & \\ 
133 & 88 Her & 6.91 & B6IIInpsh & AO & mid & 3 & N23 & 2012-02-21 & 2014-11-30 & 0 & \\ 
134 & WISE J182959.95-090837.6 & 10.76 & \nodata & New & unclassified & 1 & S13 & 2010-03-19 & 2015-03-19 & 0 & \\ 
138 & V1448 Aql & 7.57 & B2IV & AO & early & 4 & S14 & 2010-04-11 & 2014-10-23 & 6 & \\ 
139 & BD+10 3849 & 7.58 & B9Vpsh & AO & late & 4 & S14 & 2010-04-11 & 2014-10-23 & 0 & \\ 
140 & HR 7807 & 6.23 & B2Vne & AO & early & 4 & N11 & 2007-05-29 & 2013-06-13 & 0 & \\ 
141 & BD+27 3970 & 9.00 & B7Ve & ARCES & late & 4 & N12 & 2007-06-08 & 2013-06-14 & 0 & \\ 
144 & BD+30 3853 & 7.12 & B6Ve & ARCES + AO & mid & 3 & N11 & 2007-05-29 & 2013-06-13 & 0 & \\ 
146 & BD+26 1082 & 7.13 & B9IV & 18 & late & 3 & N04 & 2006-10-26 & 2014-12-31 & 0 & \\ 
147 & BD+42 3425 & 8.48 & B9Va & AO & late & 2 & N11 & 2007-05-29 & 2013-06-13 & 0 & \\ 
148 & BD+21 4007 & 9.68 & B8 & 9 & late & 3 & N11 & 2007-05-29 & 2013-06-13 & 0 & \\ 
150 & WISE J184125.48-053403.7 & 10.90 & \nodata & New & unclassified & 3 & S13 & 2010-03-19 & 2015-03-19 & 0 & \\ 
152 & SS 120 & 10.73 & B8e: & 19 & late & 4 & J06 & 2010-03-02 & 2015-05-06 & 0 & \\ 
154 & TYC 1310-2084-1 & 9.97 & B8 & 11 & late & 4 & N04 & 2006-10-26 & 2014-12-31 & 6 & \\ 
155 & TYC 3692-1671-1 & 10.61 & B3Ve & ARCES & early & 3 & N17 & 2012-09-17 & 2014-12-29 & 0 & \\ 
156 & BD+55 2992 & 8.34 & A2 & 8 & late & 3 & N16 & 2012-05-21 & 2014-12-29 & 0 & \\ 
158 & AS 478 & 9.79 & B6Ve & ARCES & mid & 3 & N24 & 2012-03-23 & 2014-12-30 & 0 & \\ 
159 & MWC 1062 & 8.80 & B5:e & 20 & mid & 3 & N24 & 2012-03-23 & 2014-12-30 & 0 & \\ 
160 & V433 Cep & 7.81 & B2.5V & AO & early & 3 & N24 & 2012-03-23 & 2014-12-30 & 4 & \\ 
161 & TYC 3968-1354-1 & 10.57 & OB- & 21 & unclassified & 3 & N24 & 2012-03-23 & 2014-12-30 & 0 & \\ 
162 & BD+27 981 & 9.96 & B8 & 13 & late & 3 & N04 & 2006-10-26 & 2014-12-31 & 1 & \\ 
163 & BD+52 3293 & 8.10 & A0 & 8 & late & 3 & N16 & 2012-05-21 & 2014-12-29 & 0 & \\ 
164 & MWC 386 & 7.70 & B0Ve & ARCES + AO & early & 3 & N24 & 2012-03-23 & 2014-12-30 & 10 & \\ 
165 & MWC 1059 & 8.68 & B2Ve & ARCES + AO & early & 3 & N24 & 2012-03-23 & 2014-12-30 & 10 & \\ 
166 & TYC 4812-2496-1 & 9.97 & \nodata & New & unclassified & 3 & S05 & 2010-02-28 & 2015-04-09 & 0 & \\ 
167 & MWC 153 & 7.84 & B1Ve & 2 & early & 4 & S05 & 2010-02-28 & 2015-04-09 & 4 & \\ 
168 & V747 Mon & 8.22 & B3IIIe & ARCES & early & 4 & J06 & 2010-03-02 & 2015-05-06 & 0 & \\ 
169 & BD+22 1147 & 8.00 & B9 & 8 & late & 6 & N04 & 2006-10-26 & 2014-12-31 & 0 & \\ 
170 & HR 2116 & 6.40 & B8VSB2 & ARCES & late & 6 & N04 & 2006-10-26 & 2014-12-31 & 0 & \\ 
171 & TYC 1326-1188-1 & 10.26 & A2 & 13 & late & 6 & N04 & 2006-10-26 & 2014-12-31 & 0 & \\ 
173 & TYC 1283-1360-1 & 10.62 & \nodata & New & unclassified & 3 & S05 & 2010-02-28 & 2015-04-09 & 0 & \\ 
176 & BD+37 1093 & 9.22 & B2Ve & ARCES & early & 11 & N04 & 2006-10-26 & 2014-12-31 & ? & \\ 
177 & BD+38 1116 & 9.65 & B2.5Vne & AO & early & 11 & N04 & 2006-10-26 & 2014-12-31 & 0 & \\ 
179 & EM* RJHA 51 & 10.56 & B5Ib & 22 & mid & 2 & S05 & 2010-02-28 & 2015-04-09 & 0 & \\ 
180 & EM* RJHA 40 & 10.61 & B3Ib & 22 & early & 2 & S05 & 2010-02-28 & 2015-04-09 & 0 & \\ 
182 & TYC 2934-118-1 & 10.24 & B7Ve & ARCES & late & 8 & N04 & 2006-10-26 & 2014-12-31 & 0 & \\ 
184 & BD+32 1046 & 9.79 & B1Ve & ARCES & early & 9 & N04 & 2006-10-26 & 2014-12-31 & 25 & \\ 
185 & BD+24 1043 & 7.56 & B8Ve & ARCES & late & 3 & N04 & 2006-10-26 & 2014-12-31 & 0 & \\ 
186 & BD+01 1699 & 9.67 & B2II & ARCES & early & 3 & J06 & 2010-03-02 & 2015-05-06 & 7 & \\ 
187 & MWC 135 & 8.92 & B1IIIe & ARCES & early & 10 & N04 & 2006-10-26 & 2014-12-31 & ? & \\ 
188 & MWC 795 & 10.44 & B8Ve & ARCES & late & 10 & N04 & 2006-10-26 & 2014-12-31 & 0 & \\ 
196 & VES 860 & 10.84 & B8 & 23 & late & 9 & N04 & 2006-10-26 & 2014-12-31 & 0 & \\ 
204 & WISE J185142.47+134817.6 & 10.70 & \nodata & New & unclassified & 1 & S13 & 2010-03-19 & 2015-03-19 & 0 & \\ 
205 & BD+03 3861 & 7.88 & B8 & 24 & late & 3 & S13 & 2010-03-19 & 2015-03-19 & 0 & \\ 
A01 & MWC 5 & 8.02 & B0.5IVe & AO & early & 3 & N16 & 2012-05-21 & 2014-12-29 & 13 & \\ 
A02 & MWC 6 & 7.52 & B3:Vne & 25 & early & 3 & N16 & 2012-05-21 & 2014-12-29 & 3 & \\ 
A03 & MWC 80 & 7.16 & B1Ve & ARCES + AO & early & 3 & N17 & 2012-09-17 & 2014-12-29 & 13 & \\ 
A04 & MWC 494 & 7.95 & B0Ve & ARCES & early & 3 & N04 & 2006-10-26 & 2014-12-31 & ? & \\ 
A05 & MWC 125 & 8.38 & B0Ve & ARCES & early & 3 & N04 & 2006-10-26 & 2014-12-31 & 0 & \\ 
A07 & MWC 799 & 7.47 & B1IV:p? & 26 & early & 3 & N04 & 2006-10-26 & 2014-12-31 & 0 & \\ 
A09 & MWC 149 & 7.78 & B1Vnne & 27 & early & 3 & S05 & 2010-02-28 & 2015-04-09 & ? & \\ 
A11 & MWC 828 & 7.88 & B0.5Ve & 2 & early & 3 & J06 & 2010-03-02 & 2015-05-06 & 5 & \\ 
A12 & MWC 541 & 8.04 & B1.5IVe & 2 & early & 3 & J06 & 2010-03-02 & 2015-05-06 & 4 & \\ 
A15 & MWC 549 & 8.70 & B1Venp & ARCES + AO & early & 3 & J06 & 2010-03-02 & 2015-05-06 & 0 & \\ 
A16 & AS 367 & 8.96 & B3Ve & 28 & early & 3 & N11 & 2007-05-29 & 2013-06-13 & 4 & \\ 
A17 & MWC 998 & 8.19 & B6Ve & AO & mid & 3 & N11 & 2007-05-29 & 2013-06-13 & 0 & \\ 
A18 & MWC 362 & 8.02 & B5V & AO & mid & 7 & N24 & 2012-03-23 & 2014-12-30 & 0 & \\ 
A19 & MWC 640 & 7.21 & B1IIIe & ARCES & early & 3 & N12 & 2007-06-08 & 2013-06-14 & 0 & \\ 
A20 & MWC 370 & 7.64 & B1.5Vnpe & AO & early & 3 & N12 & 2007-06-08 & 2013-06-14 & ? & \\ 
A21 & MWC 649 & 8.70 & B3e & 29 & early & 3 & N24 & 2012-03-23 & 2014-12-30 & 0 & \\ 
A22 & AS 483 & 9.63 & B1.5V:nne: & 26 & early & 3 & N24 & 2012-03-23 & 2014-12-30 & 0 & \\ 
A24 & MWC 752 & 7.53 & B8Ve & ARCES & late & 5 & N04 & 2006-10-26 & 2014-12-31 & 0 & \\ 
A25 & MWC 753 & 9.58 & B6Ve & ARCES & mid & 5 & N04 & 2006-10-26 & 2014-12-31 & 0 & \\ 
A26 & MWC 109 & 7.85 & B1.0II/IIIe & ARCES + AO & early & 4 & N04 & 2006-10-26 & 2014-12-31 & 40 & \\ 
A27 & EM* CDS 496 & 8.67 & OB & 30 & unclassified & 4 & N04 & 2006-10-26 & 2014-12-31 & 0 & \\ 
A28 & MWC 786 & 8.08 & B2:V:nep & 31 & early & 3 & N04 & 2006-10-26 & 2014-12-31 & ? & \\ 
A29 & MWC 127 & 7.58 & B3Ve & ARCES & early & 3 & N04 & 2006-10-26 & 2014-12-31 & 6 & \\ 
A30 & MWC 128 & 7.36 & B2:Vnne & 25 & early & 3 & N04 & 2006-10-26 & 2014-12-31 & ? & \\ 
A31 & MWC 129 & 7.69 & B2Ve & ARCES & early & 3 & N04 & 2006-10-26 & 2014-12-31 & ? & \\ 
A32 & IGR J06074+2205 & 10.19 & B0.5Ve & 32 & early & 3 & N04 & 2006-10-26 & 2014-12-31 & 3 & \\ 
A34 & AS 118 & 7.64 & B1IIIe & ARCES & early & 3 & N04 & 2006-10-26 & 2014-12-31 & 8 & \\ 
Q01 & MWC 1016 & 7.09 & B0.2III & AO & early & 4 & N11 & 2007-05-29 & 2013-06-13 & 0 & \\ 
Q02 & Hen 3-1880 & 9.39 & B8 & 19 & late & 4 & N11 & 2007-05-29 & 2013-06-13 & 0 & \\ 
Q03 & BD+36 4032 & 7.57 & O8.5III & 33 & early & 4 & N11 & 2007-05-29 & 2013-06-13 & 0 & \\ 
Q05 & BD+00 1516 & 9.32 & B9 & 34 & late & 7 & S05 & 2010-02-28 & 2015-04-09 & 0 & \\ 
Q07 & VES 95 & 10.53 & B7IIIn & 35 & late & 3 & N11 & 2007-05-29 & 2013-06-13 & 0 & \\ 
Q08 & BD+21 4017 & 9.48 & B0 & 36 & early & 3 & N11 & 2007-05-29 & 2013-06-13 & 0 & \\ 
Q09 & MWC 1120 & 7.47 & O6.5nfp & 37 & early & 3 & N16 & 2012-05-21 & 2014-12-29 & 0 & \\ 
Q11 & MWC 670 & 9.52 & B9 & 20 & late & 13 & N16 & 2012-05-21 & 2014-12-29 & 0 & \\ 
Q13 & EM* CDS 144 & 10.50 & B & 21 & unclassified & 4 & N17 & 2012-09-17 & 2014-12-29 & 0 & \\ 
Q14 & EM* CDS 427 & 10.15 & B8 & 9 & late & 3 & N03 & 2006-10-25 & 2013-03-12 & 0 & \\ 
Q15 & MWC 475 & 8.25 & B3V & AO & early & 3 & N03 & 2006-10-25 & 2013-03-12 & 0 & \\ 
Q16 & EM* CDS 468 & 8.97 & B1V & AO & early & 9 & N04 & 2006-10-26 & 2014-12-31 & 0 & \\ 
Q17 & SS 20 & 7.75 & B5III & AO & mid & 7 & N03 & 2006-10-25 & 2013-03-12 & 2 & \\ 
Q18 & AS 128 & 9.61 & B5 & 34 & mid & 12 & S05 & 2010-02-28 & 2015-04-09 & 0 & \\ 
Q20 & EM* CDS 487 & 6.63 & O7.5(f)II & AO & early & 5 & N04 & 2006-10-26 & 2014-12-31 & 0 & \\ 
Q23 & EM* CDS 1299 & 10.24 & OB-e: & 30 & unclassified & 3 & N24 & 2012-03-23 & 2014-12-30 & 0 & \\ 
\hline\hline

\\\caption{\textbf{References.} (1) \citet{1952ApJ...115..459N}; (2) \citet{Fremat2006}; (3) \citet{1999MSS...C05....0H}; (4) \citet{2013yCat....1.2023S}; (5) \citet{1977ApJS...33..459S}; (6) \citet{2003AJ....125.2531R}; (7) \citet{1968ApJS...16..275M}; (8) \citet{1980BICDS..19...74O}; (9) \citet{1995A&AS..110..367N}; (10) \citet{1985cbvm.book.....V}; (11) \citet{2001KFNT...17..409K}; (12) \citet{1958TrRig...7...33A}; (13) \citet{2002A&A...386..709F}; (14) \citet{1979RA......9..479C}; (15) \citet{1988AJ.....95.1543B}; (16) \citet{1992ApJS...81..795G}; (17) \citet{1963ArA.....3...97R}; (18) \citet{1999A&AS..137..451G}; (19) \citet{1977PW&SO...2...71S}; (20) \citet{1949ApJ...110..387M}; (21) \citet{1959LS....C01....0H}; (22) \citet{2012A&A...541A..34S}; (23) \citet{1959ApJS....4....1M}; (24) \citet{2008ApJ...685.1157U}; (25) \citet{1968PASP...80..197G}; (26) \citet{1955ApJS....2...41M}; (27) \citet{1976ApJ...210...65T}; (28) \citet{1989AN....310..223R}; (29) \citet{1942ApJ....96...15M}; (30) \citet{1970MmRAS..73..153W}; (31) \citet{1977ApJ...217..127C}; (32) \citet{2010A&A...522A.107R}; (33) \citet{2004AN....325..380N}; (34) \citet{1949AnHar.112....1C}; (35) \citet{1993A&AS...97..755T}; (36) \citet{1950ApJ...111..495P}; (37) \citet{2010AJ....139.1283W}}
\end{longtable*}

\newpage

\bibliographystyle{apj}
\bibliography{Main.bbl}

\end{document}